\pdfoutput=1

\documentclass[11pt,twoside,a4paper,cmspaper,final,collab]{cms-tdr}
\def\svnVersion{c1e00c5-D}\def\svnDate{2020/03/10}\def\cmsCernNoTag{CERN-EP-2019-151}\def\cmsCernDate{\today}\def\cmsMessage{"Published in the Journal of High Energy Physics as \href{http://dx.doi.org/10.1007/JHEP03(2020)014}{\doi{10.1007/JHEP03(2020)014}.}"}

\begin{document}\cmsNoteHeader{FSQ-15-009}

\hyphenation{had-ron-i-za-tion}
\hyphenation{cal-or-i-me-ter}
\hyphenation{de-vices}
\RCS$HeadURL: svn+ssh://svn.cern.ch/reps/tdr2/papers/FSQ-15-009/trunk/FSQ-15-009.tex $
\RCS$Id: FSQ-15-009.tex 494914 2019-05-27 04:12:39Z padula $

\newlength\cmsFigWidth
\ifthenelse{\boolean{cms@external}}{\setlength\cmsFigWidth{0.85\columnwidth}}{\setlength\cmsFigWidth{0.4\textwidth}}
\ifthenelse{\boolean{cms@external}}{\providecommand{\cmsLeft}{top\xspace}}{\providecommand{\cmsLeft}{left\xspace}}
\ifthenelse{\boolean{cms@external}}{\providecommand{\cmsRight}{bottom\xspace}}{\providecommand{\cmsRight}{right\xspace}}

\cmsNoteHeader{FSQ-15-009}
\title{Bose--Einstein correlations of charged hadrons in proton-proton collisions at $\sqrt{s} = 13\TeV$}

\date{\today}
\abstract{
Bose--Einstein correlations of charged hadrons are measured over a broad multiplicity range, from a few particles up to about 250 reconstructed charged hadrons in proton-proton collisions at $\sqrt{s} = 13\TeV$. The results are based on data collected using the CMS detector at the LHC during runs with a special low-pileup configuration. Three analysis techniques with different degrees of dependence on simulations are used to remove the non-Bose--Einstein background from the correlation functions. All three methods
give consistent results. The measured lengths of homogeneity are studied as functions of particle multiplicity as well as average pair transverse momentum and mass. The results are compared with data from both CMS and ATLAS at  $\sqrt{s} = 7\TeV$, as well as with theoretical predictions.
}

\hypersetup{
pdfauthor={CMS Collaboration},
pdftitle={Bose-Einstein correlations in pp at 13 TeV},
pdfsubject={CMS},
pdfkeywords={CMS, physics, Bose-Einstein correlations}}

\maketitle

\newcommand{\qi}{\ensuremath{q_\text{inv}}\xspace}
\newcommand{\Ntroff}{N^\text{offline}_\text{trk}}
\newcommand{\Nt}{N_\text{tracks}}
\newcommand{\Ri}{\ensuremath{R_\text{inv}}\xspace}
\newcommand{\Rl}{\ensuremath{R_\text{long}}\xspace}
\newcommand{\Ro}{\ensuremath{R_\text{out}}\xspace}
\newcommand{\Rs}{\ensuremath{R_\text{side}}\xspace}
\newcommand{\pcs}{cluster subtraction}
\newcommand{\PPb}{\ensuremath{\Pp\mathrm{Pb}}\xspace}

\section{Introduction}
\label{sec:introduction}
Bose--Einstein correlations (BECs) are quantum statistical in nature and were used for several decades to probe the size and shape of the particle emitting region in high energy collisions~\cite{gglp,mlisa-review-ARNP}. These techniques are employed to characterize the size of the emission region at the freeze-out stage of the evolving system. Such studies have been performed by the CMS Collaboration at the CERN LHC in proton-proton ($\Pp\Pp$) collisions at $\sqrt{s} = 0.9\TeV$ \cite{cms-hbt-1st,cms-hbt-2nd}, 2.36\TeV \cite{cms-hbt-1st}, 2.76\TeV \cite{fsq-14-002}  and 7\TeV \cite{cms-hbt-2nd,fsq-14-002}. In these analyses, the one-dimensional (1D) correlations were measured in terms of the invariant relative momentum $q_{\text{inv}}^2 = -q^\mu q_\mu=-(k_1 - k_2)^2=m_{\text{inv}}^2 - 4 m_\pi^2$ where $k_i$ refers to the four-momentum of each particle of the pair. The pion mass ($m_\pi$) is assumed for all of the charged particles, since pions constitute the majority of the produced hadrons. Multi-dimensional analyses of the correlation functions in $\Pp\Pp$, proton--lead ($\PPb$), and lead--lead (PbPb) collisions were performed by CMS~\cite{fsq-14-002} to explore the size of the source in different directions. Similar studies were also performed by other experiments~\cite{afs-pion,afs-kaon,afs-direction,ua1,e735,phobos,Adams:2004yc,phenix,alice,atlas2015,lhcb}. This paper uses CMS data at $\sqrt{s} = 13\TeV$ to extend the investigation of one dimensional BECs with charged hadrons produced in $\Pp\Pp$ collisions to include both a higher center-of-mass energy and higher charged particle multiplicities (up to 250 particles).

Studies using $\Pp\Pp$ (and later $\PPb$) events with very high charged particle multiplicities led to the observation of ``ridge-like'' correlations (\ie, near-side
($\Delta\phi\sim 0$) long-range ($\abs{\Delta\eta} > 2$) anisotropic azimuthal correlations)~\cite{Ridge1,Ridge2,Ridge4,Ridge5,Ridge6,Ridge7} associated with collective flow. In nucleus--nucleus collisions, such structures can be parameterized by a Fourier expansion, providing information about the initial collision geometry and its fluctuations. In hydrodynamic models, initial state anisotropies are propagated to the final state via ultrarelativistic inviscid fluid evolution up to the freeze-out stage of the system. Additional measurements employing high multiplicity (HM) events in $\Pp\Pp$ and in $\PPb$ collisions at the LHC resulted in evidence of collective behavior \cite{CMS-pp-collectivity,Ridge3} even in such small colliding systems. Altogether, these results indicate that events with high multiplicity produced in $\Pp\Pp$ collisions exhibit some properties similar to those in relativistic heavy ion collisions. The origin of such phenomena in small systems is still under debate~\cite{Dusling:2015gta}, and BEC studies supply complementary information to shed light onto the origin of the observed similarities.

In $\Pp\Pp$ collisions, dynamical correlations in the kinematic region of interest for BEC studies can also arise from processes such as resonance decays and jets. In particular, for events with a small number of particles, the relative non-BEC contribution is enhanced. On the other hand, events in the high multiplicity range in $\Pp\Pp$ collisions are more likely than events with similar multiplicities in heavy ion collisions to be affected by multi-jets. Therefore, the importance of accurate removal of these background effects is enhanced for the correlations studied in the current investigation. To address this requirement, the analysis is performed with three techniques that differ from each other in their dependence on simulations.

Correlation functions are used to find the 1D radius fit parameter  ($\Ri$, also called the length of homogeneity~\cite{Makhlin:1987gm}),  and the intercept parameter ($\lambda$), corresponding to the intensity of the correlation function at $\qi = 0$. Fitted values of $\Ri$  and $\lambda$ are presented as functions of event multiplicity as well as average pair transverse momentum ($\kt = \frac{1}{2} \abs{\vec{p}_{\mathrm{T},1} + \vec{p}_{\mathrm{T},2}} $) and mass ($\mT = \sqrt{\smash[b]{m_{\pi}^2 + \kt^2}}$). The results are also compared to both previous experimental data and to theoretical predictions.

This paper is structured as follows: Sections~\ref{sec:detector}--\ref{sec:evt-track-sel} describe the experimental setup, the datasets and Monte Carlo (MC) simulations employed in the analysis, and the event and track selections, respectively. The generation, correction, and fitting procedures for the correlation functions, and the systematic uncertainties in those procedures, are detailed in Sections~\ref{sec:bec-analysis} and \ref{sec:systematics}, respectively. Results are presented in Section~\ref{sec:results}, including comparisons with results from $\Pp\Pp$ collisions at $\sqrt{s}=7\TeV$ and theoretical predictions, and the summary is given in section~\ref{sec:summary}. Finally, Appendix \ref{sec:appendix-DR-CS} gives additional details on the analysis techniques and Appendix \ref{sec:appendix-dip} describes the study of an anticorrelation that was previously seen at lower energies~\cite{cms-hbt-2nd,fsq-14-002}. This  anticorrelation is also investigated in $\Pp\Pp$ collisions at 13\TeV over the broad multiplicity range covered by this analysis.
In particular, the depth of the dip is nonzero for events with high multiplicity. A more detailed discussion is given in an appendix because this topic is outside the main physics thrust of this paper.

\section{The CMS detector}
\label{sec:detector}

The central feature of the CMS detector is a superconducting solenoid of 6 m internal
diameter. Within the solenoid volume are a silicon pixel and strip tracker, a
lead tungstate crystal electromagnetic calorimeter (ECAL, $\abs{\eta}<3$), and a
brass and scintillator hadron calorimeter (HCAL, $\abs{\eta}<3$), each composed
of a barrel and two endcap sections, where $\eta$ is the pseudorapidity. In addition to the barrel and endcap
detectors, quartz-fiber Cherenkov hadron forward (HF) calorimeters ($3<\abs{\eta}<5$)
complement the coverage provided by the barrel and endcap detectors on both
sides of the interaction point. These HF calorimeters are azimuthally subdivided
into $20^{\circ}$ modular wedges and further segmented to form
$0.175{\times}0.175$ $(\Delta \eta{\times}\Delta \phi)$ ``towers''. A muon system located outside the
solenoid and embedded in the steel flux-return yoke is used for the reconstruction
and identification of muons up to $\abs{\eta} = 2.4$.
The silicon tracker measures charged particles within the pseudorapidity range $\abs{\eta}<2.5$.
It consists of 1440 silicon pixel and 15\,148 silicon strip detector modules.
For nonisolated particles of $1<\pt<10\GeV$ and $\abs{\eta}<1.4$, the track resolutions are
typically 1.5\% in \pt and 25--90 (45--150)\mum in the transverse (longitudinal) impact parameter \cite{Chatrchyan:2014fea}.
The BPTX (Beam Pickup for Timing for the eXperiments) devices are used to trigger the detector readout.
They are located around the beam pipe at a distance of 175\unit{m} from the IP on either side,
and are designed to provide precise information on the LHC
bunch structure and timing of the incoming beams.
Events of interest are selected using a two-tiered trigger system~\cite{Khachatryan:2016bia}. The first level (L1), composed of custom hardware processors, uses information from the calorimeters and muon detectors to select events at a rate of around 100\unit{kHz} within a time interval of less than 4\mus. The second level, known as the high-level trigger (HLT), consists of a farm of processors running a version of the full event reconstruction software optimized for fast processing, and reduces the event rate to around 1\unit{kHz} before data storage. A detailed description of the CMS  detector, together with a definition of the coordinate system and kinematic variables, can be found in Ref.~\cite{Chatrchyan:2008zzk}.

\section{Data and simulated samples}
\label{subsec:data-MC}

This analysis uses $\Pp\Pp$ data at $\sqrt{s}=13\TeV$ collected at the LHC in 2015. The data were taken using
a special LHC configuration providing an average of 0.1 interactions per bunch crossing,
resulting in a very low probability of simultaneous $\Pp\Pp$ collisions (pileup).
The events were selected using minimum-bias (MB) and HM triggers, with samples corresponding to
total integrated luminosities (\lumi) of 0.35 and 459\nbinv, respectively.
The different luminosities of the MB and HM samples are due to different prescale factors applied to
the number of events that pass the selection criteria of the respective triggers.

The MB events are selected using  a trigger that requires at least one tower on either side of the HF to have a deposited energy above 1\GeV. This trigger mainly reduces effects from detector noise, beam backgrounds, and cosmic rays, while maintaining a high efficiency (greater than 99\% for reconstructed track multiplicities above 10, as estimated with simulated samples) for events coming from inelastic proton-proton collisions.

To increase the number of HM events, three triggers with different multiplicity thresholds are used. At L1, these triggers require the scalar sum of the transverse energy in the ECAL, HCAL, and HF towers to be larger than 40, 45, or 55\GeV. At the HLT stage, events  are selected by requiring track multiplicities larger than 60, 85, or 110, pre-selected by L1 at 40, 45, or 55\GeV, respectively.
In the HLT, tracks are reconstructed using pixel detector information. The low pileup configuration is critical in ensuring a high purity of single $\Pp\Pp$ collisions in the HM dataset.

Monte Carlo simulated event samples are generated using \PYTHIA 6.426~\cite{Sjostrand:2006za} and \PYTHIA 8.208 \cite{Sjostrand:2007gs} with tunes Z2*~\cite{Chatrchyan:2013gfi,Khachatryan:2015pea} and CUETP8M1~\cite{Khachatryan:2015pea}, respectively. For events with generated charged particle multiplicity above 95, \PYTHIA 8 simulations used the 4C~\cite{Corke2011} tune. The event generator \textsc{epos} 1.99  with the LHC tune (\textsc{epos lhc})~\cite{PhysRevC.92.034906} is also used, primarily for systematic uncertainty studies. Interactions of longer-lived unstable particles and the detector response is simulated using {\GEANTfour}~\cite{Agostinelli2003250}. The number of simulated events is 10--20 million for MB and 3--6 million for HM.

\section{Event and track selections}
\label{sec:evt-track-sel}

Events are selected offline by requiring all of the following conditions:

\begin{itemize}
\item At least one reconstructed vertex with a distance with respect to the center of the nominal interaction region of less than 15\cm in the longitudinal (along the beam) direction and of less than  0.15\cm transverse  to the beam.
\item Beam-related background suppression by rejecting events for which less than 25\%
of all reconstructed tracks pass the high-purity selection as defined in Ref. ~\cite{Chatrchyan:2014fea}.
\item Coincidence of at least one tower with total energy above 3\GeV in both of the HF calorimeters,
a criterion that selects primarily nondiffractive events.
\end{itemize}

Reconstructed tracks are required to have $\abs{\eta}<2.4$  and $\pt > 0.2\GeV$ as well as the following selections:

\begin{itemize}
\item $\abs{\sigma(\pt)/\pt} < 0.1$, where $\sigma(\pt)$ is the uncertainty in the $\pt$ measurement.
\item $\abs{d_{xy}/\sigma(d_{xy})} < 3.0$ and $\abs{d_{z}/\sigma(d_{z})} < 3.0$, where the transverse ($d_{xy}$) and longitudinal ($d_{z}$) distances are measured with respect to the primary vertex (defined as the one with the highest track multiplicity
in the event), while $\sigma(d_{xy})$ and $\sigma(d_{z})$ are the uncertainties in the $d_{xy}$ and $d_{z}$ measurements, respectively.
\end{itemize}

In addition, each track is required to have at least one valid hit in one of the pixel detector layers in order to reduce the contamination from processes such as electron pairs from photon conversion and tracks from decays of long-lived resonances.

When determining the reconstructed charged particle multiplicity of an event, slightly different track requirements than those listed above are imposed to be consistent with the criteria used by the HLT to determine this event multiplicity. The quantity $N^\text{offline}_\text{trk}$ includes tracks within $\abs{\eta} < 2.4$ with $\pt > 0.4\GeV$, selected without the requirement on the number of valid pixel detector hits.
Variable bin widths are used, from 3 to 10 units of multiplicity, depending on the value of $\Ntroff$.
The corresponding particle level multiplicity, $N_\text{tracks}$, is corrected for the acceptance and efficiency, as described below, and is used for comparisons with other experiments.

\begin{linenumbers}
For characterizing the performance of the track reconstruction, the following quantities have been checked using MC simulations:
(i) absolute efficiency (track selection and detector acceptance);
(ii) fraction of misreconstructed tracks; (iii) probability of reconstructing multiple tracks
from a single primary particle; (iv) fraction of nonprimary reconstructed tracks.
The total efficiency is almost constant at 80\% for the range $1< \pt <10\GeV$ and is above 50\% for all $\eta$ and $\pt$ ranges investigated. The misreconstructed track rate (\ie, the rate of reconstructed tracks that do not share at least 75\%
of their hits with any track at the generator level) tends to slightly increase in the lower (${\lesssim}1\GeV$) \pt region, but it
is always below 1\%. A similar \pt dependence is observed for the fraction of nonprimary reconstructed tracks,
but the rate is always below 2\%. The probability of reconstructing multiple tracks from a single primary particle
is of the order of $10^{-3}$ and is negligible compared to the other quantities. Using these estimates,
correction factors for each track in a given ($\eta$,$\pt$) bin are determined~\cite{Khachatryan:2015lha}.
\end{linenumbers}

\section{Bose--Einstein correlation analysis}
\label{sec:bec-analysis}

\subsection{Definitions of signal and background}

For each event, the sample containing the BEC signal is formed
by pairing all same-sign tracks  (\ie $++$ or $- -$ and denoted ``SS")
with $\abs{\eta} < 2.4$ and $\pt > 0.2\GeV$. Opposite sign pairs (\ie $+ -$ and denoted ``OS"), within the same
 $\abs{\eta}$ and  $\pt$ ranges, are used by two of the analysis methods for removing non-BEC contributions to the correlation functions.
The distributions in terms of
the relative momentum of the pair $q_{\text{inv}}$
are divided into bins of the reconstructed charged particle
multiplicity, $N^\text{offline}_\text{trk}$, and of $\kt$.

Although no particle identification is used, the contamination by particles other than charged pions is
expected to be small, since pions are the dominant hadron species in
the sample. For instance, the ratio of kaons to pions is about 12\%, and protons
to pions is roughly 6\% for 7\TeV $\Pp\Pp$ collisions~\cite{Chatrchyan:2012qb}; for $\Pp\Pp$ collisions at
13\TeV~\cite{PhysRevD.96.112003}, the ratios are 11--12\% and  5--6\%, respectively, depending on the track multiplicity range.
The unidentified kaons and protons contaminate the correlation function mainly in the
low-$q$ region, where the BEC effect is stronger
(the contribution of nonidentical particle pairs depletes the signal). 
This contamination may lead to a reduction of the intercept parameter $\lambda$ as shown in Fig. 6 of Ref.~\cite{fsq-14-002} for $\Pp\Pp$ collisions at 7 TeV, which compares analyses of the same data using correlation functions generated with pairs of identical charged pions or pairs of unidentified charged hadrons. In contrast, as also seen in Fig. 6 in Ref.~\cite{fsq-14-002}, the BEC radius parameter shows consistent results for the two analyses, and is therefore not significantly affected by the use of unidentified charged particles instead of identified pion pairs.  

Ideally, the background distribution (reference sample) should contain all the physics effects that are present in the signal
distribution, except for the BECs. This reference sample can be constructed in several ways, most commonly by mixing tracks from different events,
as in this analysis. The default reference sample (called $\eta$-mixing) is constructed by pairing SS tracks from different events using the same procedure as Refs.~\cite{cms-hbt-1st,cms-hbt-2nd}. Two events are mixed only if they have similar reconstructed charged particle multiplicity in each of three pseudorapidity ranges: $-2.4 < \eta < -0.8$, $-0.8 < \eta < 0.8$, and $0.8 < \eta < 2.4$. For determining this matching criterion, a weight is assigned to each track of the event, depending on the $\eta$ range in which it occurs, and these weights are summed for each event. Then, the events are ordered according to this sum and the mixing is done by selecting two adjacent events in the ordered list and pairing each track in one event with all of the tracks in the other one. After being combined in this way, both events are discarded and not included in other pairings.

After choosing the reference sample, a correlation function is defined as a single ratio (SR) having the signal distribution, \ie, the $\qi$ distribution of pairs from the same event as the numerator, and the reference distribution as the denominator:
\begin{linenomath*}
\begin{equation}
  \text{SR} (\qi) \equiv C_2 (\qi) =
  \left(\frac{\mathcal{N}_{\text{ref}}}{\mathcal{N}_{\text{sig}}}\right)
  \;
  \left(\frac{\rd N(\qi)_\text{sig}/\rd \qi}
             {\rd N(\qi)_\text{ref}/\rd \qi}\right),
 \label{eq:1-d-singleratios-gen}
\end{equation}
\end{linenomath*}
\noindent where $C_2 (\qi)$ refers to the two-particle correlation defined in Eq.~(\ref{eq:1-d-singleratios-gen}) by a SR, $\mathcal{N}_{\text{sig}}$ and $\mathcal{N}_{\text{ref}}$ correspond to the number of pairs estimated by the value of the integral of the signal and reference distributions, respectively. Refinements of this definition are presented in Section~\ref{subsec:analysistech} and in Appendix~\ref{sec:appendix-DR-CS}.

\subsection{Coulomb interactions and correction}
\label{subsec:coulomb-effects}

The correlation functions include the effect of the quantum statistics obeyed by the pair of identical particles, but are also sensitive to final-state interactions  between the charged hadrons.
The Coulomb final-state interaction~\cite{Sinyukov:1998fc} affects the two-particle relative momentum distribution in different ways for SS or  OS
pairs, creating a depletion (enhancement) in the low $\qi$ range of the correlation function caused by repulsion (attraction) for SS (OS) pairs.
The effect of the mutual Coulomb interaction is incorporated in the factor $K$, the squared average of the relative wave function $\Psi$, as
$K(\qi) = \int \rd^3\vec{r} \; f(\vec{r}) \; \abs{\Psi(\vec{k},\vec{r})}^2$,
where $f(\vec{r})$ is the source intensity of emitted particles, with $\vec{r}$ and $\vec{k}$ representing the pair relative separation and relative momentum, respectively~\cite{fsq-14-002}.
For point-like sources, $f(\vec{r}) = \delta(\vec{r})$ and
the integral gives the Gamow factor, which in the
case of SS and OS pairs is given by:
\begin{linenomath*}
\begin{equation}
\begin{aligned}
G^\text{SS}_w(\zeta) &= \abs{\Psi^\text{SS}(0)}^2
                        = \frac{2\pi\zeta}{\re^{2\pi\zeta}-1}, &
G^\text{OS}_w(\zeta) &= \abs{\Psi^\text{OS}(0)}^2
                        = \frac{2\pi\zeta}{1-\re^{-2\pi\zeta}},
\label{eq:gamow}
\end{aligned}
\end{equation}
\end{linenomath*}
\noindent where $\zeta=\alpha m/\qi$ is the Landau parameter, $\alpha$ is the
fine-structure constant, and $m$ is the particle mass~\cite{GyuKaufWil:1979}.

In a previous CMS analysis~\cite{fsq-14-002}, no significant differences in the final results were observed in the case of pions
when correcting with the Gamow factor or with the full estimate derived for extended (as opposed to point-like) sources~\cite{{PhysRevD.33.72,Biyajima:1994br,Bowler:1991vx,Sinyukov:1998fc}}. Therefore, in this analysis the corrections for the final state Coulomb interaction are performed using the Gamow  factor.

\subsection{Fitting the correlation function}
\label{sub:fitting_functions}

Ideally, as in the case of static systems, the two-particle correlation function can be related to the Fourier transform of the emitting source distribution at decoupling. Because of their simplicity, parameterizations of the Gaussian type have been used
to relate the source distribution and the measured correlation function. In Ref.~\cite{cms-hbt-1st,cms-hbt-2nd,fsq-14-002}, the Gaussian distribution was studied and
yielded fit results much worse than for an exponential function
or the L\'evy  class of parameterizations.

In this analysis, the fits performed to the data correlation functions employ symmetric L\'evy stable distributions,
\begin{linenomath*}
  \begin{equation}
   C_{2, \text{BE}} (\qi) = C [1 + \lambda \re^{ - (\qi R_{\text{inv}})^a } ]  \; ( 1 + \epsilon \;\qi ),
    \label{eq:1d-levy}
    \end{equation}
\end{linenomath*}
where $C_{2, \text{BE}} (\qi)$ refers to the two-particle BEC, $C$ is a constant, $\Ri$ and $\lambda$ are the (BEC) radius and intercept parameters, respectively.
The exponent  $a$ is the L\'evy index of stability satisfying the inequality $0 < a \le 2$. If treated as a free parameter when fitting the correlation functions, this exponent usually returns a number between the value characterizing an exponential function ($a=1$) and that for a Gaussian distribution ($a = 2$).
More details can be found in Ref. \cite{csorgohegyizajc}.  The additional term, linear in $\qi$ (proportional to the constant $\epsilon$), is introduced to account for long-range correlations that may  be  absent in the reference sample. The fit values for $\epsilon$ depend on the multiplicity range$\colon$ negligible for high multiplicity bins (above 100 tracks), and reaching ${\sim}0.2\GeV^{-1}$ for low multiplicity bins (below 20 tracks).

The  L\'evy distribution with $a$ as a free parameter returns the best quality fits, but it is not adopted for extracting the results because it does not allow a direct interpretation of the shape of the source distribution by means of a Fourier transform.
Fitting with a pure Gaussian distribution ($a = 2$) returns very poor quality fits for all of the multiplicity and \kt bins.
Therefore, an exponential form (with $a$ fixed at 1.0) is used for the default fit function. For this parameterization, the functional form for the correlation function, $ \re^{- \qi R_{\text{inv}}}$, is the Fourier transform of a source function $\rho(t,\vec{x})$ corresponding to a Cauchy--Lorentz distribution. A Laguerre-extended exponential fit function \cite{csorgohegyi} returns a $\chi^2/\mathrm{dof}$ of the order of unity (where $\mathrm{dof}$ is the number of degrees of freedom) and yields the same $\Ri$ values as the simple exponential case, with the caveat that the resulting BEC fits depend on additional parameters from the Laguerre polynomial expansion.

A $\chi^2$ test is used in the fitting procedure.
A negative log-likelihood ratio, assuming that the bin content of the signal and  that of the reference sample
histograms are Poissonian distributions
(as implemented in Refs.~\cite{Ahle:2002mi,ATLAS-paper2017}) returned
results consistent with the $\chi^2$ approach.

\subsection{Analysis techniques}
\label{subsec:analysistech}

As discussed in the Introduction, both low and high multiplicity correlation functions in $\Pp\Pp$ collisions are particularly sensitive to the influence of non-BEC effects such as resonance decays and jets. To ensure an accurate accounting of these background contributions, and especially to investigate any variability of the final results, the background removal is performed with three techniques that have different degrees of dependence on simulations. Since two of these methods were used in previous CMS BEC studies ~\cite{cms-hbt-1st,cms-hbt-2nd,fsq-14-002}, and are described in detail in Ref.~\cite{fsq-14-002}, they are only briefly summarized here. Additional details can be found in Appendix~\ref{sec:appendix-DR-CS}.

For the double-ratio (DR) technique~\cite{cms-hbt-1st,cms-hbt-2nd, fsq-14-002}, the numerator is an SR as in  Eq.~(\ref{eq:1-d-singleratios-gen}) applied to the data, and the denominator is an SR computed similarly with MC events simulated without BEC effects. In each case, the reference samples for data and simulation are constructed in the same way by considering all charged particles instead of only charged pions in the MC. The DR correlation function is fitted using Eq.~(\ref{eq:1d-levy}) to obtain $\Ri$ and $\lambda$. This procedure reduces any bias due to the construction of the reference sample, with the advantage of directly removing non-BEC effects remaining in the data SR. However, it requires the use of well-tuned MC simulations to describe the overall behavior of the data.

The \pcs\ (CS)~\cite{fsq-14-002} technique uses only SRs from data.  Correlation functions for OS pairs are used for parameterizing the contamination from resonances and jet fragmentation (called ``cluster contribution''), which are still present in the correlation function~\cite{Alver:2008aa,Ridge1,Ridge4}. The amount of these contributions that is present in the SS pairs is evaluated by using the same shape found for OS pairs and varying the overall scale to fit the data (details are given in Ref.~\cite{fsq-14-002}). To find $\Ri$ and $\lambda$, the correlation function is fitted with a function combining the cluster and Bose--Einstein contributions, with the latter parameterized using Eq.~(\ref{eq:1d-levy}).

The hybrid CS (HCS) method, is similar to a method used for $\PPb$ data by the ATLAS experiment~\cite{ATLAS-paper2017}, has less dependence on MC simulations than the DR method and a smaller number of fit parameters to be adjusted to data than the CS method. In contrast to the CS procedure, which is fully based on the control samples  in data, the HCS technique uses MC simulations for converting the fit parameters from OS single ratios into those for SS.

\begin{figure}[hptb]
\centering
  \includegraphics[width=0.45\textwidth]{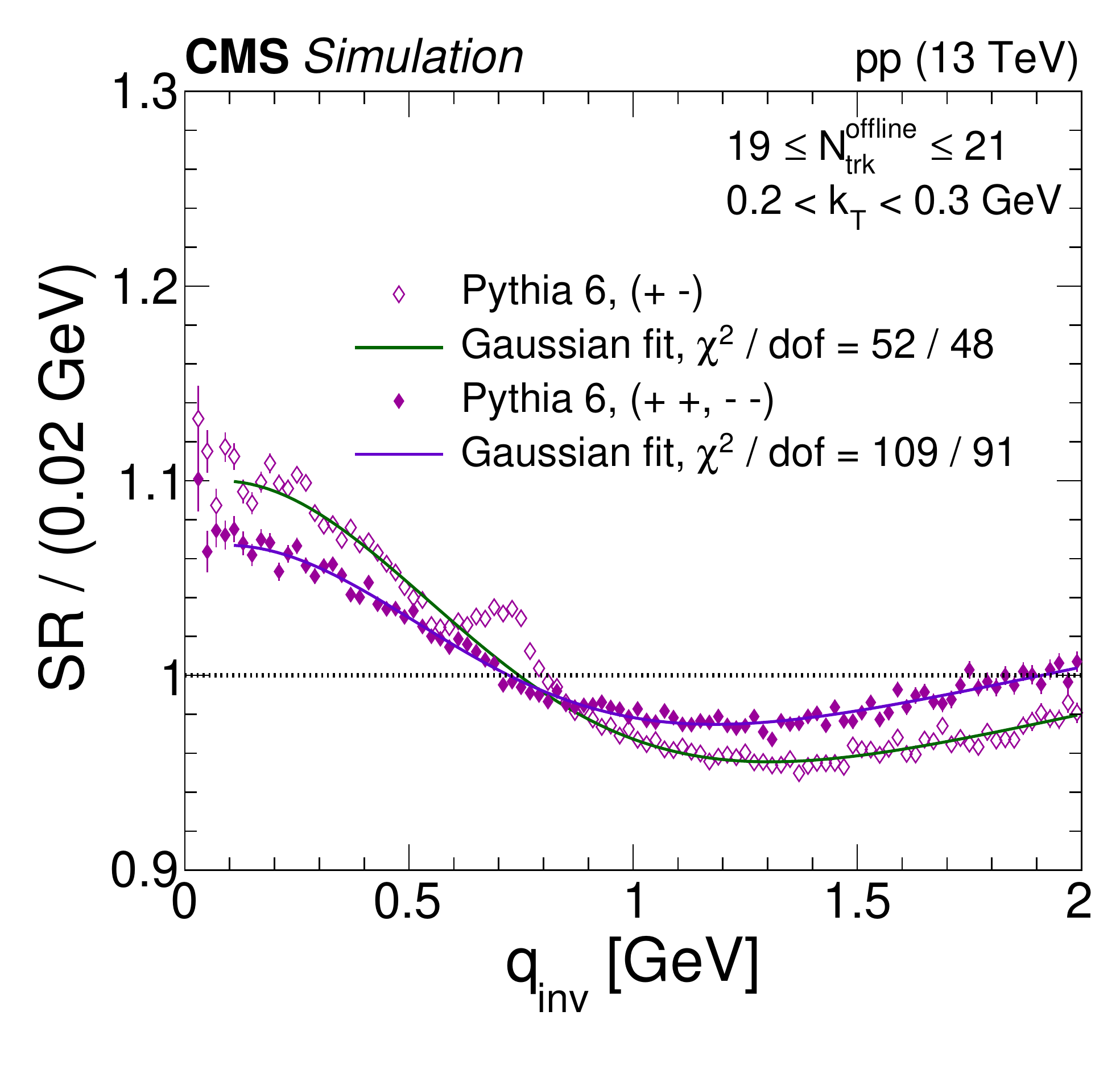}
  \includegraphics[width=0.45\textwidth]{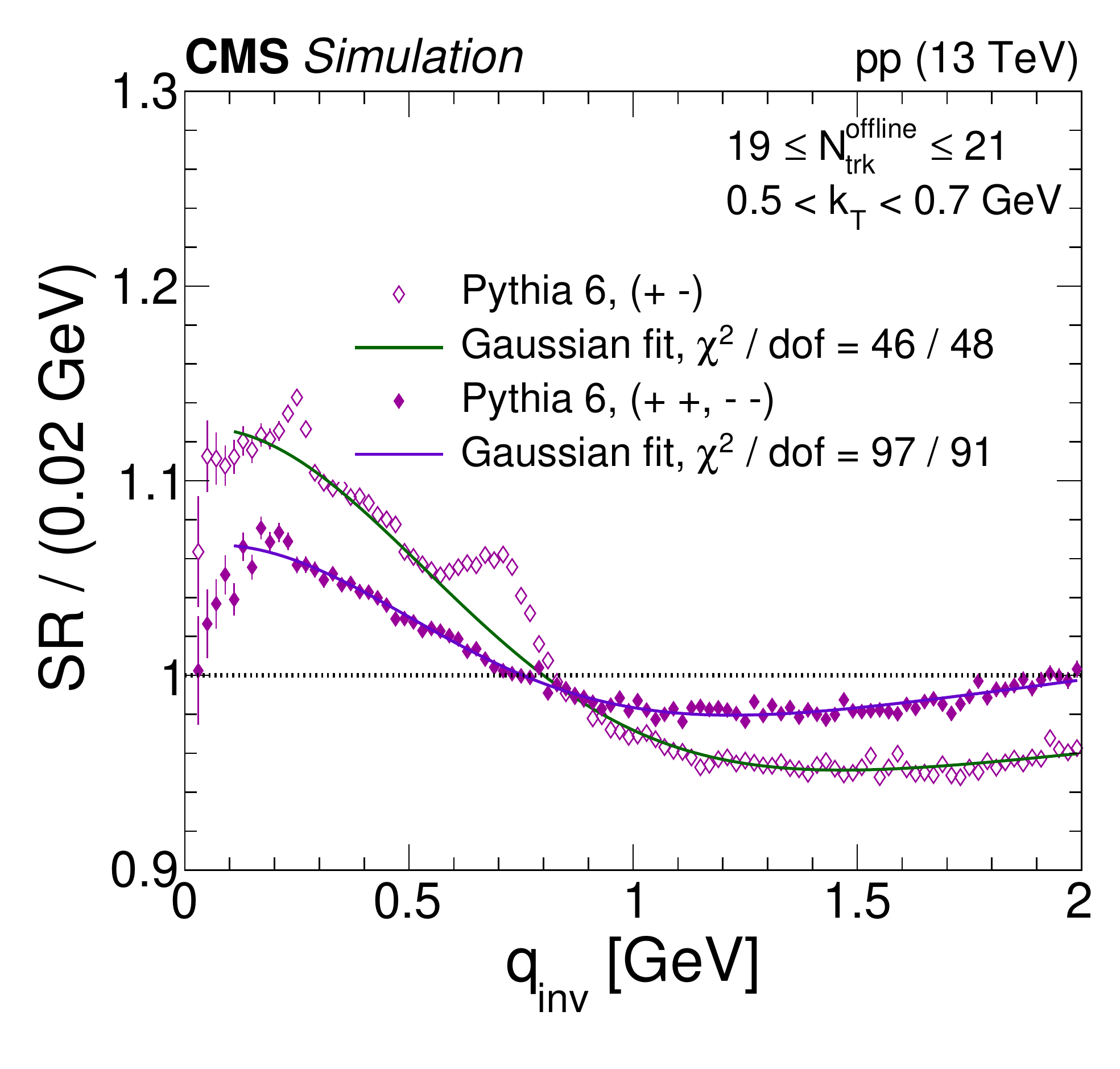} \\
  \includegraphics[width=0.45\textwidth]{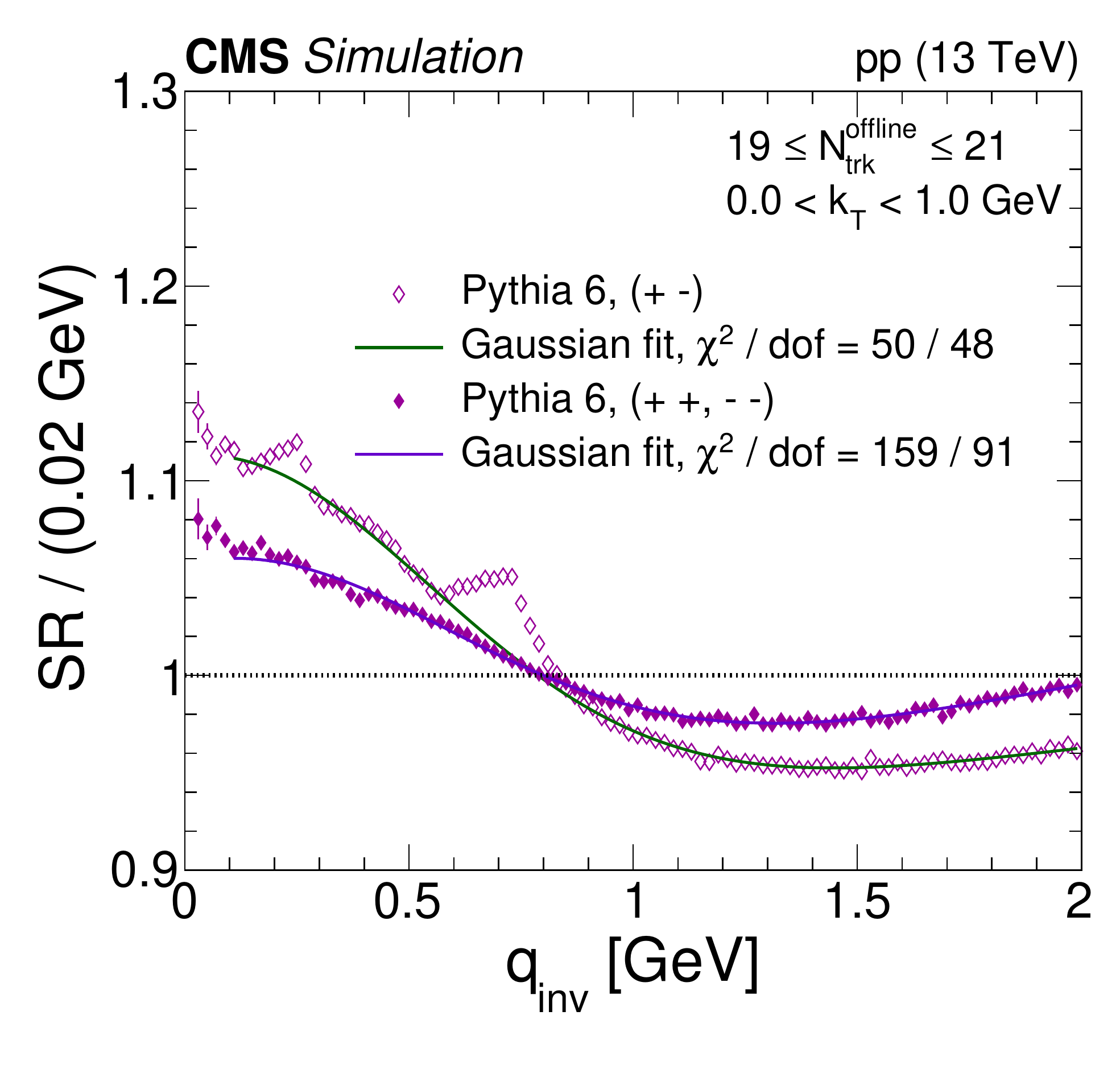}
  \includegraphics[width=0.45\textwidth]{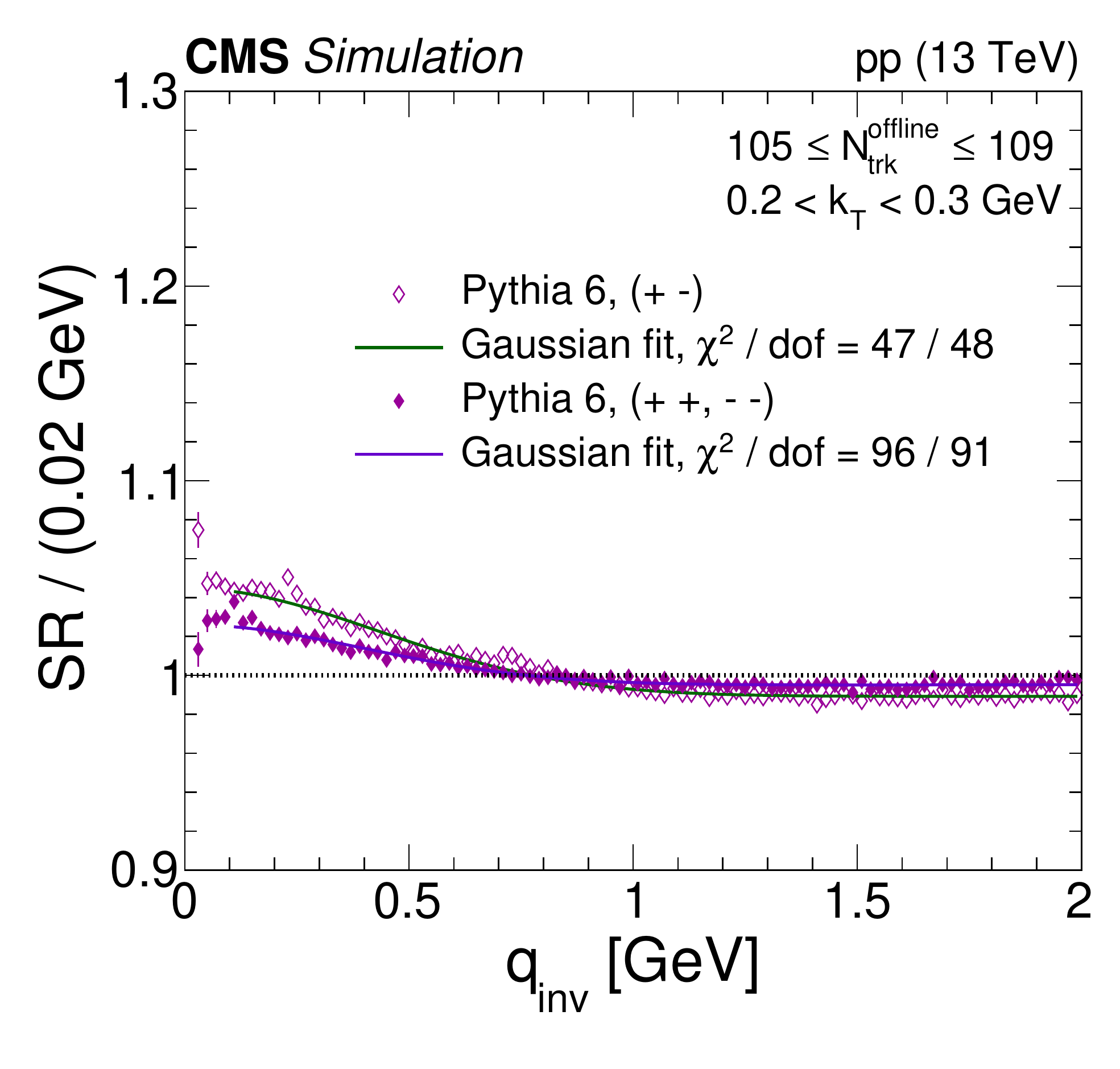} \\
  \includegraphics[width=0.45\textwidth]{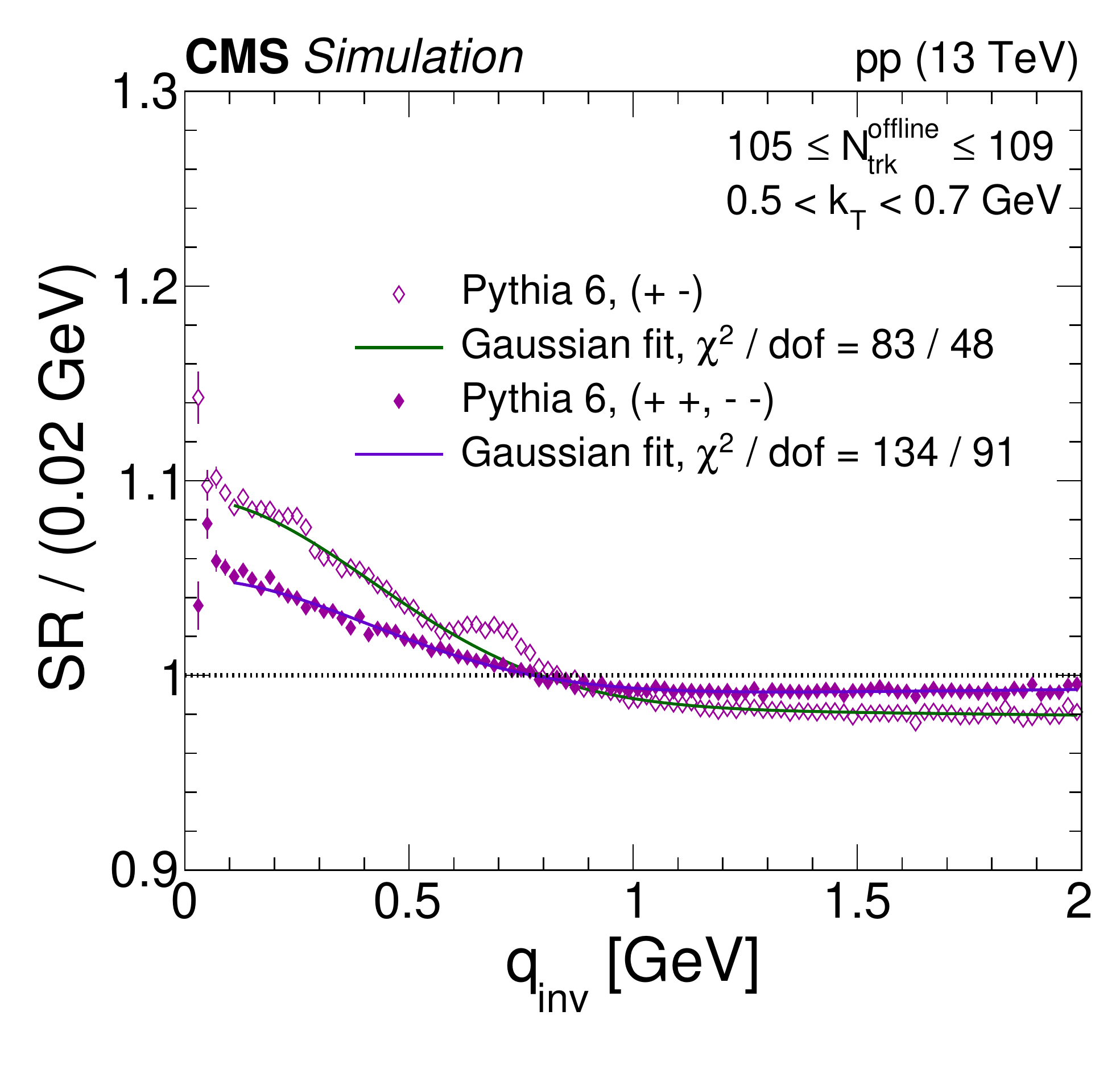}
  \includegraphics[width=0.45\textwidth]{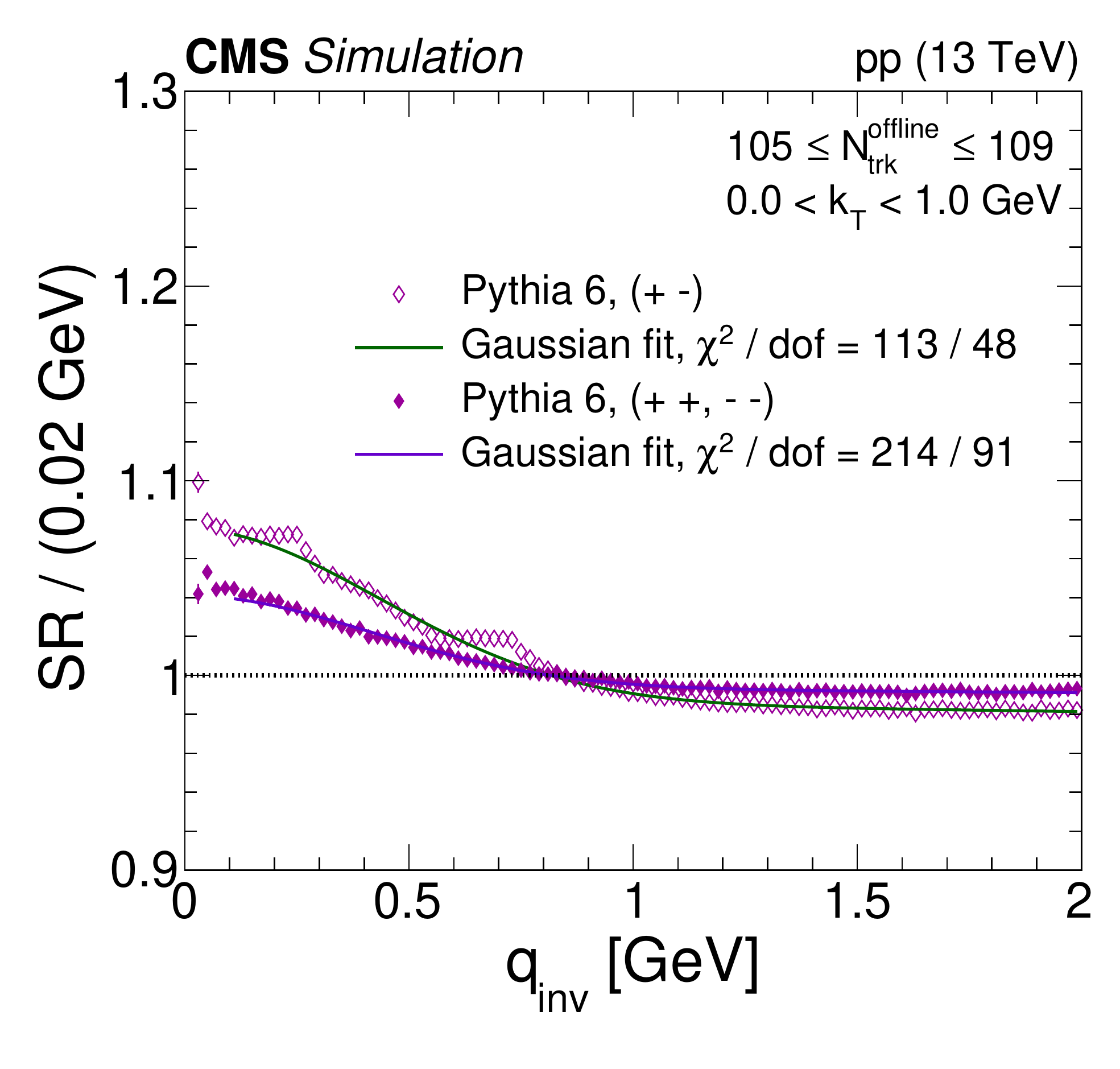}
 \caption{The same-sign ($++,--$) and opposite-sign ($+-$) single ratios employing \PYTHIA 6 (Z2* tune) in different bins of $\Ntroff$ and $\kt$ with the respective Gaussian fit from Eq.~(\ref{eq:bkg_form_ATLASmethod}). The following $\qi$ ranges are excluded from the fits to the opposite sign single ratios: $0.2 < \qi < 0.3$, $0.4 < \qi < 0.9$, and $0.95 < \qi < 1.2\GeV$. Coulomb interactions are not included in the simulation. The error bars represent statistical uncertainties and in most cases are smaller than the marker size.
   }
 \label{fig:py6_os_ss_sr_ATLASmethod}
\end{figure}

\begin{figure}[hptb]
  \centering
    \includegraphics[width=0.45\textwidth]{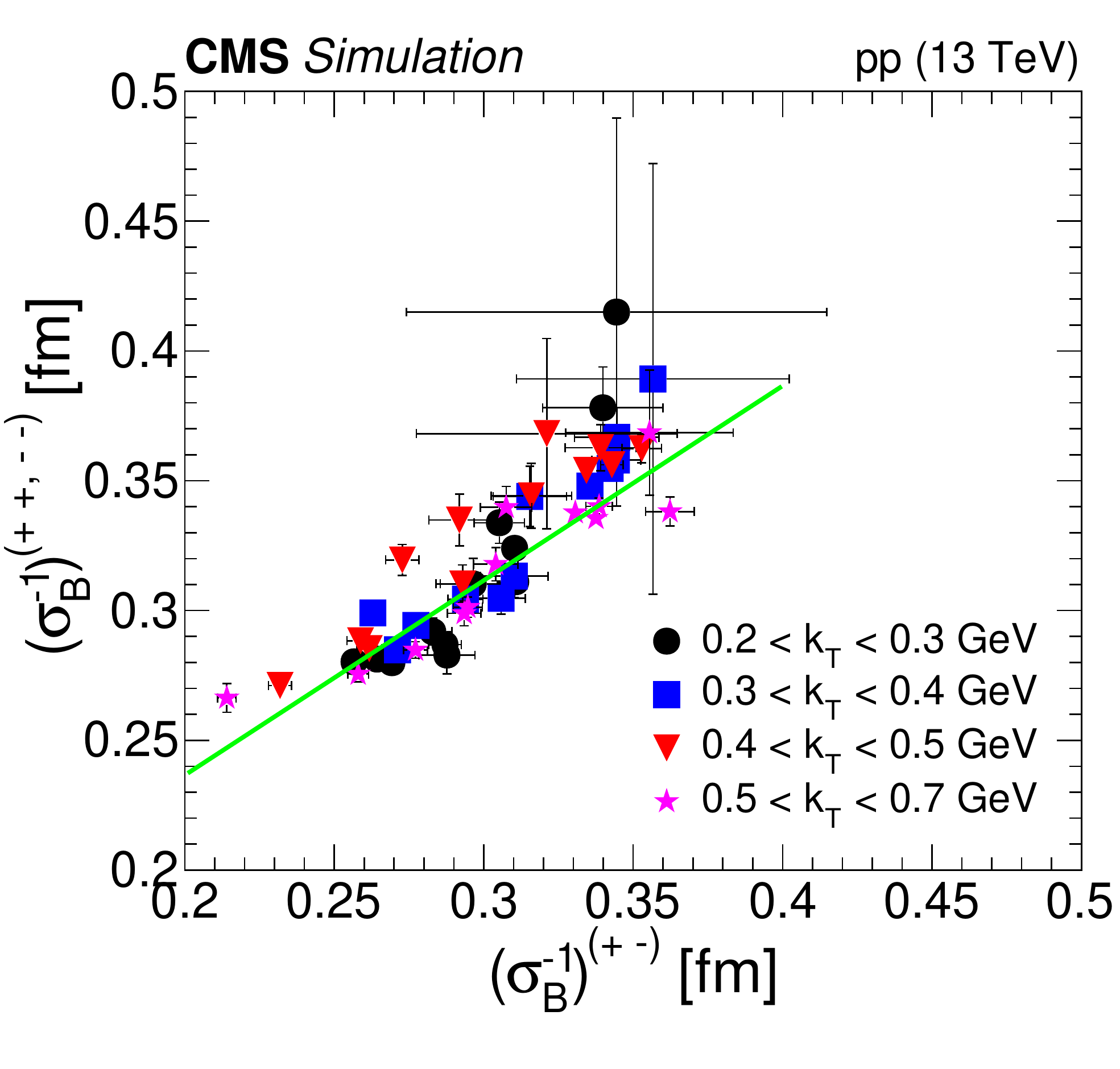}
    \includegraphics[width=0.45\textwidth]{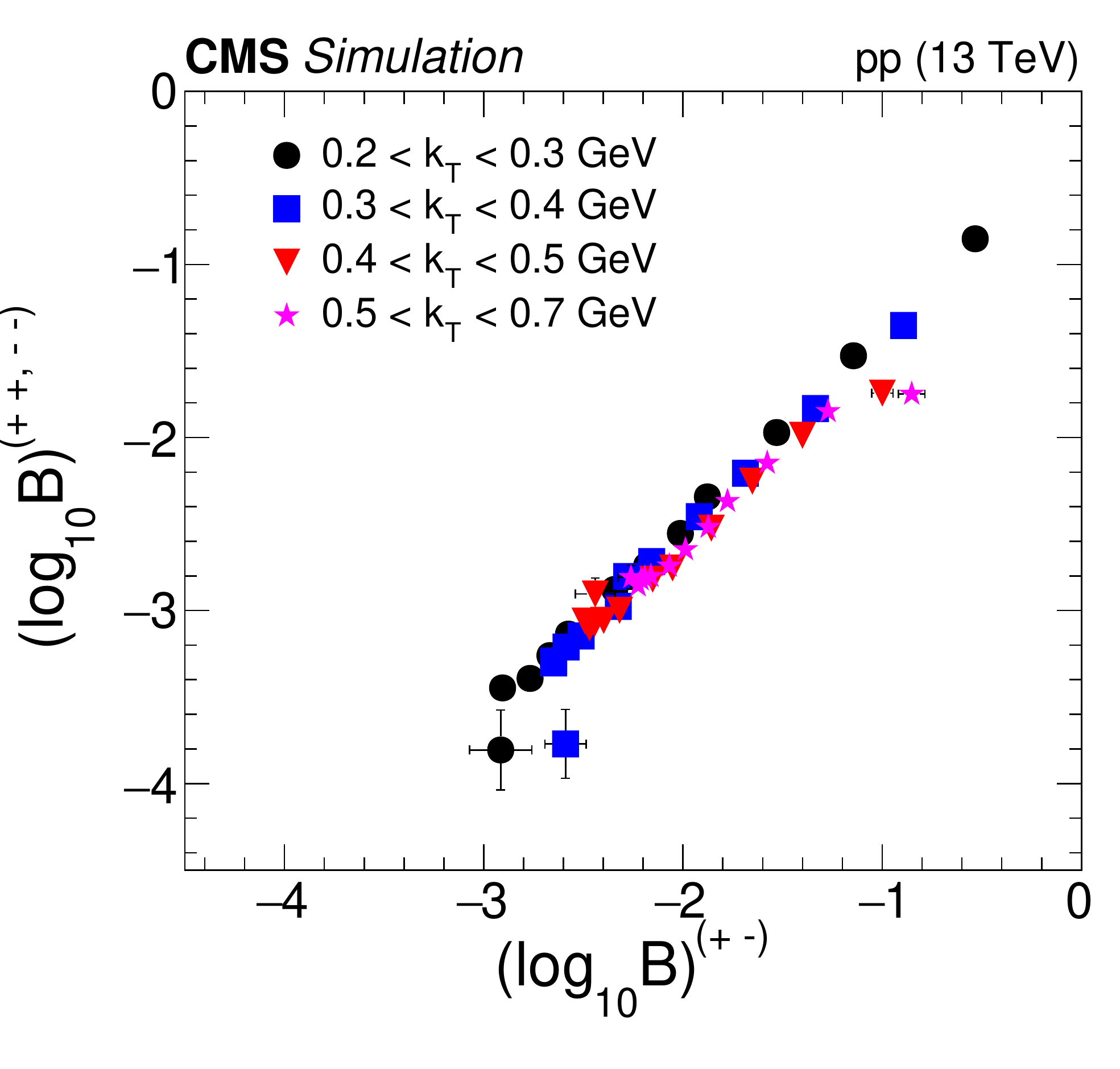}
  \caption{Relations between same-sign ($++,--$) and opposite-sign ($+-$) fit parameters from Eq.~(\ref{eq:bkg_form_ATLASmethod}), as a function of $\kt$ and $\Ntroff$ for events in MB (\ie, higher $\sigma_{\text{B}}^{-1}$ and lower $\text{log}_{10}\text{B}$) and HM (\ie, lower $\sigma_{\text{B}}^{-1}$ and higher $\text{log}_{10}\text{B}$) ranges. The fit values found for the parameters corresponding to the peak's width (left) and the amplitude (right) of the same-sign and opposite-sign correlations are shown. For a given $\kt$ range, each point represents an $\Ntroff$ bin. The line in the left plot is a linear fit to all the data. The error bars represent statistical uncertainties.
           }
  \label{fig:sigmaB_relation_ATLASmethod}
\end{figure}

The first step is to fit the SR constructed using MC simulations separately for SS and OS track pairs
using:
\begin{linenomath*}
\begin{equation}
\Omega(\qi) = \mathcal{N}\left[1 + B \exp{\left(-\left|\frac{\qi}{\sigma_{B}}\right|^{\alpha_{B}}\right)}\right],
\label{eq:bkg_form_ATLASmethod}
\end{equation}
\end{linenomath*}
where $B$ and $\sigma_{B}$ are fit parameters used to describe the  amplitude and the width of the peak near $\qi = 0$
and $\alpha_{B}$ defines the overall shape of the function. For OS pairs, the following regions in $\qi$ are excluded from the fitting process due to
the contamination of resonance decays: 0.2--0.3\GeV,
0.4--0.9\GeV,
and 0.95--1.2\GeV.
In addition, the range  $\qi < 0.1\GeV$ is excluded because this region has a large contribution from three-body decays.
A Gaussian shape ($\alpha_{B} = 2$) provides a reasonable overall
description of the distributions in $\Ntroff$ and \kt bins.
The $\chi^2/\mathrm{dof}$ values are, in general, not compatible with unity. This is expected because of the distorted
shape of the distributions, which are dependent on the MC simulation.
Examples of SRs using \PYTHIA 6 (Z2* tune) are shown in Fig.~\ref{fig:py6_os_ss_sr_ATLASmethod}.
Two conversion functions relate the amplitudes and the widths of the fits to SS ($++$ and $--$) to those found for OS ($+-$) MC correlations:
\begin{linenomath*}
\begin{equation}
\left[(\sigma_{B})^{-1}\right]^{(++,--)}=\rho\left[(\sigma_{B})^{-1}\right]^{(+-)} + \beta,
\label{eq:sigmaBArelation_ATLASmethod}
\end{equation}
\end{linenomath*}
\begin{linenomath*}
\begin{equation}
B^{(++,--)} = \mu(\kt)\left[B^{+-}\right]^{\nu(\kt)}.
\label{eq:BArelation_ATLASmethod}
\end{equation}
\end{linenomath*}
The parameters found when fitting the conversion function for the widths, $\rho = 0.82 \pm 0.04$ and $\beta = 0.077 \pm 0.013$, are independent of $\kt$ for the \PYTHIA 6 (Z2* tune),  whereas the parameters relating the amplitudes, $\mu$ and $\nu$, are functions of $\kt$. In Fig.~\ref{fig:sigmaB_relation_ATLASmethod}, the relations between the fit parameters for different bins in $\kt$ and $\Ntroff$ are shown for MB and HM events (for a given \kt range, each point in Fig.~\ref{fig:sigmaB_relation_ATLASmethod} represents an
$\Ntroff$ bin).

\begin{figure}[hptb]
\centering
   \includegraphics[width=0.45\textwidth]{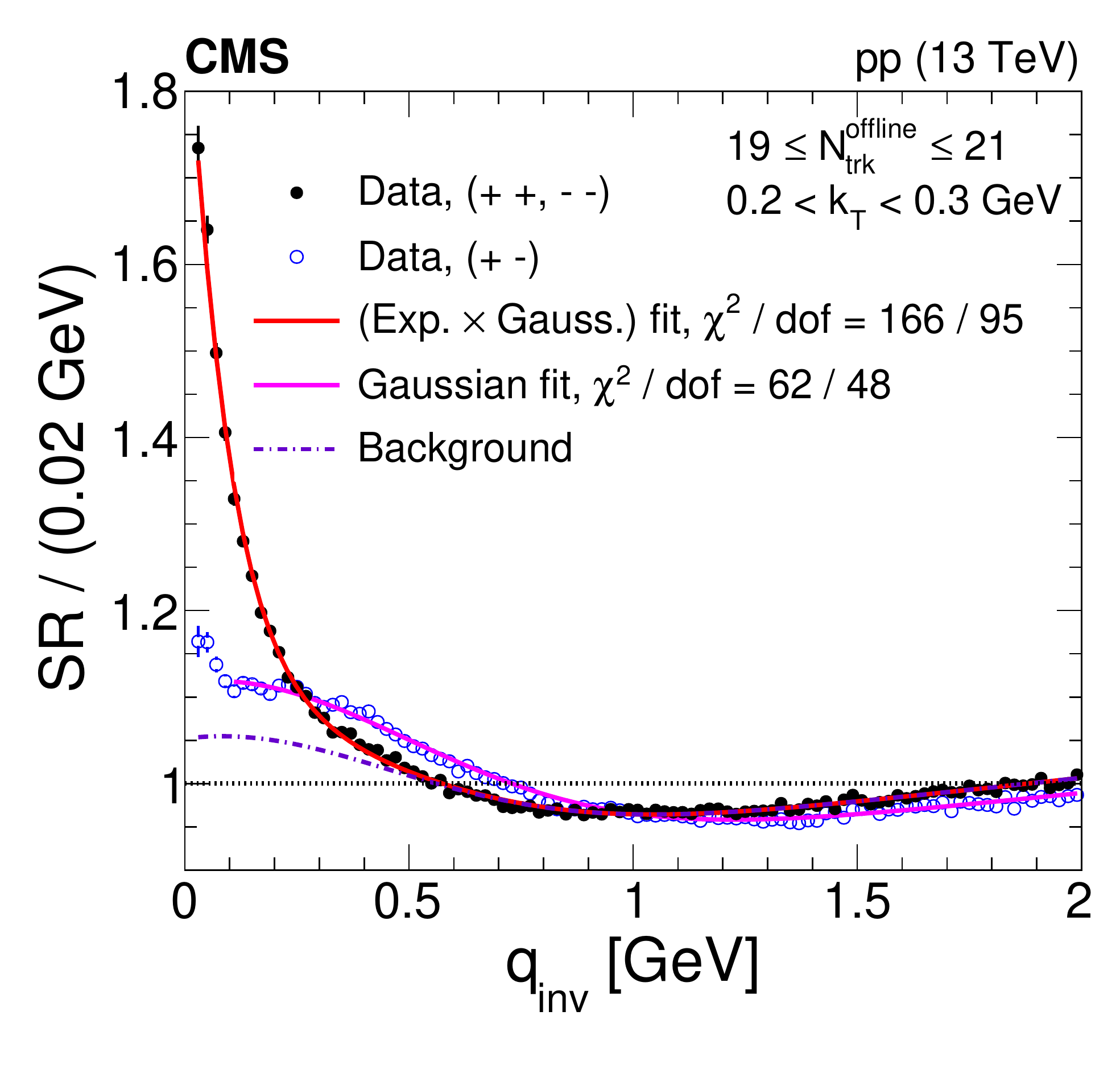}
   \includegraphics[width=0.45\textwidth]{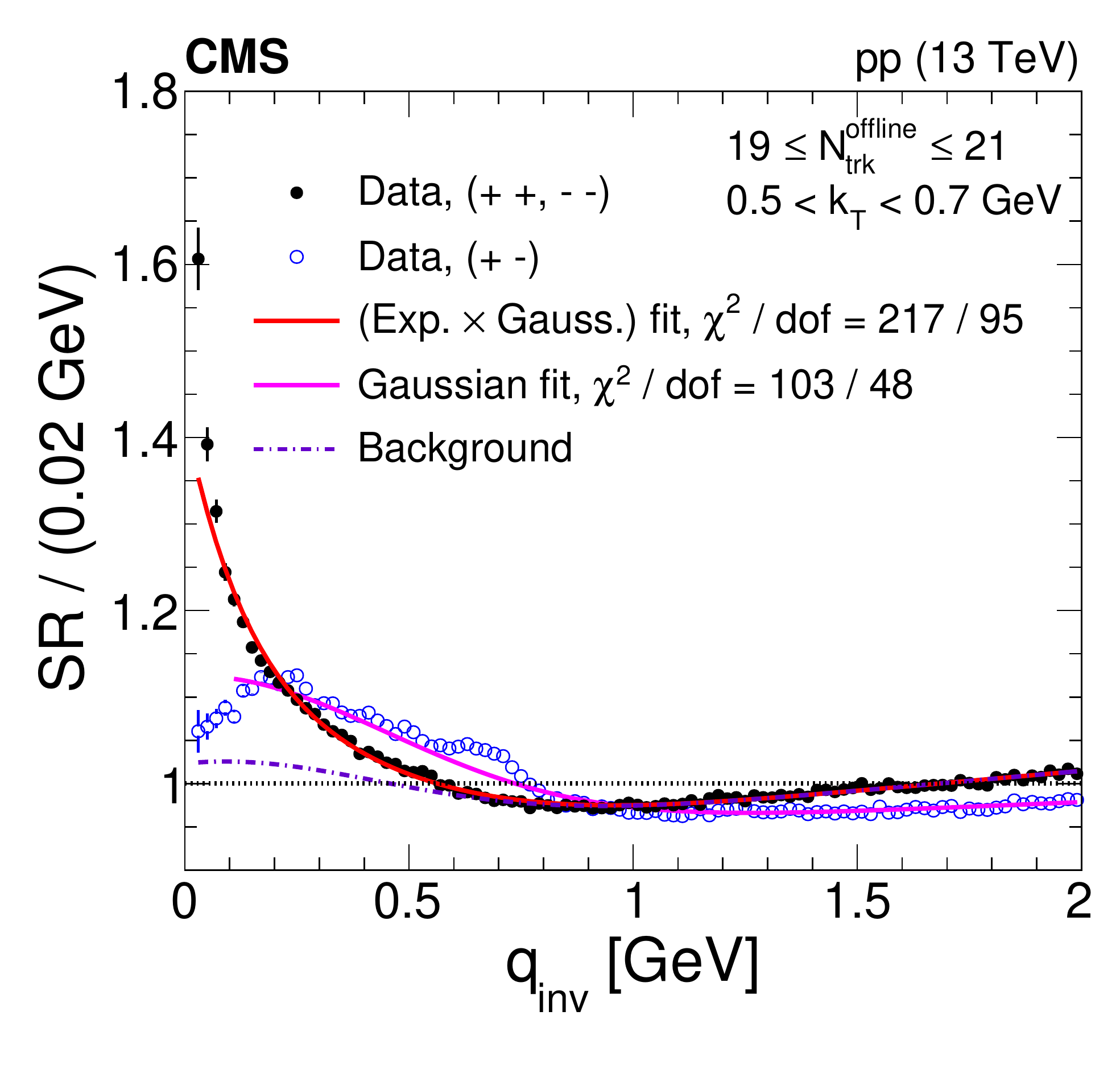} \\
   \includegraphics[width=0.45\textwidth]{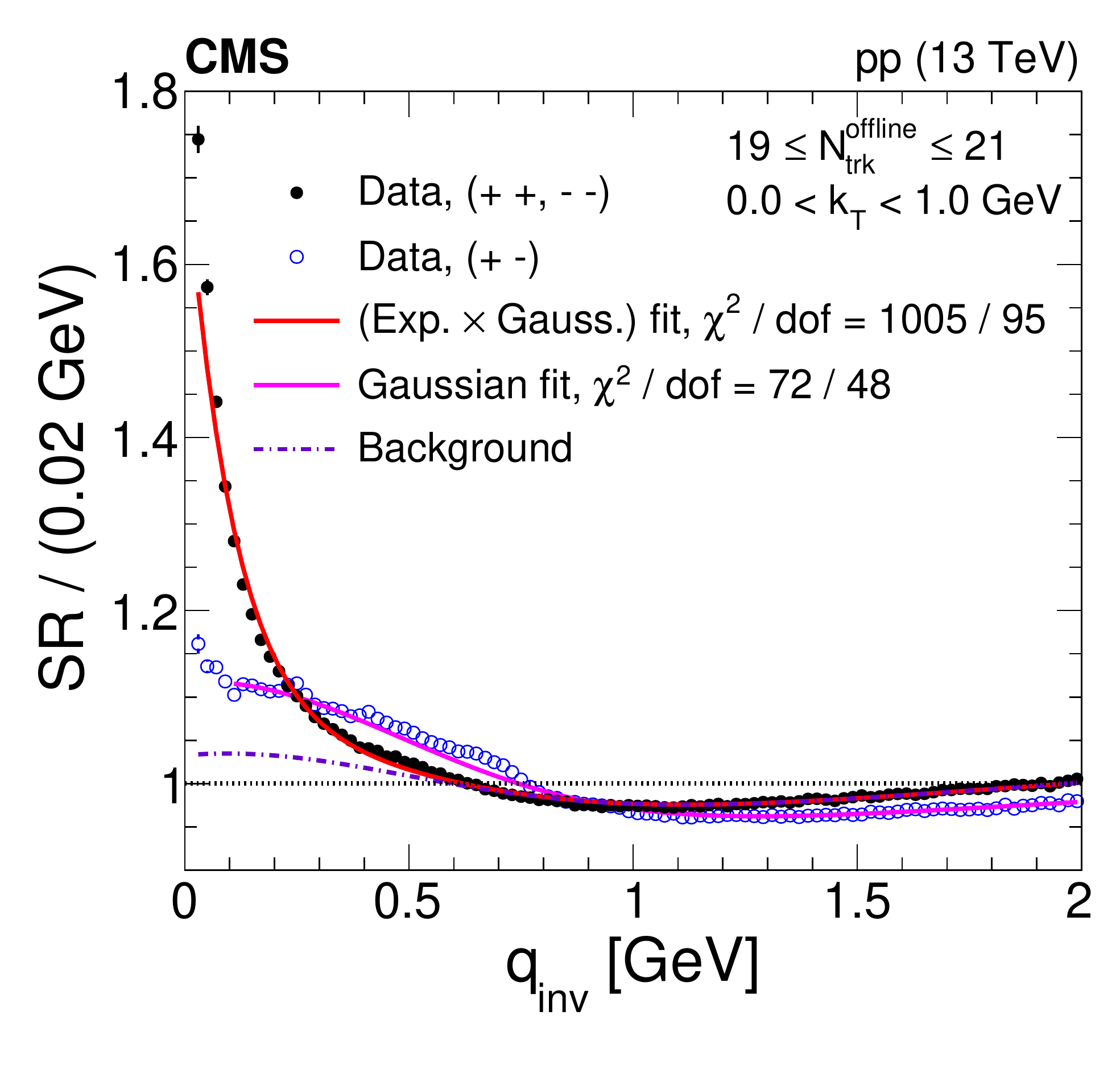}
   \includegraphics[width=0.45\textwidth]{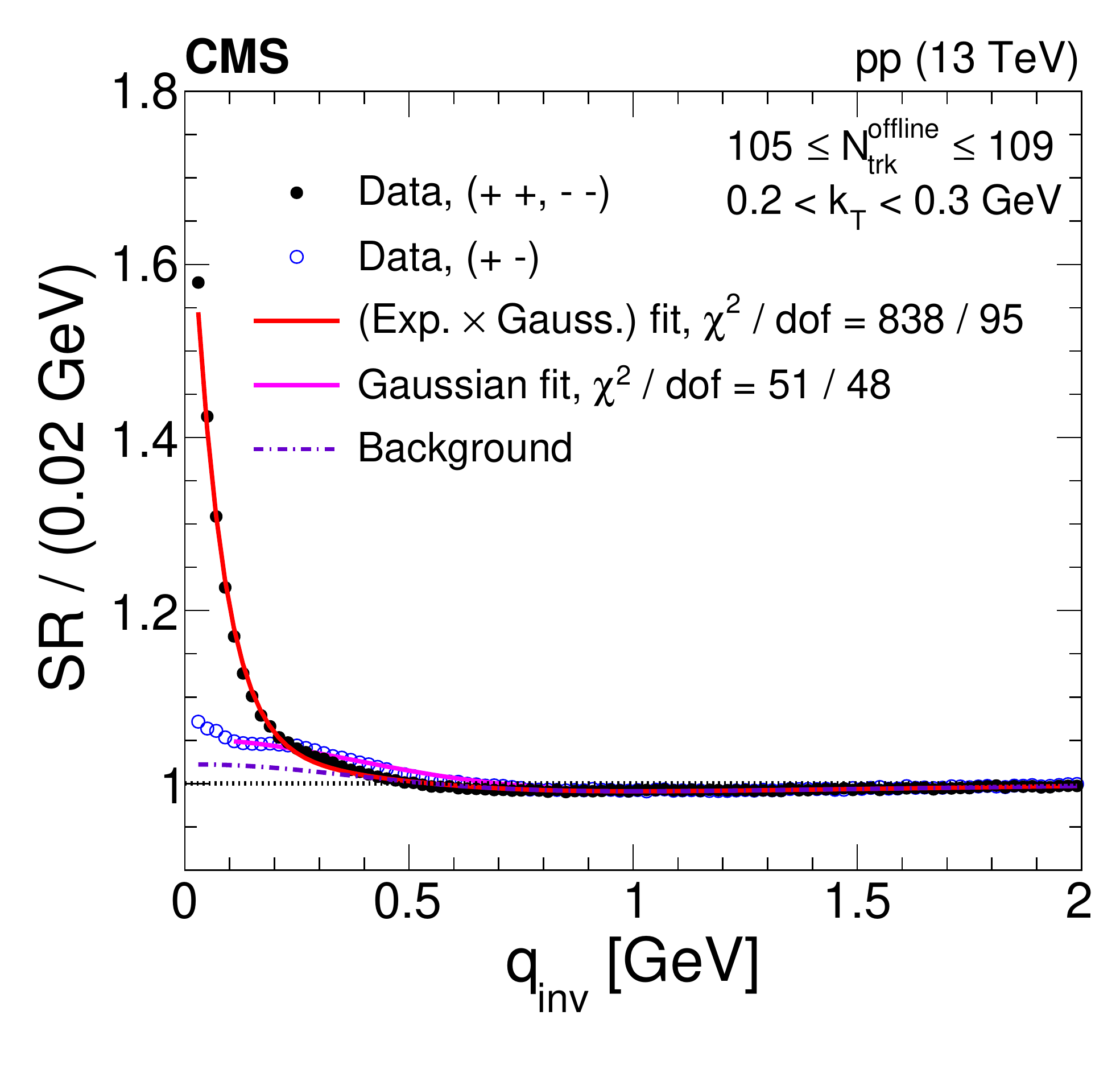} \\
   \includegraphics[width=0.45\textwidth]{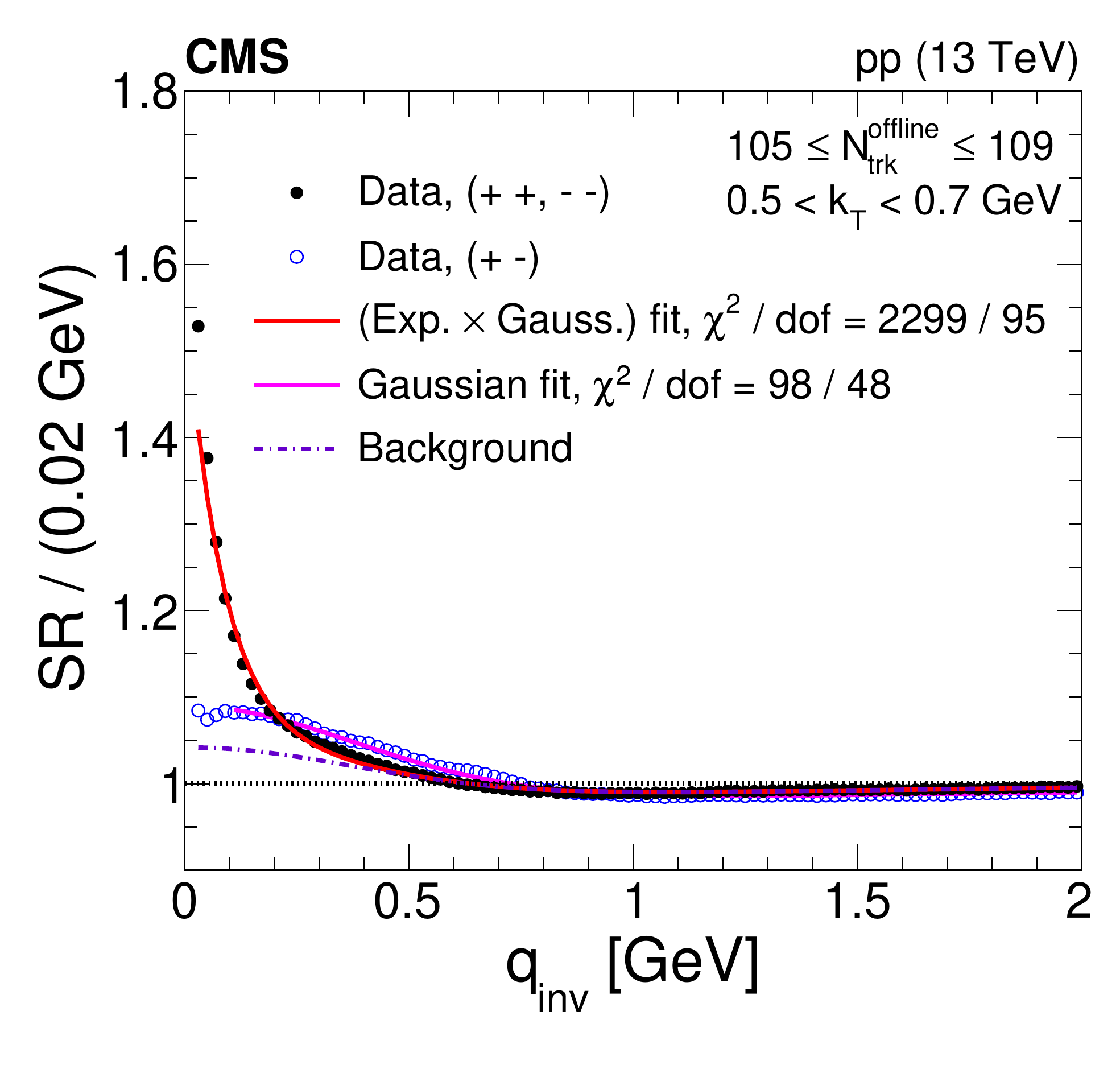}
   \includegraphics[width=0.45\textwidth]{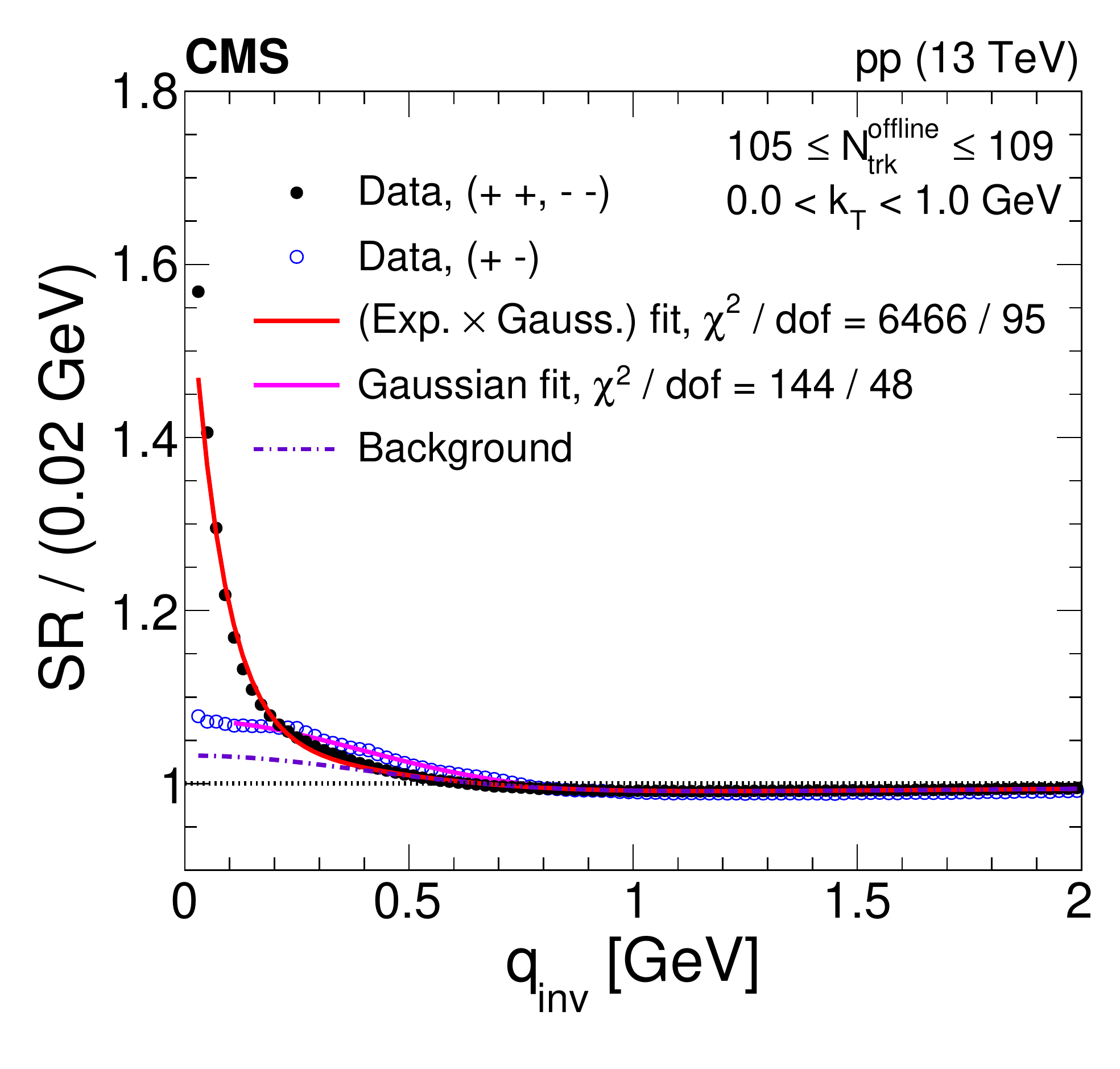}
  \caption{The same-sign ($++,--$) and opposite-sign ($+-$) single ratios in data for different bins of $\Ntroff$ and $\kt$, with their respective fits. The label ``(Exp. $\times$ Gauss.) fit'' refers to the same-sign data and is given by Eq.~(\ref{eq:sig_plus_bkg_form_ATLASmethod}). The label ``Gaussian fit'' corresponds to  Eq.~(\ref{eq:bkg_form_ATLASmethod}) applied to opposite-sign data and ``Background''  is the component of Eq.~(\ref{eq:sig_plus_bkg_form_ATLASmethod}) that is found from the Gaussian fit using Eqs.~(\ref{eq:sigmaBArelation_ATLASmethod}) and (\ref{eq:BArelation_ATLASmethod}) to convert the fit parameters. Coulomb corrections are accounted for using the Gamow factor. The error bars represent statistical uncertainties and in most cases are smaller than the marker size.
      }
  \label{fig:data_os_ss_sr_ATLASmethod}
\end{figure}

Similarly, single ratios for OS pairs in data are constructed and fitted with the function given by Eq.~(\ref{eq:bkg_form_ATLASmethod}), yielding the parameters $B$ and $\sigma_B$ for OS data. The conversion functions based on simulation are used to calculate $B$ and $\sigma_B$ for SS pairs in data. The resulting estimated background, found using $\Omega(\qi)$ in Eq.~(\ref{eq:bkg_form_ATLASmethod}), is included in Eq.~(\ref{eq:sig_plus_bkg_form_ATLASmethod})
to fit SS data:
\begin{linenomath*}
\begin{equation}
C_2 (\qi) = \Omega(\qi) C_{2, \text{BE}} (\qi),
\label{eq:sig_plus_bkg_form_ATLASmethod}
\end{equation}
\end{linenomath*}
where $C_{2, \text{BE}} (\qi)$ describes the BEC component as in Eq.~(\ref{eq:1d-levy}) (with a=1). So, the final fit function has two components: one with fixed parameters to describe non--BEC effects and another with fitted BEC parameters. In Fig.~\ref{fig:data_os_ss_sr_ATLASmethod}, examples of fits using this combined function are shown. The shape of the correlation function is not trivial and cannot be described by a simple function, since it is distorted by many components, such as resonances, jets, etc., and the individual contribution of each of these components is not known. Furthermore, when fitting the correlation function, only statistical uncertainties are considered, which are, in general, smaller than 0.5\%, depending on the bin. Therefore, it is expected that fitting with a simple exponential function would not necessarily result in a $\chi^2/\mathrm{dof}$ near unity.

\section{Systematic uncertainties}
\label{sec:systematics}
The following sources of systematic uncertainties are investigated: the bias from the particular choice of reference sample with respect to all other constructed possibilities,
the bias from the MC simulation tune adopted in the analysis, the influence of the track selection and relevant corrections, the interference of tracks from pileup collisions, the bias from $z$- and $xy$-vertex positions, the efficiency of the HM triggers, and the effect from the Coulomb correction. The more important biases are those due to the reference samples and the MC simulations. The track selection lead to a nonnegligible effect, mainly in higher-$\kt$ bins, where larger contamination from jets is expected. In addition, the Coulomb correction has a significant contribution. To compute the systematic uncertainties associated with the nonnegligible effects mentioned above, samples of measurements of $R_{\text{inv}}$ and $\lambda$ are generated by varying the corresponding source of bias in bins of $\Ntroff$ and $\kt$, and the root mean square (RMS) spread of each sample is associated with the systematic uncertainty in that bin. The other potential sources of bias listed above returned maximal deviations of the order of the statistical uncertainties (${\sim}1$--5\%), and are not included in the estimate of the total systematic uncertainty.

To estimate the systematic uncertainties associated with the construction of the reference sample, additional samples are constructed with alternative techniques. Within the category of mixed events, tracks are randomly combined  from samples of 25, 40, or 100 events, all of which are in the same range of $\Ntroff$. Reference samples are also constructed with tracks from the same event used to form the signal sample, but making pair combinations such that only one of the two tracks has its three-momentum reflected with respect to the origin, \ie, $(p_x,p_y,p_z) \to (-p_x,-p_y,-p_z)$.  Another case corresponds to rotating one of these two tracks by 180 degrees in the plane transverse to the beam, \ie, $(p_x,p_y,p_z) \to (-p_x,-p_y,p_z)$.

The uncertainties related to the reference samples and to the MC samples are estimated by associating these two sources and repeating the measurements eighteen times, (6 reference samples)$\times$ (3 MC samples). For the reference samples, the default $\eta$-mixing method and the five reference samples described above are used. For MC simulation, samples are generated using \PYTHIA 6 (Z2* tune), \PYTHIA 8 (CUETP8M1 tune for MB and 4C tune for HM), and \textsc{epos lhc}.

For the track selection, in addition to the default definition in Section~\ref{sec:evt-track-sel}, five additional different configurations were considered, changing combinations of looser and tighter criteria on the track variables. These six configurations were used to build a sample of measurements for different track selections.

For the Coulomb correction, a procedure similar to Ref.~\cite{cms-hbt-1st} is adopted, where the Gamow factor is multiplied by a strength parameter $\kappa$. Fitting the correlation function by allowing $\kappa$ to vary yields values consistent with unity within a statistical uncertainty of $\pm 15\%$. A conservative uncertainty of 15\% applied to the Gamow factor is then propagated, repeating the measurements by varying the magnitude of the Gamow factor up and down by this amount.

\begin{table}[h!]
 \caption{Total systematic uncertainties in different \kt bins for the hybrid cluster subtraction technique. The ranges in the uncertainties indicate the minimum and maximum values found for all multiplicity bins.}
 \label{tab:tot_syst}
 \begin{center}
 \begin{tabular}{lcc}
  \hline
  $\kt$ (\GeVns{})  & \multicolumn{2}{c}{Relative uncertainties}  \\
  \hline
               & $R_{\text{inv}} (\%)$  & $\lambda (\%)$   \\
  Integrated   & 5--20         & 5--20          \\
  (0.2, 0.3)   & 4--8          & 5--8           \\
  (0.3, 0.4)   & 4--7          & 4--7           \\
  (0.4, 0.5)   & 4--8          & 4--8          \\
  (0.5, 0.7)   & 6--26         & 9--22         \\
  \hline
 \end{tabular}
 \end{center}
\end{table}

The uncertainties from the three sources listed above are computed independently and then added in quadrature to obtain the total systematic uncertainty. The largest contribution originates from the reference samples and the MC tune, reaching values of ${\sim}20\%$ in the final measurement. The other sources are always smaller than ${\sim}6\%$. Table~\ref{tab:tot_syst} shows the ranges of systematic uncertainties (variation in $\Ntroff$) for each $\kt$ bin and integrated in $\kt$. Lower multiplicities and higher $\kt$ ranges have larger systematic uncertainties because the contamination from the jet fragmentation background is higher in those regions.

\section{Results}
\label{sec:results}

\begin{figure}[hptb]
  \centering
    \includegraphics[width=0.45\textwidth]{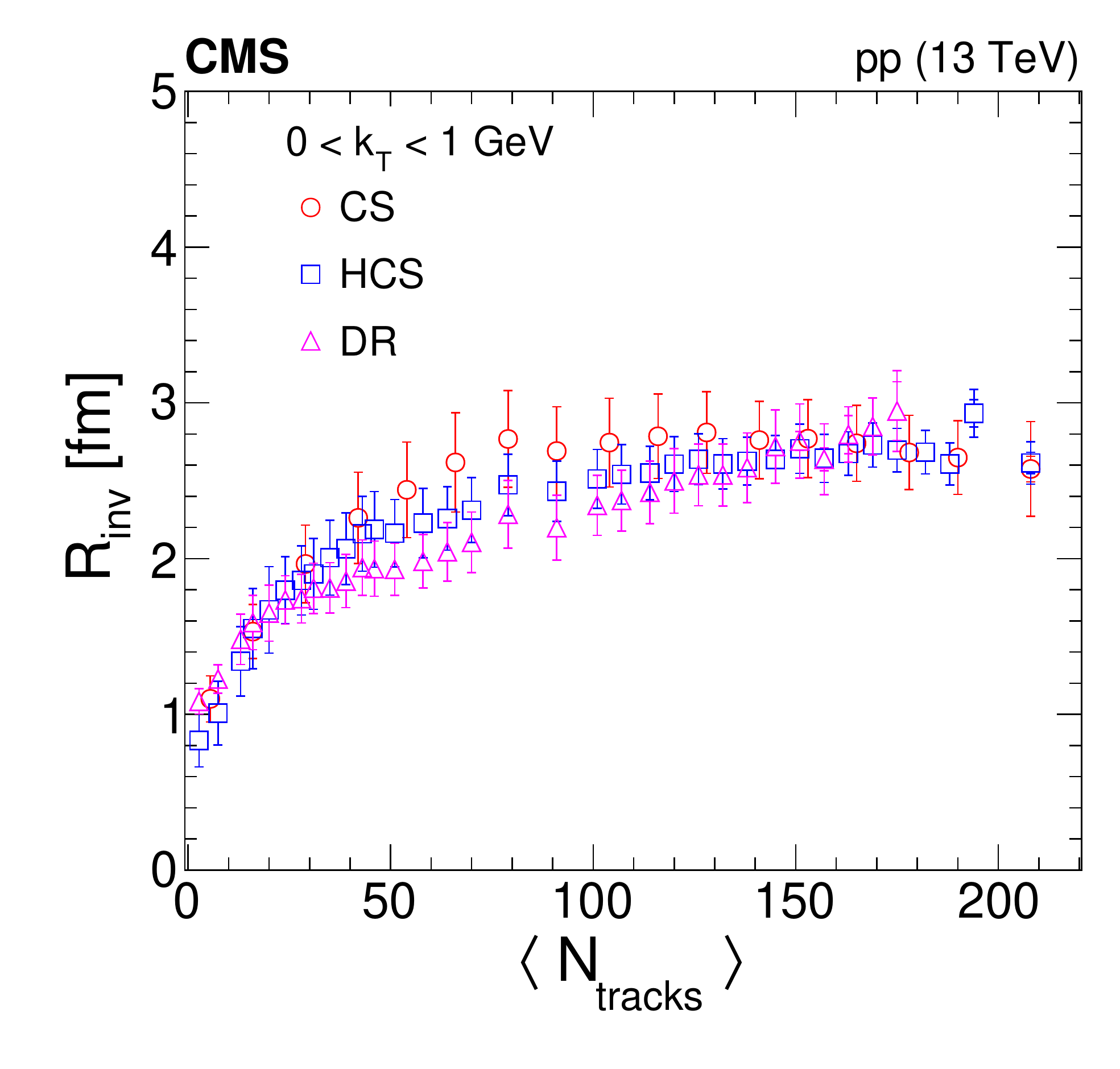}
    \includegraphics[width=0.45\textwidth]{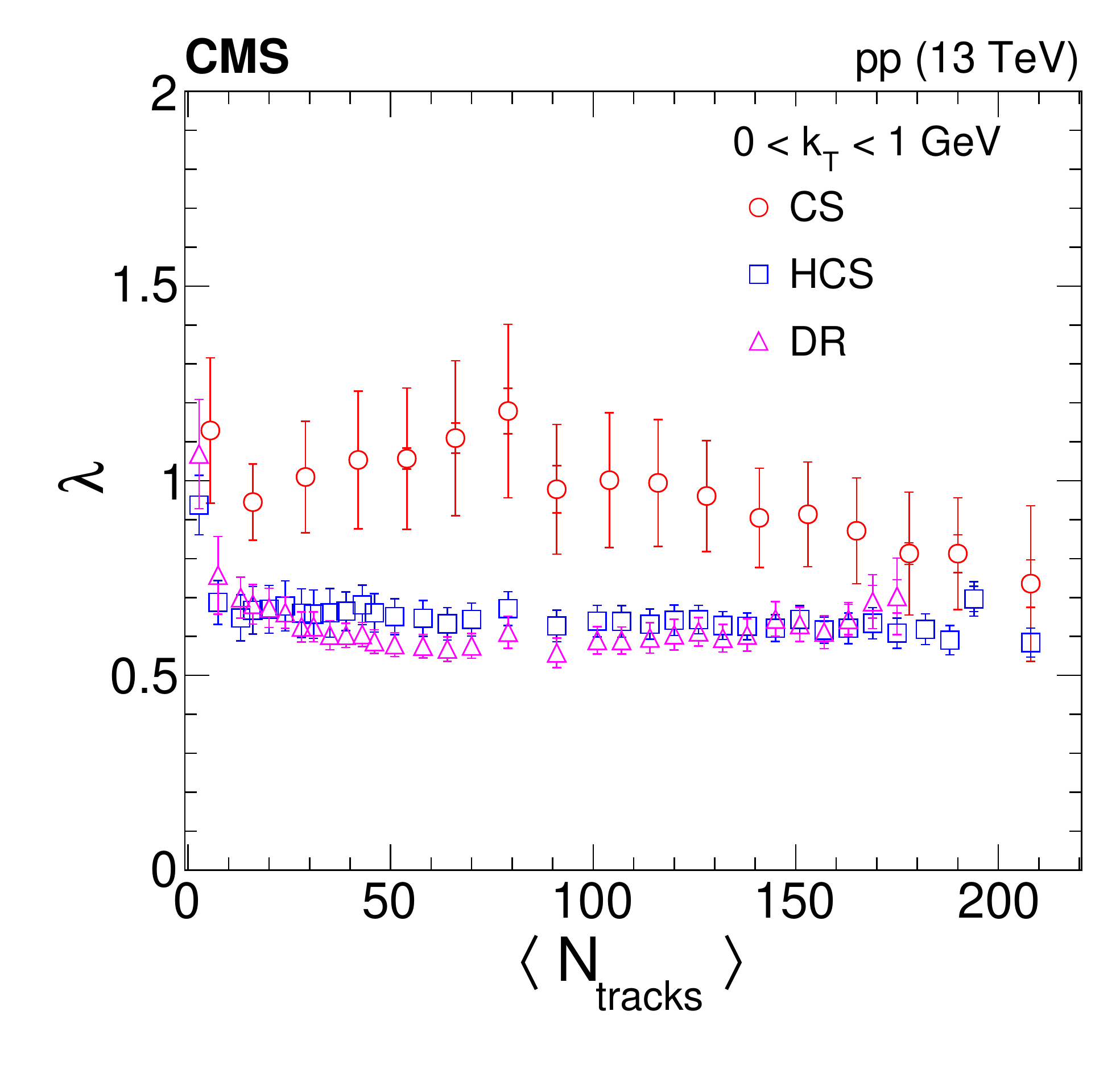} \\
    \includegraphics[width=0.45\textwidth]{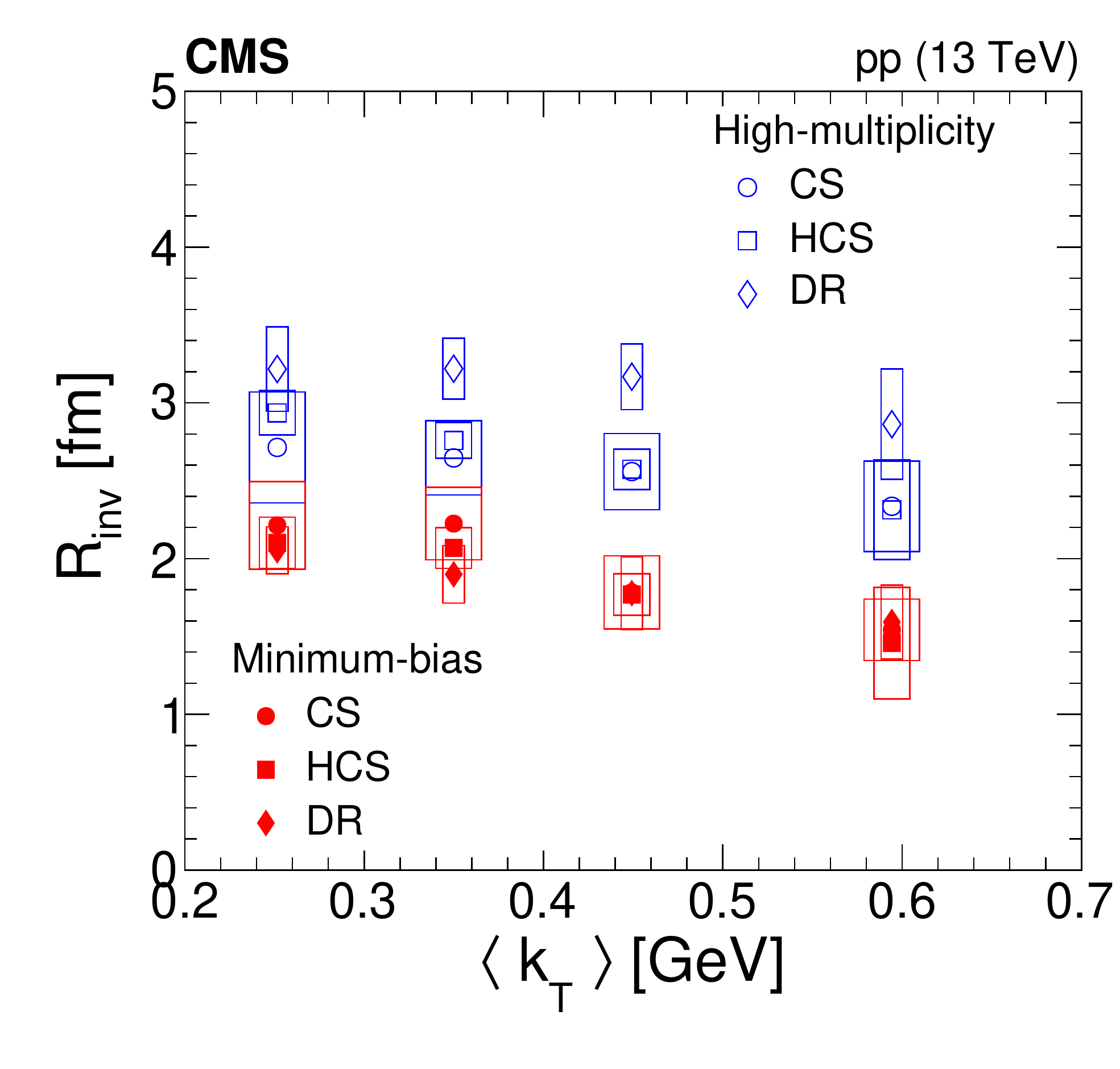}
    \includegraphics[width=0.45\textwidth]{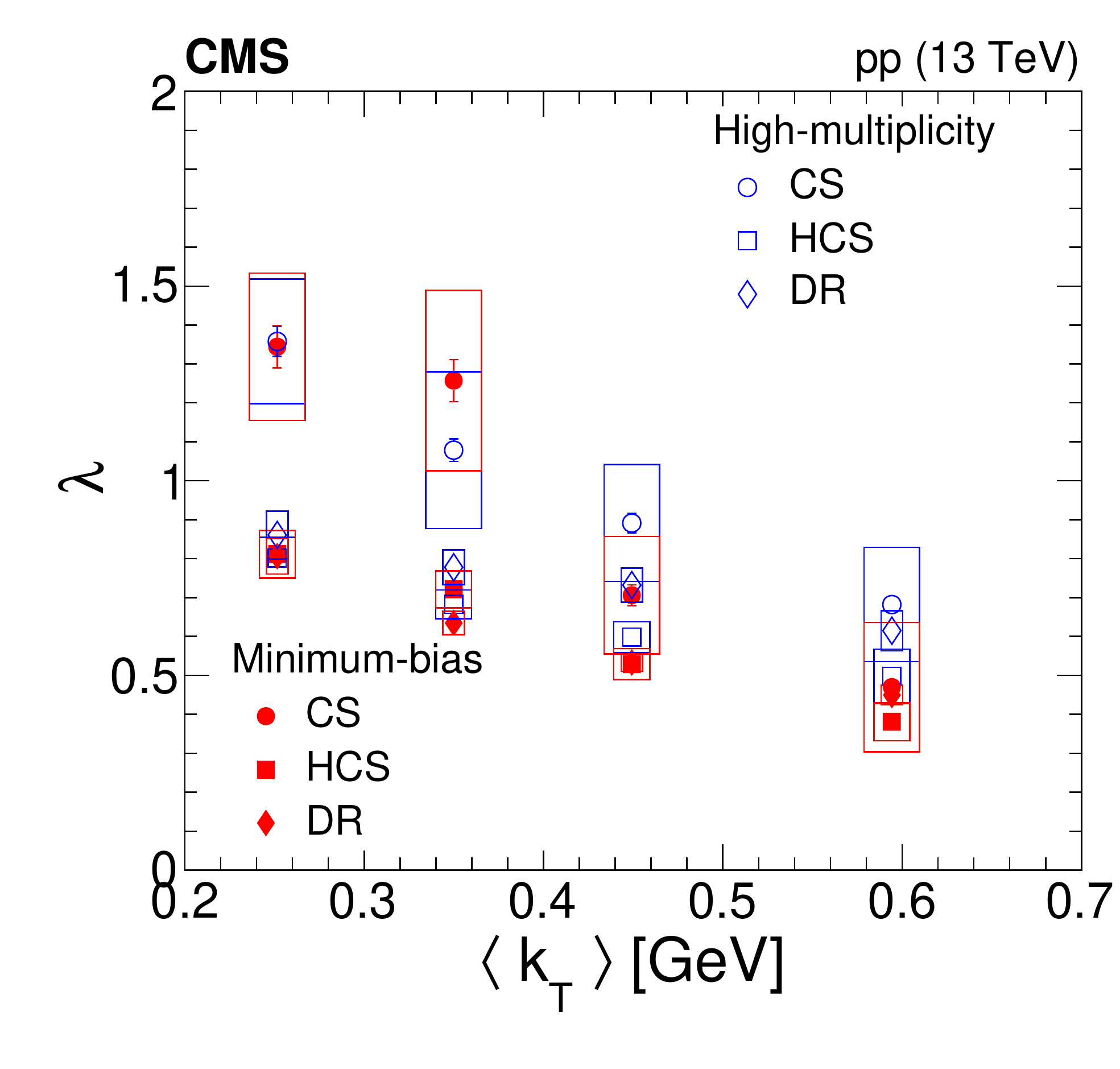}
  \caption{Results for $\Ri$ (left) and  $\lambda$ (right) from the three methods as a function of
  multiplicity (upper) and \kt (lower). In the upper plots, statistical and systematic uncertainties are represented by internal and external error bars, respectively. In the lower plots, statistical and systematic uncertainties are shown as error bars and open boxes, respectively.
             }
  \label{fig:comparativeresults}
\end{figure}

The values of $\Ri$ and $\lambda$  obtained with each of the three methods as functions of $\langle \Nt \rangle$ and $\kt$ are shown in Fig.~\ref{fig:comparativeresults},
where $\langle \Nt \rangle$ is the average multiplicity at particle level corrected for acceptance and efficiency.
The top panel presents the results  for $\Ri$ and $\lambda$ versus $\langle \Nt \rangle$,
for integrated values of the pair transverse momentum in the range $0 < \kt < 1\GeV$.
The radius fit parameter $\Ri$ increases as a function of multiplicity, showing a change
in slope around $\langle \Nt \rangle\sim 20$--30 and a tendency to saturate at higher multiplicities.
For the DR and HCS methods, the intercept parameter $\lambda$ rapidly decreases for increasing multiplicities in the very small $\langle \Nt \rangle$ region, whereas for multiplicities  ${\gtrsim}10$, it shows an almost constant value with increasing $\langle \Nt \rangle$. The systematic uncertainties are larger for $\lambda$ using the CS method, with the fit values of $\lambda$ fluctuating and decreasing at higher multiplicities. This happens because  $\lambda$ is very sensitive to the background modeling (non-BEC effects), which
leads to larger uncertainties associated with its determination.

In the bottom panel of Fig.~\ref{fig:comparativeresults}, fit parameters $\Ri$ and $\lambda$ are shown in two
multiplicity bins, MB ($\Ntroff < 80$) and HM ($\Ntroff \ge 80$), as a function of $\langle \kt \rangle$, where the average is performed in each $\kt$ bin, whose width is variable.
The length of homogeneity $\Ri$ tends to decrease with increasing \kt, more so at lower multiplicity. This behavior is compatible with an
emitting source that was expanding prior to decoupling~\cite{Makhlin:1987gm,hamapad,piotrwojtek,hydrokin}. The correlation intensity $\lambda$ also decreases
with increasing values of $\kt$, with a more pronounced slope than that for $\Ri$.

The increase of $\Ri$ with the event multiplicity and decrease with the average pair momentum in $\Pp\Pp$ collisions were predicted in Ref.~\cite{hamapad}. These predictions were based on the assumption that a quark-gluon plasma (QGP) \cite{Collins:1974ky,Cabibbo:1975ig,Freedman:1976xs,Shuryak:1977ut} could be formed in high energy collisions of  small systems in events with multiplicities similar to those investigated here. In the model, high multiplicities are related to large fireball masses formed in the collision, corresponding to a one-dimensional expansion based on Khalatnikov's solution~\cite{khalatnikov} of the Landau hydrodynamical model~\cite{landau}. These model predictions were also compared to BEC data for pions~\cite{afs-pion} and kaons~\cite{afs-kaon}  measured in $\Pp\Pp$,  $\PAp\Pp$ and $\PGa\PGa$ collisions at the CERN ISR, and described the overall behavior of the correlation functions more closely than the Gaussian fit adopted in the analysis of the data.

 As shown in Fig.~\ref{fig:comparativeresults}, the three methods produce results for $\Ri$ that are compatible  within the experimental uncertainties. Compatibility tests among the three methods, based on the variations of the measured values, assume the experimental uncertainties are either fully uncorrelated or fully correlated. For the fully correlated, most conservative case, the results agree within two standard deviations for most of the bins investigated.
For the CS method, larger deviations are observed for the $\lambda$ parameter and the associated systematic uncertainties are also larger.
This parameter is particularly sensitive to the analysis procedure adopted to remove the non-Bose--Einstein effects present in the signal, as observed in Ref.~\cite{fsq-14-002}. Therefore,  when comparing to other energies and  theoretical models, only the values found using the HCS method are shown.
This technique is less sensitive to the MC tune than the DR method, mainly in the HM region, and has smaller systematic uncertainties than the CS method.
The ratio of RMS over mean for the differences amongst the values of
the radius fit parameters obtained with the three methods is adopted as the relative uncertainty due to the variation between techniques (here called "intramethod variation").

\begin{figure}[hptb]
\centering
  \includegraphics[width=0.45\textwidth]{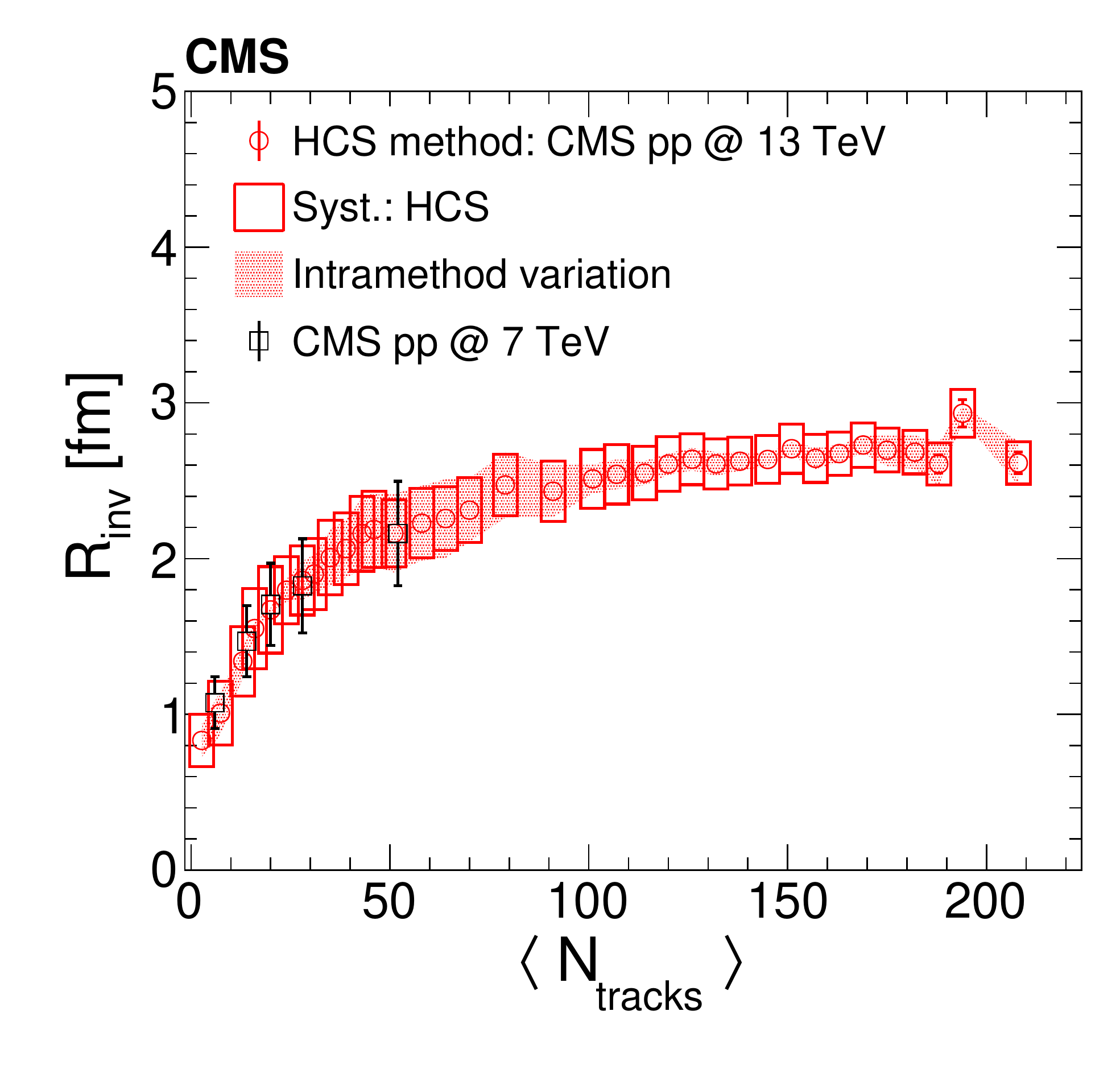}
  \includegraphics[width=0.45\textwidth]{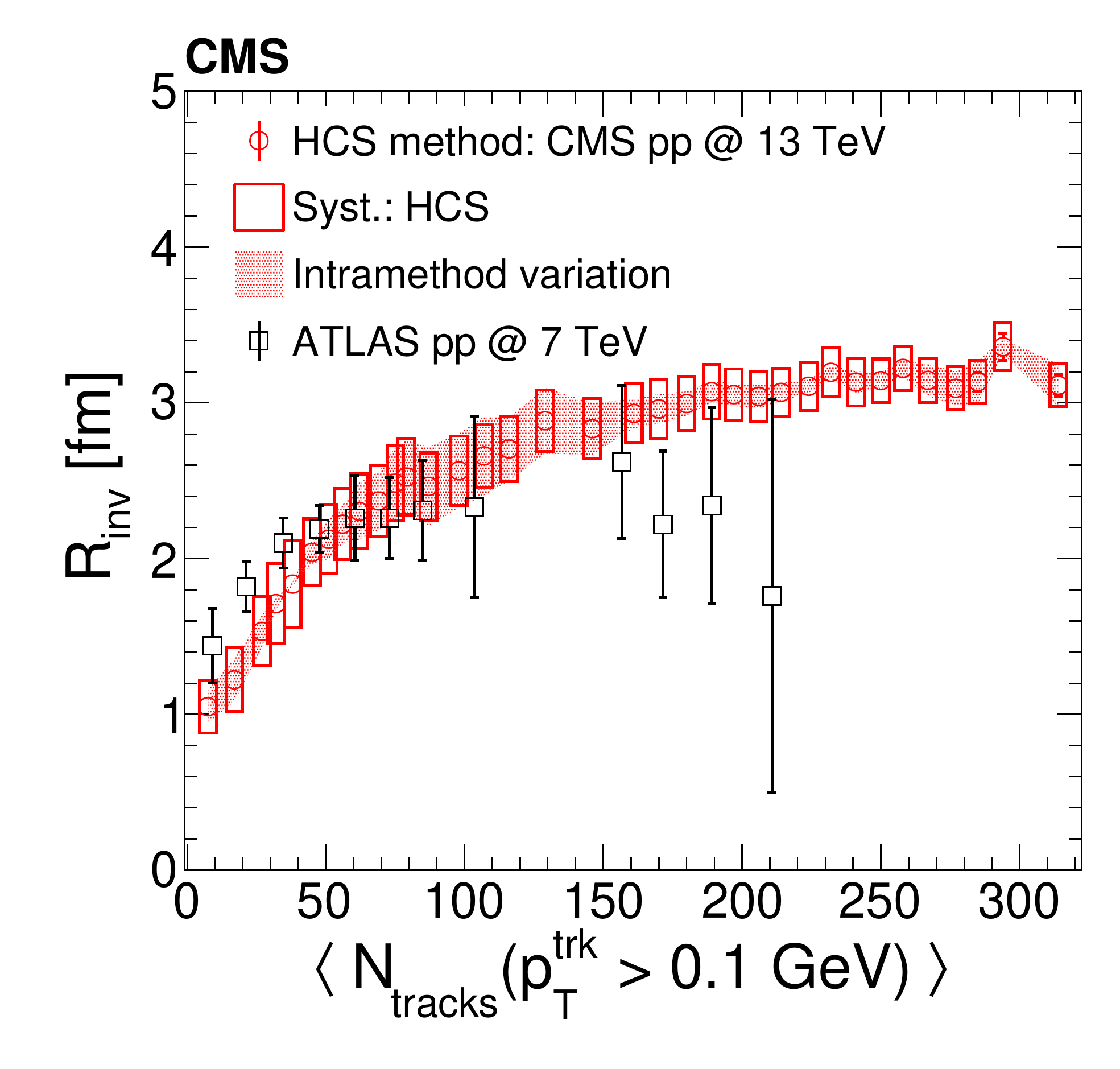}
 \caption{The $\Ri$  fit parameters as a function of particle-level multiplicities using the
HCS method in $\Pp\Pp$ collisions at 13\TeV compared to results
for $\Pp\Pp$ collisions at  7\TeV from CMS (left) and ATLAS (right).
Both the ordinate and abscissa for the CMS data in the right plot have been adjusted for compatibility with the ATLAS analysis procedure, as explained in the text. The error bars in the CMS~\cite{fsq-14-002} case
represent systematic uncertainties (statistical uncertainties are smaller than the marker size) and in the ATLAS~\cite{atlas2015} case,
 statistical and systematic uncertainties added in quadrature.
 }
 \label{fig:comparisonDiffEnergiesAndExperiments}
\end{figure}

The $\Ri$ parameters for $\Pp\Pp$ collisions at 13\TeV are shown in Fig.~\ref{fig:comparisonDiffEnergiesAndExperiments} as a function of multiplicity and compared with the corresponding results obtained in $\Pp\Pp$ collisions at  7\TeV by CMS~\cite{fsq-14-002}  (left) and ATLAS~\cite{atlas2015}  (right). In the ATLAS measurement, tracks with $\pt > 0.1\GeV$ are included in the multiplicity and the correction to particle-level multiplicity is done using an unfolding procedure, as described in Ref.~\cite{atlas2015}. To have consistent multiplicities, the $\Nt$ values for CMS data in this figure are corrected for the extrapolation down to $\pt = 0.1\GeV$. In addition, the opposite-sign ($+ -$) reference sample adopted by ATLAS causes distortions in the same-sign ($+ +, - -$) correlation functions due to resonance contamination. To circumvent this problem, the ATLAS correlation functions are fitted excluding ranges around the resonance peaks. Therefore, for establishing a more consistent comparison, the analysis with the CMS data was repeated excluding those regions in the fits to both the OS and SS correlation functions. The $\Ri$  values for $\Pp\Pp$ collisions at 13\TeV are compatible with those obtained by both CMS and ATLAS at 7\TeV over the entire multiplicity range investigated.

\begin{figure}[hptb]
  \centering
    \includegraphics[width=0.45\textwidth]{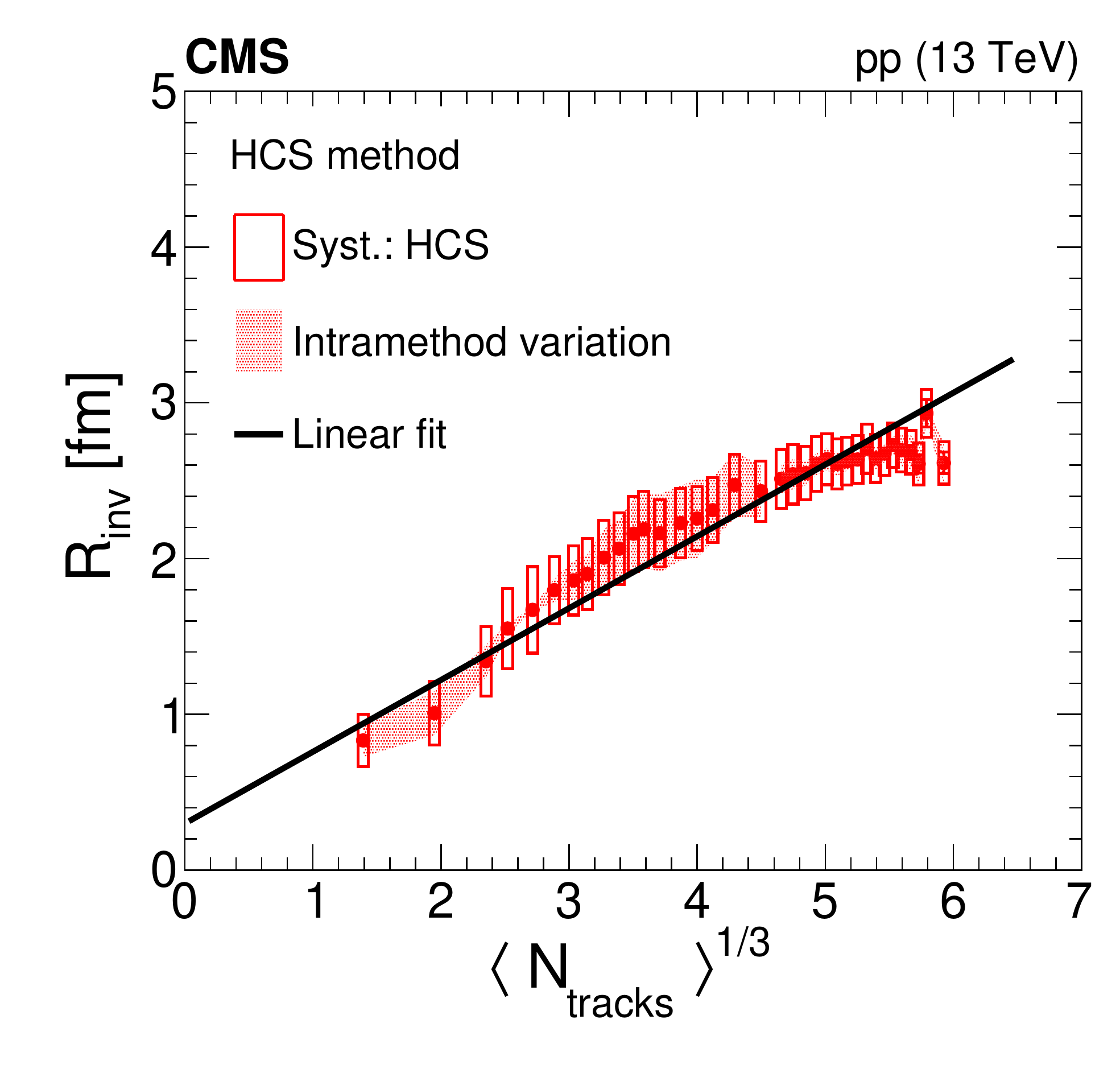}
    \includegraphics[width=0.45\textwidth]{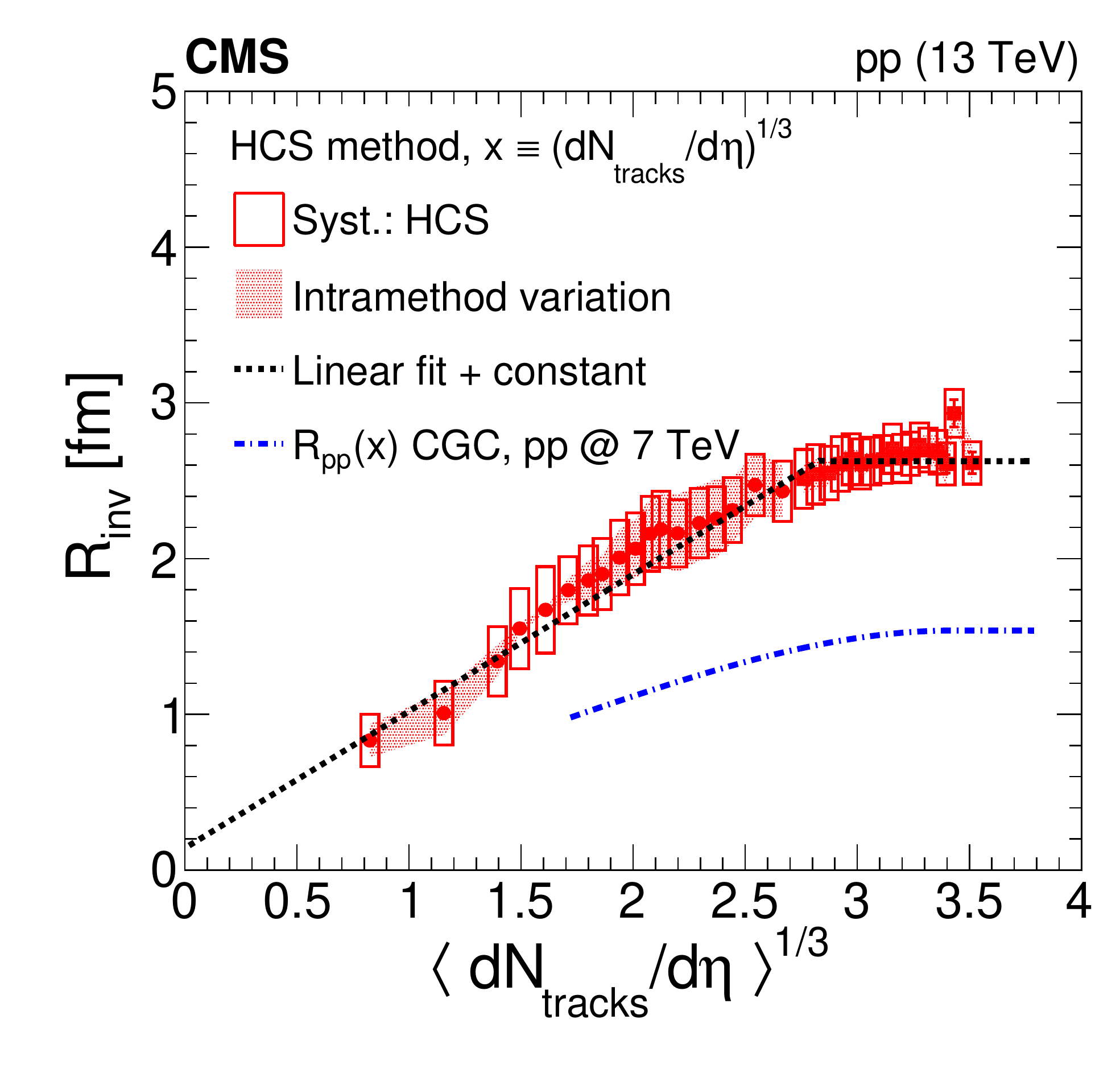}
 \caption{ Comparison of $\Ri$ obtained with the HCS method with theoretical expectations.
Values of $\Ri$ as a function of $\langle N_\text{tracks}\rangle^{1/3}$ (left) are shown
with a linear fit to illustrate the expectation from hydrodynamics.
Values of $\Ri$ are compared with predictions from the CGC as a function of $\langle \rd\Nt/\rd\eta \rangle^{1/3}$ (right).
The dot-dashed blue line is the result of the parameterization in Eq.~(\ref{eq:CGC}).
The linear plus constant function (dashed black lines) for $\langle \rd\Nt/\rd\eta \rangle^{1/3}$ is shown to
illustrate the qualitative behavior suggested by the CGC (the matching point of the two lines is the result of a fit).
Only statistical uncertainties are considered.
}
 \label{fig:RadiusVsNtoOnethrid}
\end{figure}

In Fig.~\ref{fig:RadiusVsNtoOnethrid}, the linear dependence of $\Ri$ on $N_\text{tracks}^{1/3}$ reflects the growth of the number of produced particles with the system volume (or equivalently, $N_\text{tracks}\propto \Ri^{3}  $).  This dependence is investigated for a range of energies and colliding systems, with the results that, $\Ri$ is independent of collision energy when compared at the same multiplicity. Radii can also be studied as a function of $(\rd N/\rd\eta)^{1/3}$ for investigating the dependence of the final-state geometry on the multiplicity density at freeze-out. Values of $\Ri$ are plotted as a function of $\langle N_\text{tracks}\rangle^{1/3}$ and $(\rd\Nt/\rd\eta)^{1/3}$ in Fig.~\ref{fig:RadiusVsNtoOnethrid}. Statistical uncertainties are represented by the error bars, systematic uncertainties related to the HCS method are shown as the open boxes, and relative uncertainties from the variation between methods are represented by the shaded bands.

Data for $\Ri$ and average particle transverse momentum $\langle\pt\rangle$ versus multiplicity were investigated in Ref.~\cite{campanini2011} to deduce approximate equations of state from experimental measurements in $\Pp\Pp$ and $\PAp\Pp$ collisions and search for possible signatures of the phase transition from hadrons to the QGP. A phase transition would cause a change in slope for both observables in the same region of multiplicity per unit pseudorapidity ($\rd\Nt/\rd\eta$). In Ref.~\cite{campanini2011}, the authors claim that their compilation of $\Ri$ values as a function of $\rd\Nt/\rd\eta$ for data from several experiments at different center-of-mass energies shows a common behaviour that is independent of the energy. In particular, for $\Ri$ obtained by CMS~\cite{cms-hbt-1st} and ALICE~\cite{alice}, they claim that a linear function in $(\rd\Nt/\rd\eta)^{1/3}$ for $\rd\Nt/\rd\eta > 7.5$, matched with a fifth degree polynomial for smaller $\rd\Nt/\rd\eta$ values, fits the data better than a single function of $(\rd\Nt/\rd\eta)^{1/3}$ over the entire range. The $\rd\Nt/\rd\eta$ values in Ref.~\cite{campanini2011} are for spectra extrapolated to \pt of zero and therefore correspond more closely to the right panel of Fig.~\ref{fig:comparisonDiffEnergiesAndExperiments}. For the CMS acceptance, a value of $\rd\Nt/\rd\eta\sim 7.5$ corresponds to $\langle \Nt^{(\pt > 0.1)} \rangle\sim 35$ and  $\langle \Nt \rangle\sim 23$. For comparison to Fig.~\ref{fig:RadiusVsNtoOnethrid}, $\langle \Nt \rangle\sim 23$ is equivalent to $\langle \Nt \rangle^{1/3}\sim 2.8$ and $\langle \rd\Nt/\rd\eta \rangle^{1/3}\sim 1.7$. This overall qualitative behavior of $\Ri$ vs. $\langle \Nt \rangle$ or $\langle \rd\Nt/\rd\eta \rangle$ seems  compatible with the present results shown in Figs.~\ref{fig:comparisonDiffEnergiesAndExperiments} and \ref{fig:RadiusVsNtoOnethrid}, but the value of $\Ri$ around which the data could change slope is not evident, since it is also dependent on the lowest value of \pt considered in data.

Although theoretical predictions based on hydrodynamics are not available for $\Pp\Pp$ collisions at 13\TeV yet,
expectations for qualitative trends can be found in the literature.
A framework based on event-by-event (3+1)-dimensional viscous hydrodynamics,
found that the three components of the radius fit parameters continuously grow with $\langle \Nt \rangle^{1/3}$ for $\Pp\Pp$
collisions~\cite{piotrwojtek}. Calculations using a hydrokinetic model also show a linear growth of the lengths of homogeneity with $\langle \Nt \rangle^{1/3}$~\cite{hydrokin}.
Such a continuous increase is consistent with the results shown in  Fig.~\ref{fig:RadiusVsNtoOnethrid} and was also observed for
different collision systems (CuCu, AuAu, PbPb, and $\Pp\Pp$) and
energies (ranging from 62.4\GeV to 7\TeV) in Fig.~1 of Ref.~\cite{piotrwojtek}.
To illustrate this expectation from hydrodynamics, a fit with a single linear function over the entire $\langle \Nt \rangle^{1/3}$
range is shown in the left panel of Fig.~\ref{fig:RadiusVsNtoOnethrid}.  
A similar trend of increasing length of homogeneity with multiplicity based on the UrQMD microscopic model for several colliding nuclei at different energies and for $\Pp\Pp$ collisions at 7\TeV is found in Ref.~\cite{PhysRevC.85.044901}. Instead of the one-dimensional $\Ri$ used in the present analysis, which combines temporal and spatial information, Fig. 1 of  Ref.~\cite{PhysRevC.85.044901} shows the three-dimensional radius parameters $\Rl$ (along the beam) as well as $\Ro$ and $\Rs$ (both in the plane transverse to the beam direction), as functions of the charged particle multiplicity at midrapidity,  $(\rd \Nt/\rd\eta)^{1/3}$. Although the behavior of each radius component varies, a trend of increasing with multiplicity is seen for all colliding systems. In the case of $\Pp\Pp$ collisions, both $\Rl$ and $\Rs$ show a clear increasing trend, whereas the $\Ro$ values estimated by UrQMD show an almost flat behavior or a slight tendency to decrease with increasing multiplicity. 

The color glass condensate (CGC) effective theory predicts an increase of the interaction radius (resulting from the initial overlapping of the two protons) as a function of the rapidity density $\rd N/\rd y$~\cite{larry2013}. This dependence is parametrized by a third order polynomial in  terms of $x=(\rd N/\rd y)^{1/3}$ for $x < 3.4$;
beyond this point, the radius tends to a constant value,
the so-called ``saturation'' radius. In the case of $\Pp\Pp$ collisions at 7\TeV, this can be expressed by~\cite{larry2013}
\begin{linenomath*}
\begin{equation}
R_{\Pp\Pp} (x) = \begin{cases} (1\unit{fm}) [a + bx + cx^2 +dx^3] & \text{(for  $x < 3.4$)} \\
e (\unit{fm}) & \text{(for  $x\geq 3.4$)} \end{cases}
\label{eq:CGC}
\end{equation}
\end{linenomath*}
with parameter values of $a = 0.387$, $b = 0.0335$, $c = 0.274$, $d = -0.0542$, and $e = 1.538$.
According to Ref.~\cite{larry2013}, the minimum multiplicity where the computation in Eq.~(\ref{eq:CGC}) could be reliable is around $ \rd N/\rd\eta\sim 5$ in cases where no $\pt$ cut is applied to the data. This condition is better illustrated by the right plot in Fig.~\ref{fig:comparisonDiffEnergiesAndExperiments}, and considering that the $\eta$ coverage in CMS and ATLAS is about 4.8, this minimum value would correspond to $\langle \Nt^{(\pt > 0.1)} \rangle\sim 25$.

This prediction for the radius behavior is based on a calculation relating particle multiplicity to a saturation scale using computations
of the interaction radius determined from the CGC~\cite{raju2013}. The parameterization given in Eq.~(\ref{eq:CGC})
is compared with the results from the HCS method in the right panel of Fig.~\ref{fig:RadiusVsNtoOnethrid}.
The predictions from the CGC are well below the data and the predicted saturation
radius~\cite{larry2013,raju2013} is almost half the size of that seen in the data. In Ref.~\cite{raju2013}, no explicit calculation
of the BEC radii (corresponding to emission after the last interaction) is performed.  Instead, only the initial size of the $\Pp\Pp$ interaction region and the initial energy density are used, without considering any fluid dynamic evolution, which may explain the underestimated values for the CGC radius parameter seen in Fig.~\ref{fig:RadiusVsNtoOnethrid}.
Since the CGC calculation does not include the evolution of the system, it is not unexpected that it underestimates the measured radii.
However, the CGC dependence of the radii on particle density, consisting of a rise followed by saturation, is similar to the shape seen in the data.
To illustrate this behavior, a linear plus constant function is fitted to the data using statistical uncertainties only (see the right panel of Fig.~\ref{fig:RadiusVsNtoOnethrid}).

The tendency to a constant value of $\Ri$
at higher $\langle N_\text{tracks}\rangle$ was suggested by ATLAS in Ref.~\cite{atlas2015}, based on their data shown in the right panel of Fig.~\ref{fig:comparisonDiffEnergiesAndExperiments}, although their uncertainties were large. In their case, this saturating behavior is considered to be achieved for $\langle \Nt^{(\pt > 0.1)} \rangle\sim 55$, at a much smaller value (less than $1/3$) than that suggested by the CGC calculations, where it is claimed that the saturation radius would be reached for $(\rd N/\rd\eta)^{1/3}\sim 3.4$, or $\langle N_\text{tracks}\rangle\sim 190$ charged particles. The new CMS data, with their smaller uncertainties, appear to be more consistent with a saturation at this higher region of multiplicity.

\begin{figure}[hptb]
\centering
  \includegraphics[width=0.45\textwidth]{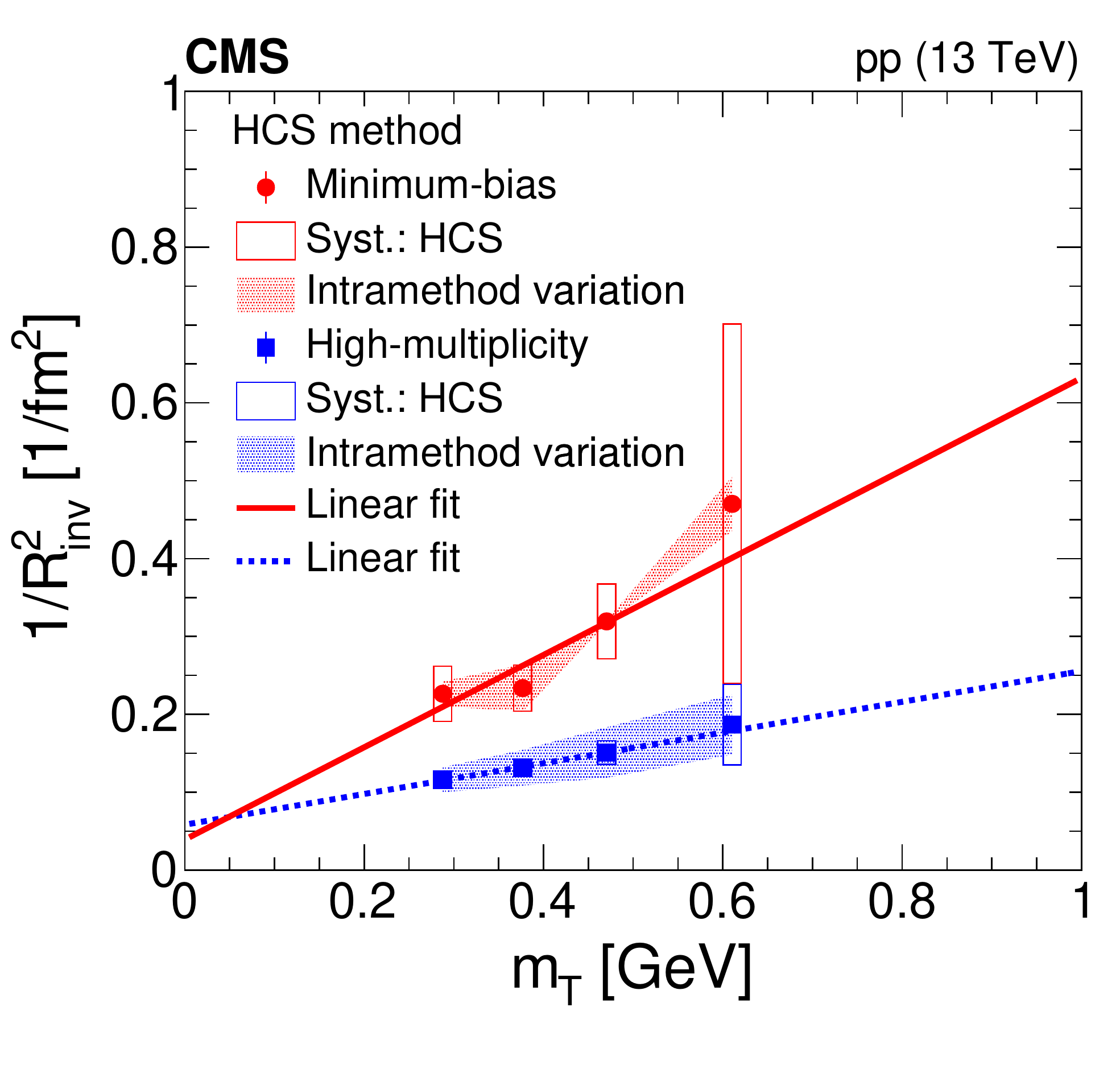}
  \includegraphics[width=0.45\textwidth]{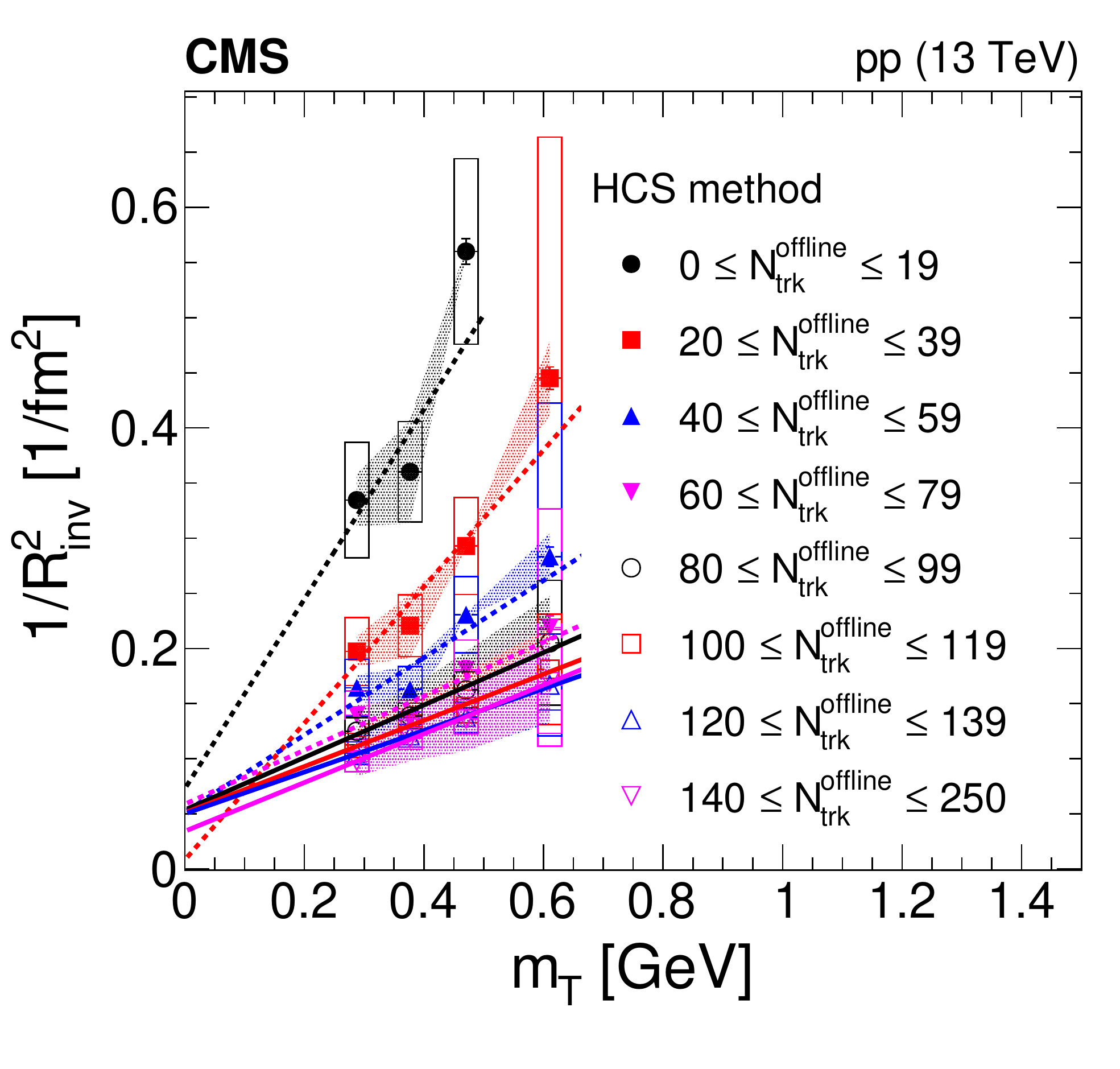}
 \caption{The distribution $1/R_{\text{inv}}^2$ as a function of $\mT$ for the HCS
 method. Results corresponding to the MB range ($0 \le N_\text{trk}^\text{offline} \le 79$) and to the HM one ($80 \le N_\text{trk}^\text{offline} \le 250$) are shown (left). Results are also shown in more differential bins of multiplicity (right). Statistical uncertainties are represented by the error bars, systematic uncertainties related to the HCS method are shown as open boxes, and the relative uncertainties from the intramethods variation are represented by the shaded bands. Only statistical uncertainties are considered in all the fits.
    }
 \label{fig:OneOverR2_vs_mT}
\end{figure}

Finally, in hydrodynamic models, valuable information about the collective transverse expansion of the system (transverse flow) can be obtained from the slope of a linear fit to $1/R_{\text{inv}}^2$ versus the transverse mass, $\mT$. In addition, the value of $1/R_{\text{inv}}^2$ at $\mT = 0$ reflects the final-state geometrical size of the source. Figure~\ref{fig:OneOverR2_vs_mT} shows $1/R_{\text{inv}}^2$ versus $\mT$ for a variety of multiplicity ranges. The left plot shows that the expansion in the low-multiplicity region is faster than in the HM region. The corresponding geometrical sizes are $R_{\mathrm{G}}^{\mathrm{MB}} = 5.1 \pm 0.4\stat\unit{fm}$ and $R_{\mathrm{G}}^{\mathrm{HM}} = 4.2 \pm 1.1\stat\unit{fm}$, for the low and high multiplicity regions, respectively. The right panel of Fig.~\ref{fig:OneOverR2_vs_mT} shows the variations of $1/R_{\text{inv}}^2$ with $\mT$ in finer multiplicity ranges, showing in more detail the evolution of the slope: the collective flow decreases with increasing multiplicity, but this trend seems to saturate around a reconstructed multiplicity of ${\sim} 80$.

This dependence $R_{\text{inv}}^{-2}\propto a+b \mT$ (which was universally observed in nucleus--nucleus collisions, for different colliding system sizes, collision energy, and centrality) implies that the radiiare driven by the initial geometry, as well as by the transverse flow (in a 3D analysis, also by the longitudinal flow). The present data suggest that a similar phenomenological modeling  is appropriate for $\Pp\Pp$ collisions at LHC energies. Theoretical models indicate that the intercept of the linear fit to $R_{\text{inv}}^{-2}(\mT)$ versus $\mT$ equals the geometrical size at freeze-out, whereas the slope gives the square of a Hubble-type flow parameter~\cite{Csorgo-etal-hubble}, divided by the freeze-out temperature, \ie, $H^2/T_\mathrm{f}$~\cite{Makhlin:1987gm,Csorgo:1995bi}. Assuming $T_\mathrm{f}\sim 150\MeV$ for the freeze-out temperature, the values of the Hubble-type parameter can be estimated as $H_{_{\mathrm{HM}}} = 0.17 \pm 0.04\stat \unit{fm}^{-1}$ and $H_{_{\mathrm{MB}}} = 0.298 \pm 0.004\stat \unit{fm}^{-1}$ in the HM and MB regions, respectively.

These values are comparable to those estimated for RHIC AuAu collisions, \ie,  ${\sim} 0.1- 0.2 \unit{fm}^{-1}$ \cite{Adler:2004rq,Adams:2004yc,Abelev:2009tp,Adare:2015bcj}, and can also be converted into a flow velocity by multiplying by the measured size. Finally, it should be noted that the extracted Hubble-type parameter is larger in the MB case than in the HM case. These findings are again qualitatively consistent with the measurements in nucleus--nucleus collisions, where the slope parameter of $R_{\text{inv}}^{-2}$ vs $\mT$ usually shows larger Hubble-type parameters for peripheral than for central collisions. These observations suggest yet another similarity between HM events in high energy $\Pp\Pp$ collisions and relativistic nucleus--nucleus collisions.

\section{Summary}
\label{sec:summary}

A Bose--Einstein correlation measurement is reported using data collected with the CMS detector at the LHC in proton-proton collisions at $\sqrt{s}=13\TeV$, covering a broad range of charged particle multiplicity, from a few particles up to 250 reconstructed charged hadrons. Three analysis methods, each with a different dependence on Monte Carlo simulations, are used to generate correlation functions, which are found to give consistent results. One dimensional studies of the radius fit parameter, $\Ri$, and the intercept parameter, $\lambda$, have been carried out for both inclusive events and high multiplicity events selected using a dedicated online trigger. For multiplicities in the range $0 < \Ntroff < 250$ and average pair transverse momentum $0 < \kt < 1\GeV$, values of the radius fit parameter and intercept are in the ranges $0.8 < \Ri < 3.0\unit{fm}$ and $0.5 < \lambda \lesssim 1.0$, respectively.

Over most of the multiplicity range studied, the value of $\Ri$ increases with increasing event multiplicities and is proportional to $\langle\Nt\rangle^{1/3}$, a trend which is predicted by hydrodynamical calculations. For events with more than ${\sim}100$ charged particles, the observed dependence of $\Ri$ suggests a possible saturation, with the lengths of homogeneity also consistent with a constant value. Comparisons of the multiplicity dependence are made with predictions of the color glass condensate effective theory, by means of a parameterization of the radius of the system formed in $\Pp\Pp$ collisions. The values of the radius parameters in the model are much lower than those in the data, although the general shape of the dependence on multiplicity is similar in both cases. The radius fit parameter $\Ri$ is also observed to decrease with  increasing $\kt$, a behavior that is consistent with emission from a system that is expanding prior to its decoupling.

Inspired by hydrodynamic models, the dependence of $\Ri^{-2}$ on the average pair transverse mass was investigated and the two are observed to be proportional, a behavior similar to that seen in nucleus--nucleus collisions. The proportionality constant between $\Ri^{-2}$ and transverse mass can be related to the flow parameter of a Hubble-type expansion of the system. For $\Pp\Pp$ collisions at 13\TeV, this expansion is slower for larger event  multiplicity, a dependence that was also found in nucleus--nucleus collisions. Therefore, the present analysis reveals additional similarities between the systems produced in high multiplicity $\Pp\Pp$ collisions and those found using data for larger initial systems. These results may provide additional constraints on future attempts using hydrodynamical models and/or the color glass condensate framework to explain the entire range of similarities between $\Pp\Pp$ and heavy ion interactions.

\begin{acknowledgments}
We congratulate our colleagues in the CERN accelerator departments for the excellent performance of the LHC and thank the technical and administrative staffs at CERN and at other CMS institutes for their contributions to the success of the CMS effort. In addition, we gratefully acknowledge the computing centers and personnel of the Worldwide LHC Computing Grid for delivering so effectively the computing infrastructure essential to our analyses. Finally, we acknowledge the enduring support for the construction and operation of the LHC and the CMS detector provided by the following funding agencies: BMBWF and FWF (Austria); FNRS and FWO (Belgium); CNPq, CAPES, FAPERJ, FAPERGS, and FAPESP (Brazil); MES (Bulgaria); CERN; CAS, MoST, and NSFC (China); COLCIENCIAS (Colombia); MSES and CSF (Croatia); RPF (Cyprus); SENESCYT (Ecuador); MoER, ERC IUT, PUT and ERDF (Estonia); Academy of Finland, MEC, and HIP (Finland); CEA and CNRS/IN2P3 (France); BMBF, DFG, and HGF (Germany); GSRT (Greece); NKFIA (Hungary); DAE and DST (India); IPM (Iran); SFI (Ireland); INFN (Italy); MSIP and NRF (Republic of Korea); MES (Latvia); LAS (Lithuania); MOE and UM (Malaysia); BUAP, CINVESTAV, CONACYT, LNS, SEP, and UASLP-FAI (Mexico); MOS (Montenegro); MBIE (New Zealand); PAEC (Pakistan); MSHE and NSC (Poland); FCT (Portugal); JINR (Dubna); MON, RosAtom, RAS, RFBR, and NRC KI (Russia); MESTD (Serbia); SEIDI, CPAN, PCTI, and FEDER (Spain); MOSTR (Sri Lanka); Swiss Funding Agencies (Switzerland); MST (Taipei); ThEPCenter, IPST, STAR, and NSTDA (Thailand); TUBITAK and TAEK (Turkey); NASU (Ukraine); STFC (United Kingdom); DOE and NSF (USA).

\hyphenation{Rachada-pisek} Individuals have received support from the Marie-Curie program and the European Research Council and Horizon 2020 Grant, contract Nos.\ 675440, 752730, and 765710 (European Union); the Leventis Foundation; the A.P.\ Sloan Foundation; the Alexander von Humboldt Foundation; the Belgian Federal Science Policy Office; the Fonds pour la Formation \`a la Recherche dans l'Industrie et dans l'Agriculture (FRIA-Belgium); the Agentschap voor Innovatie door Wetenschap en Technologie (IWT-Belgium); the F.R.S.-FNRS and FWO (Belgium) under the ``Excellence of Science -- EOS" -- be.h project n.\ 30820817; the Beijing Municipal Science \& Technology Commission, No. Z181100004218003; the Ministry of Education, Youth and Sports (MEYS) of the Czech Republic; the Lend\"ulet (``Momentum") Program and the J\'anos Bolyai Research Scholarship of the Hungarian Academy of Sciences, the New National Excellence Program \'UNKP, the NKFIA research grants 123842, 123959, 124845, 124850, 125105, 128713, 128786, and 129058 (Hungary); the Council of Science and Industrial Research, India; the HOMING PLUS program of the Foundation for Polish Science, cofinanced from European Union, Regional Development Fund, the Mobility Plus program of the Ministry of Science and Higher Education, the National Science Center (Poland), contracts Harmonia 2014/14/M/ST2/00428, Opus 2014/13/B/ST2/02543, 2014/15/B/ST2/03998, and 2015/19/B/ST2/02861, Sonata-bis 2012/07/E/ST2/01406; the National Priorities Research Program by Qatar National Research Fund; the Ministry of Science and Education, grant no. 3.2989.2017 (Russia); the Programa Estatal de Fomento de la Investigaci{\'o}n Cient{\'i}fica y T{\'e}cnica de Excelencia Mar\'{\i}a de Maeztu, grant MDM-2015-0509 and the Programa Severo Ochoa del Principado de Asturias; the Thalis and Aristeia programs cofinanced by EU-ESF and the Greek NSRF; the Rachadapisek Sompot Fund for Postdoctoral Fellowship, Chulalongkorn University and the Chulalongkorn Academic into Its 2nd Century Project Advancement Project (Thailand); the Nvidia Corporation; the Welch Foundation, contract C-1845; and the Weston Havens Foundation (USA).
\end{acknowledgments}

\bibliography{auto_generated}

\clearpage
\appendix

\section{Double ratios and cluster subtraction techniques}
\label{sec:appendix-DR-CS}

\begin{figure}[hptb]
\centering
  \includegraphics[width=0.45\textwidth]{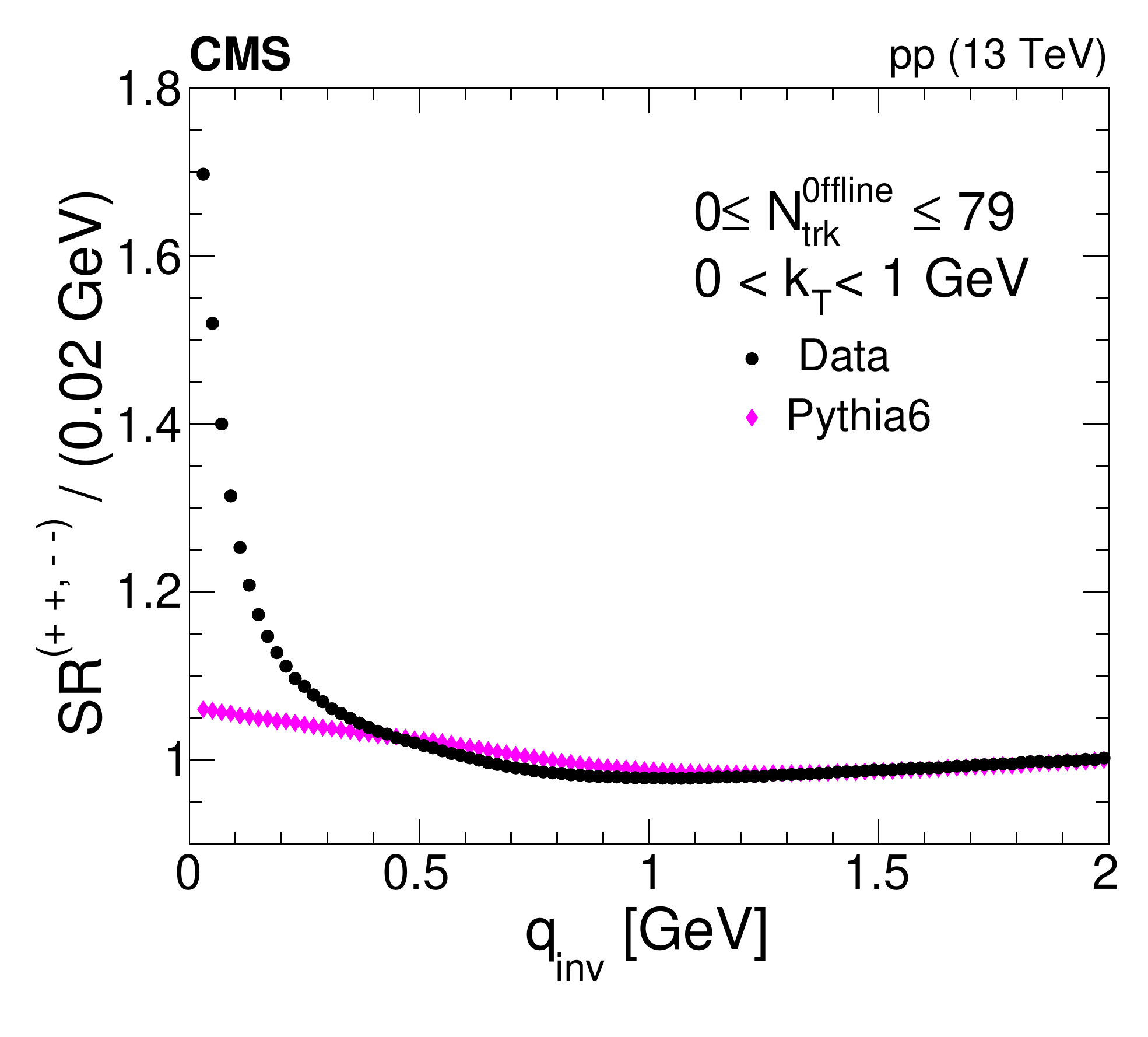}
  \includegraphics[width=0.45\textwidth]{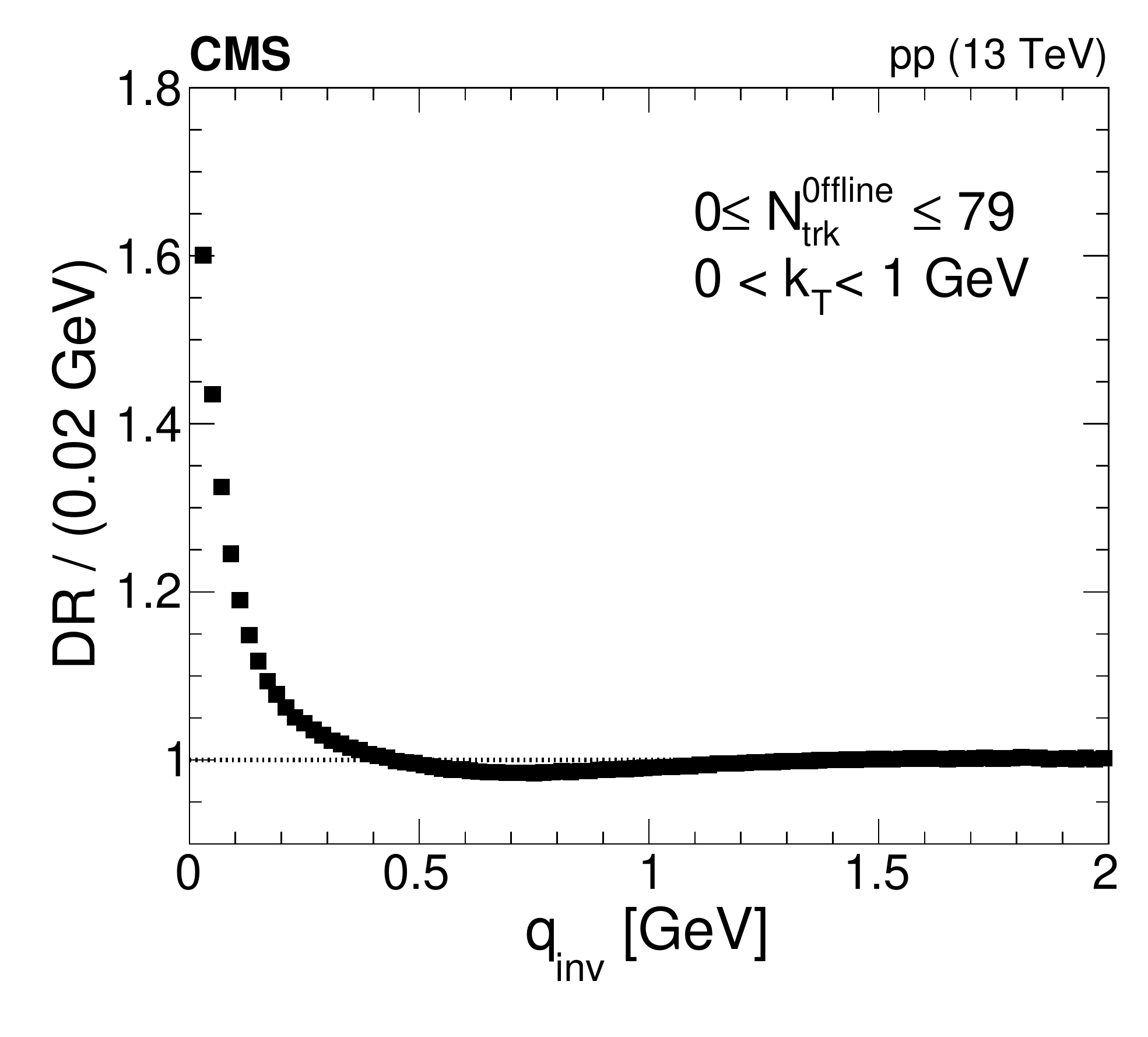}
  \caption{Illustration of the steps in the double ratio method. The single ratio in data is constructed (left), followed by a similar procedure with simulated events (\PYTHIA 6, Z2* tune). The ratio of the two curves on the left defines the double ratio (right). The reference sample is obtained with the $\eta$-mixing procedure. All results correspond to integrated values in $N^\text{offline}_\text{trk}$ and \kt.
  }
  \label{fig:illustration-single-double-ratios}
\end{figure}

The analysis procedure using the double ratio (DR) technique follows Refs.~\cite{cms-hbt-1st,cms-hbt-2nd, fsq-14-002}. For illustrating the procedure, Fig.~\ref{fig:illustration-single-double-ratios} shows the single ratio (SR) defined in data, the one defined in simulation, as well as the DR. The reference sample used is the $\eta$-mixing sample defined previously, and the MC tune adopted is \PYTHIA 6 Z2*. Both these choices are the default ones employed for obtaining the results in the DR method.

For obtaining the parameters of the BEC effect in this method, a DR is defined as
\begin{linenomath*}
\begin{equation}
 \text{DR} (\qi) \equiv C_{2, \text{BE}}(\qi) = \frac{\text{SR} (\qi)}{\text{SR} (\qi)_{\text{MC}}} =
   \frac{\left[\left(\frac{\mathcal{N}_{\text{ref}}}
                   {\mathcal{N}_{\text{sig}}}\right)
  \left(\frac{\rd N_\text{sig}/\rd\qi}
             {\rd N_{\text{ref}}/\rd\qi}\right)\right]}
  {\left[\left(\frac{\mathcal{N}_{\text{ref}}}
                   {\mathcal{N}_{\text{sig}}}\right)_{\text{MC}}
        \left(\frac{\rd N_{\text{MC}}/\rd\qi}
                   {\rd N_{\text{MC, ref}}/\rd\qi}\right)\right]},
\label{eq:doubleratio}
\end{equation}
\end{linenomath*}
\noindent where   $C_{2, \text{BE}}$ refers to the two-particle BEC, $\text{SR} (\qi)_{\text{MC}}$ is the SR in Eq.~(\ref{eq:1-d-singleratios-gen}), but computed with simulated events without BEC effects. In each case, the reference samples for data and simulation are constructed in the same way by considering all charged particles instead of only charged pions in the MC. 

The DR method was used in Refs.~\cite{cms-hbt-1st,cms-hbt-2nd,fsq-14-002} to reduce the bias due to the construction of the reference sample. The DR technique also has the advantage of reducing the non-BEC background that could remain in the SR (\ie, correlations coming from resonance decays and from energy-momentum conservation are not included in the reference sample, which is constructed with the event mixing technique adopted throughout this analysis). Therefore, by selecting a MC simulation that describes well the general properties of the data, those residual correlations can, in principle, be removed by taking the DRs with an SR defined similarly in MC events from non-BEC contributions \cite{ZEUS,L3}).

\begin{figure}[hptb]
\centering
    \includegraphics[width=0.45\textwidth]{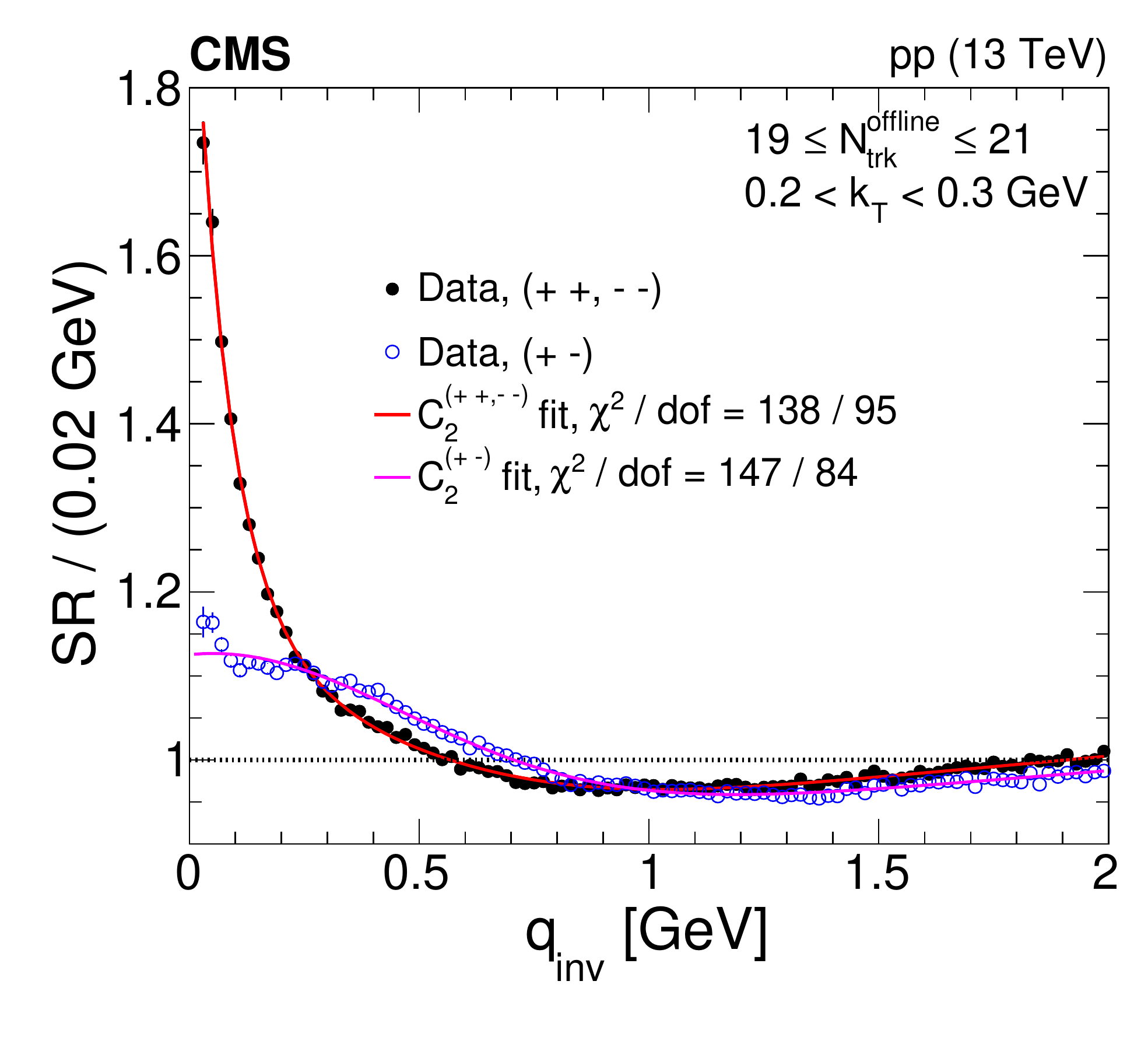}
    \includegraphics[width=0.45\textwidth]{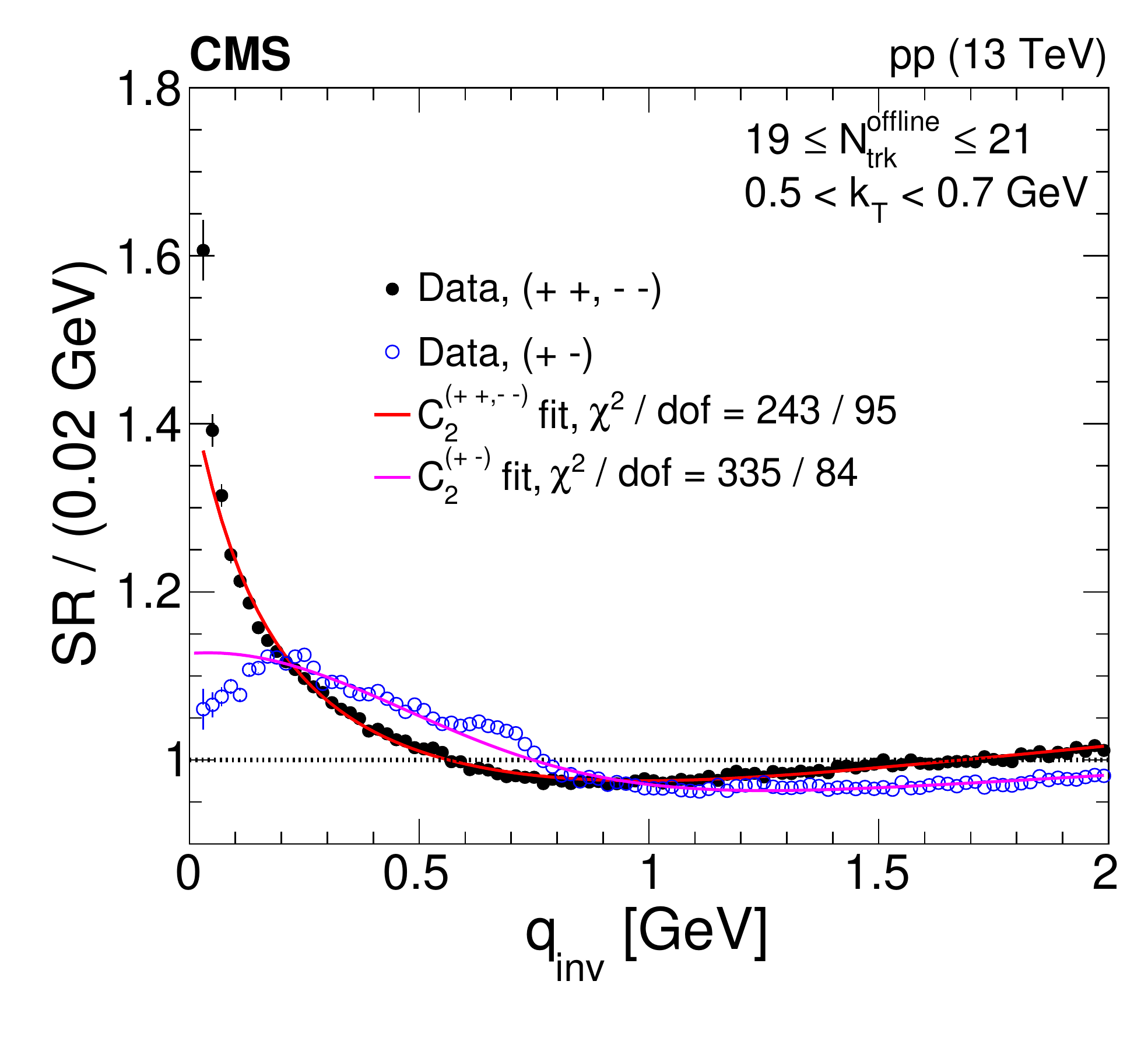} \\
    \includegraphics[width=0.45\textwidth]{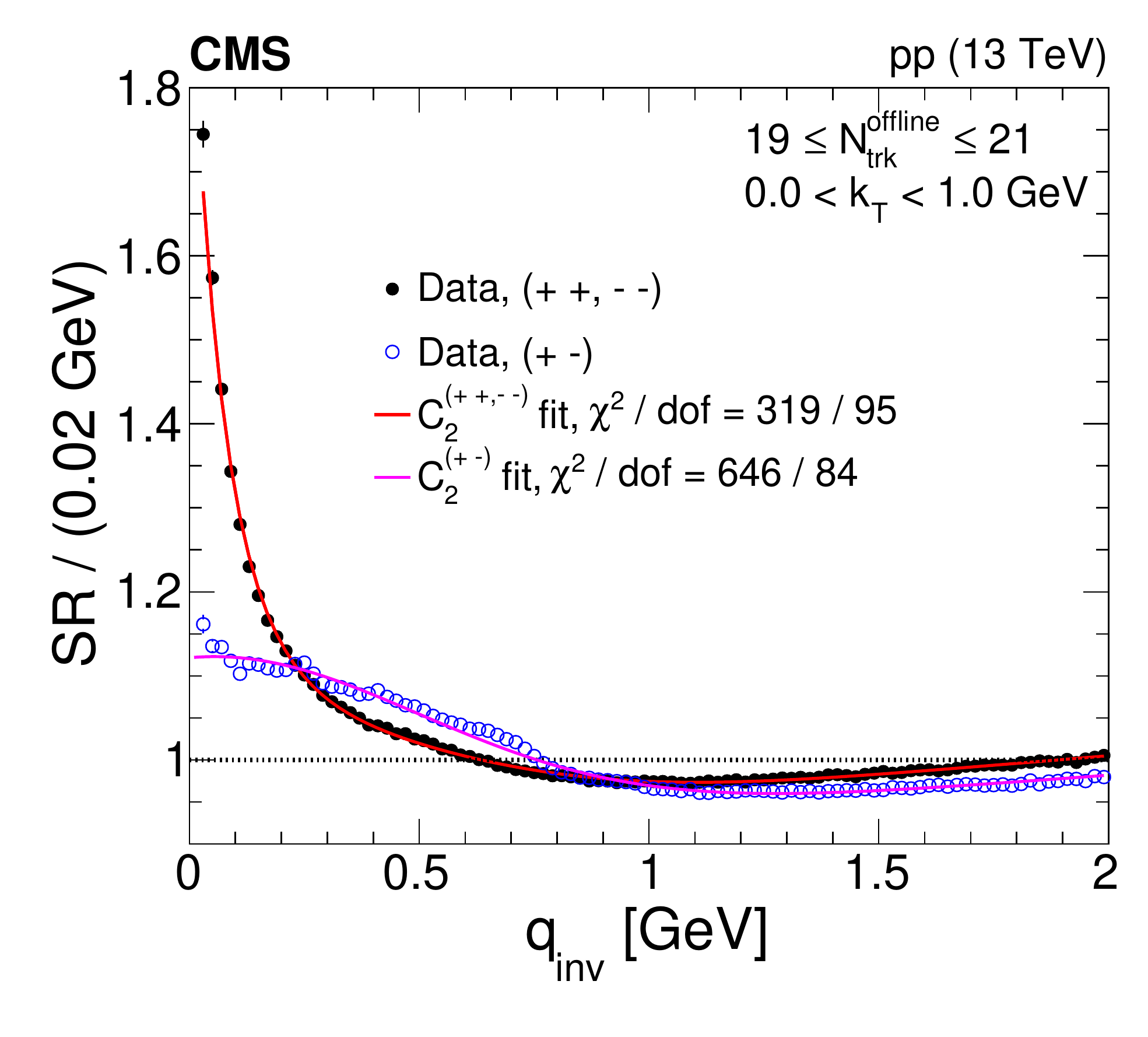}
    \includegraphics[width=0.45\textwidth]{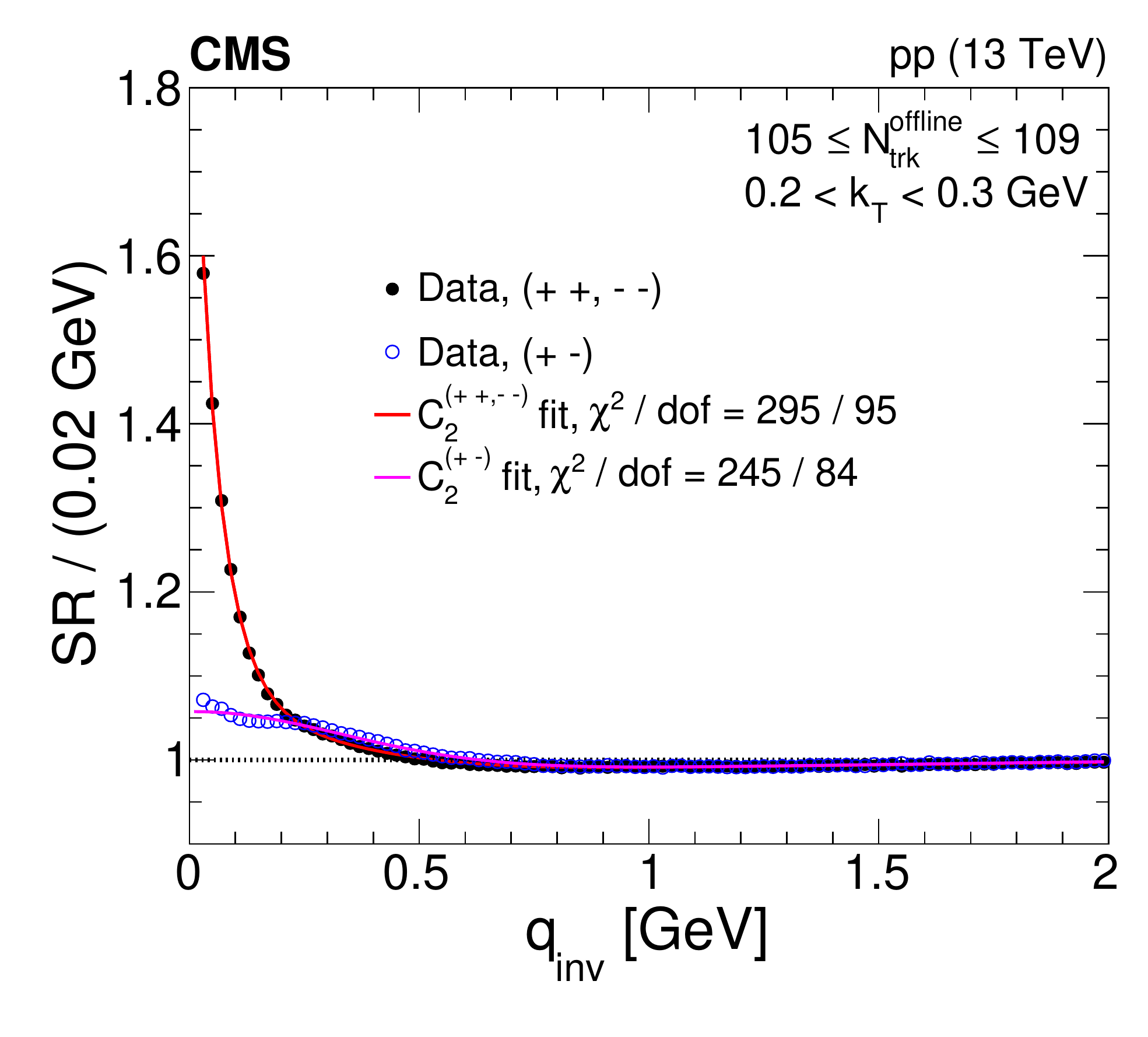} \\
    \includegraphics[width=0.45\textwidth]{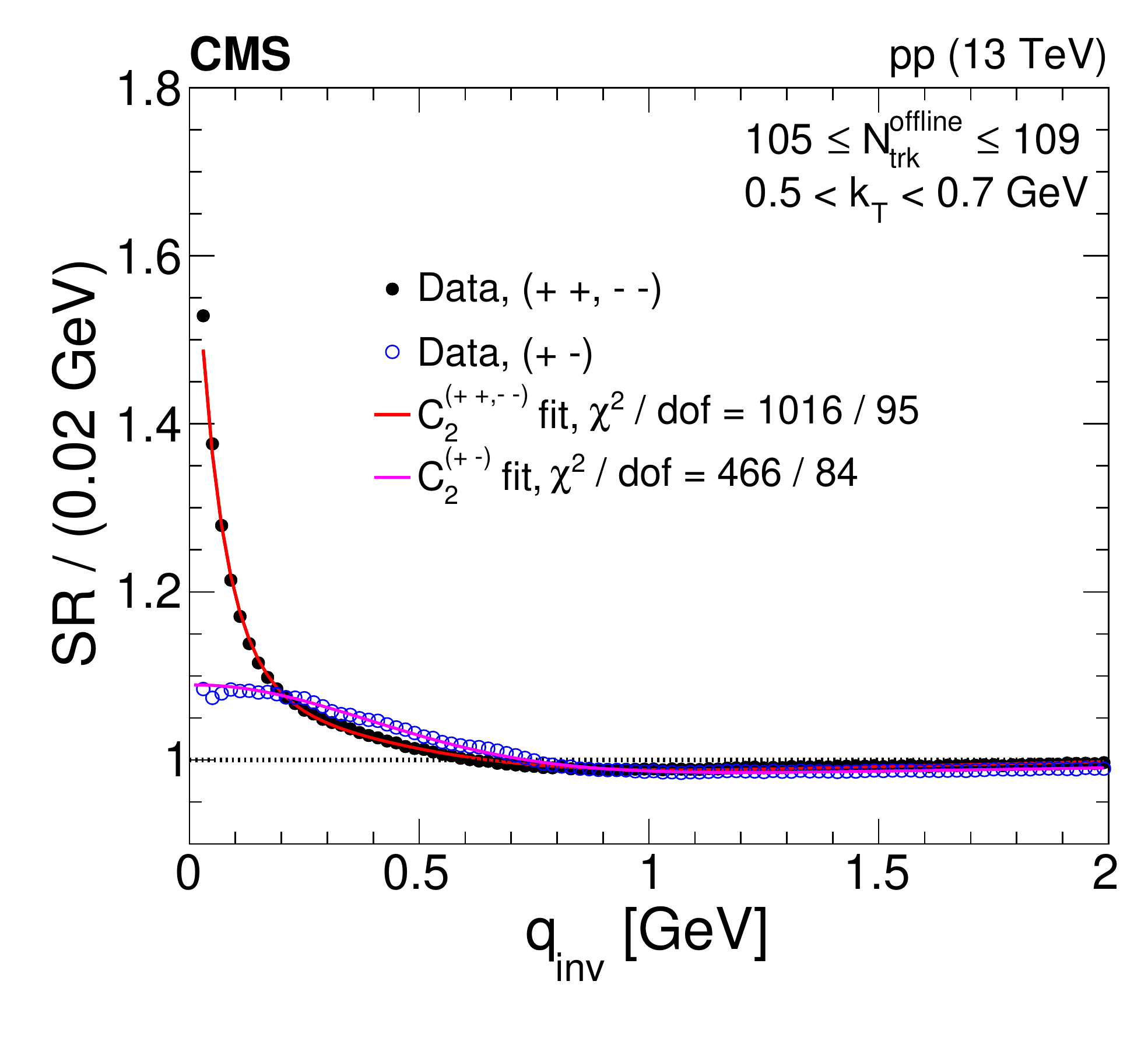}
    \includegraphics[width=0.45\textwidth]{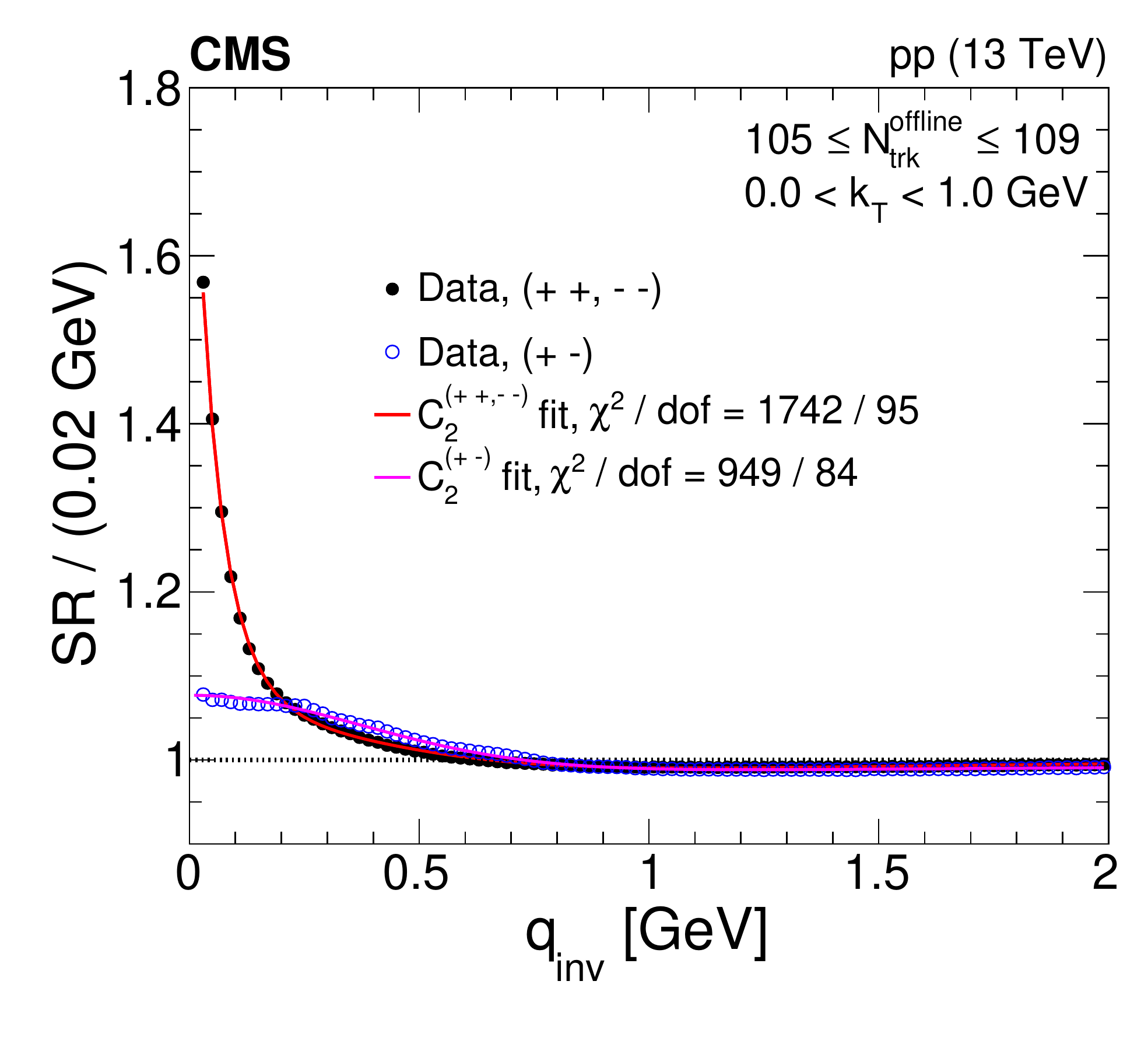}
  \caption{ The same-sign ($++,--$) and opposite-sign ($+-$) single ratios are shown in different $\Ntroff$ and $\kt$ bins, together with the full fits (continuous curves) given in Eqs.~(\ref{eq:minijet}) and (\ref{eq:pid_c2_1d}), for minimum-bias ($19 \le \Ntroff \le 21$) and HM  ($105 \le \Ntroff \le 109$) events.  The error bars represent statistical uncertainties and in most cases are smaller than the marker size.
  }
  \label{fig:same_sign_minijet_vs_Ntroffec_and_kt}
\end{figure}

The CS method, as described in Ref.~\cite{fsq-14-002}, employs a different approach by dealing with only SRs, where contaminations (called ``cluster contribution'') from resonances and jet fragmentation are still present~\cite{Alver:2008aa,Ridge1,Ridge4}. For the purpose of evaluating and removing such cluster effects, the one-dimensional opposite-sign (OS) SR correlation functions are fitted with the expression in Eq.~(\ref{eq:minijet}), which  describes the data in all $\kt$ and $\Ntroff$ ranges (examples illustrating such fits are shown in Fig.~\ref{fig:same_sign_minijet_vs_Ntroffec_and_kt}),
\begin{linenomath*}
\begin{equation}
  C_2^{(+-)} (\qi) = c \left[1 + \frac{b}{\sigma_b \sqrt{2\pi}}
    \exp\left(- \frac{\qi^2}{2\sigma_b^2}\right)\right] (1+ \epsilon \qi ),
\label{eq:minijet}
\end{equation}
\end{linenomath*}
\noindent where $b$ and $\sigma_b$ are $\Ntroff$- and \kt- dependent parameters, and $c$ is an overall normalization constant, and $C_2^{(+-)} (\qi)$ refers to the opposite-sign correlation function. To avoid regions with substantial resonance contamination in the OS correlation function, the ranges $0.60 < \qi < 0.80\GeV$ and $\qi < 0.04\GeV$ are not used in the fits due to the \Pgr and photon conversion contributions, respectively; $b$ and $\sigma_b$ are parametrized as in Eqs.~(\ref{eq:sigma_b}) and~(\ref{eq:b-andsigmab}), respectively. Results of the fits using these parameterizations are shown in Table~\ref{tab:minijet1D_pars}.
\begin{linenomath*}
\begin{equation}
 b(\Ntroff,\kt) = \frac{b_0}{\big({\Ntroff}\big)^{n_b}} \exp\left(\frac{\kt}{k_0}\right) ;
  \label{eq:sigma_b}
\end{equation}
\end{linenomath*}
\begin{linenomath*}
\begin{equation}
  \sigma_b(\Ntroff,\kt) =
  \left[\sigma_0 + \sigma_1 \exp\left(-\frac{\Ntroff}{N_0}\right)\right] \kt^{n_\mathrm{T}} .
  \label{eq:b-andsigmab}
\end{equation}
\end{linenomath*}
\begin{table}[h]
  \caption{Values of the fit parameters from Eqs.~(\ref{eq:sigma_b}) and~(\ref{eq:b-andsigmab}), describing the cluster contribution in the data OS correlation function. The estimated uncertainty in the parameters is about 10\%.}
  \label{tab:minijet1D_pars}
  \begin{center}
    \begin{tabular}{ccccccc}
      \hline
       $b_0$ & $n_b$ & $k_0$ & $\sigma_0$ & $\sigma_1$ & $N_0$ & $n_{\mathrm{T}}$ \\
      \hline
       $ 0.90  $ & $ 0.55 $ & $ 0.8  $ & $ 0.35  $ & $  0.30  $ & $ 64  $ & $ 0.081 $  \\
      \hline
    \end{tabular}
  \end{center}
\end{table}

Once the values of $b$ and $\sigma_b$ are determined from OS SR, the cluster contamination in the SS SR correlation function can be estimated. However, an important element, the conservation of electric charge in both minijet  and multibody resonance decays, results in a stronger correlation (reflected in the parameter b), for the OS pairs than for the corresponding SS ones. Therefore, the form of the cluster-related contribution obtained from OS pairs is used to fit the SS correlations, but multiplied by an amplitude $z(\Ntroff, \kt)$.

The expression in Eq.~(\ref{eq:pid_c2_1d}) is first used to fit the same-sign (SS) SRs for finding $z(\Ntroff, \kt)$. The values obtained for the cluster amplitude are fitted  for each $\kt$ bin by a theoretically-motivated parametrization $[ (a_{1} \Ntroff + b_{1})/(1 + a_{1}\Ntroff +b_{1}) ]$~\cite{fsq-14-002}, based on the ratio of the combinatorics of SS and OS particle pairs. Finally, this parametrization is employed in Eq.~(\ref{eq:pid_c2_1d}) for expressing  $z(\Ntroff, \kt)$ and refitting the SRs. For this fit, all the other variables (\ie, $b$, $\sigma_b$, $z$), given by the parameters obtained in earlier steps of the procedure, are kept fixed and only the overall normalization, $c$, and the parameters of the BEC function, $C_\text{BEC}(\qi)$, are fitted. In Fig.~\ref{fig:same_sign_minijet_vs_Ntroffec_and_kt} some examples of correlation functions with the respective full fits, as in Eq.(\ref{eq:pid_c2_1d}), are shown:
\begin{linenomath*}
\begin{equation}
  C_2^{(++,--)}(\qi) = c \left[1 + z(\Ntroff, \kt) \frac{b}{\sigma_b \sqrt{2\pi}}
    \exp\left(- \frac{\qi^2}{2\sigma_b^2}\right)\right] C_{2, \text{BE}}(\qi),
 \label{eq:pid_c2_1d}
\end{equation}
\end{linenomath*}
where  $C_2^{(++,--)} (\qi)$  and $ C_{2, \text{BE}} (\qi)$ refer to the same-sign correlation function and to the BEC, respectively.

\section{Investigation of an observed anticorrelation}
\label{sec:appendix-dip}

In previous CMS analyses~\cite{cms-hbt-2nd,fsq-14-002}, the presence of an anticorrelation (dip) in the BEC functions was reported in $\Pp\Pp$ collisions with characteristics that did not show a clear dependence on the center-of-mass energy. The DR technique is used for studying this behavior for $\Pp\Pp$ collisions at 13\TeV since the two methods based on control samples in data include both the BEC and non-BEC contributions together in the fits, making it harder to disentangle these two components.

\begin{figure}[hptb]
\centering
    \includegraphics[width=0.45\textwidth]{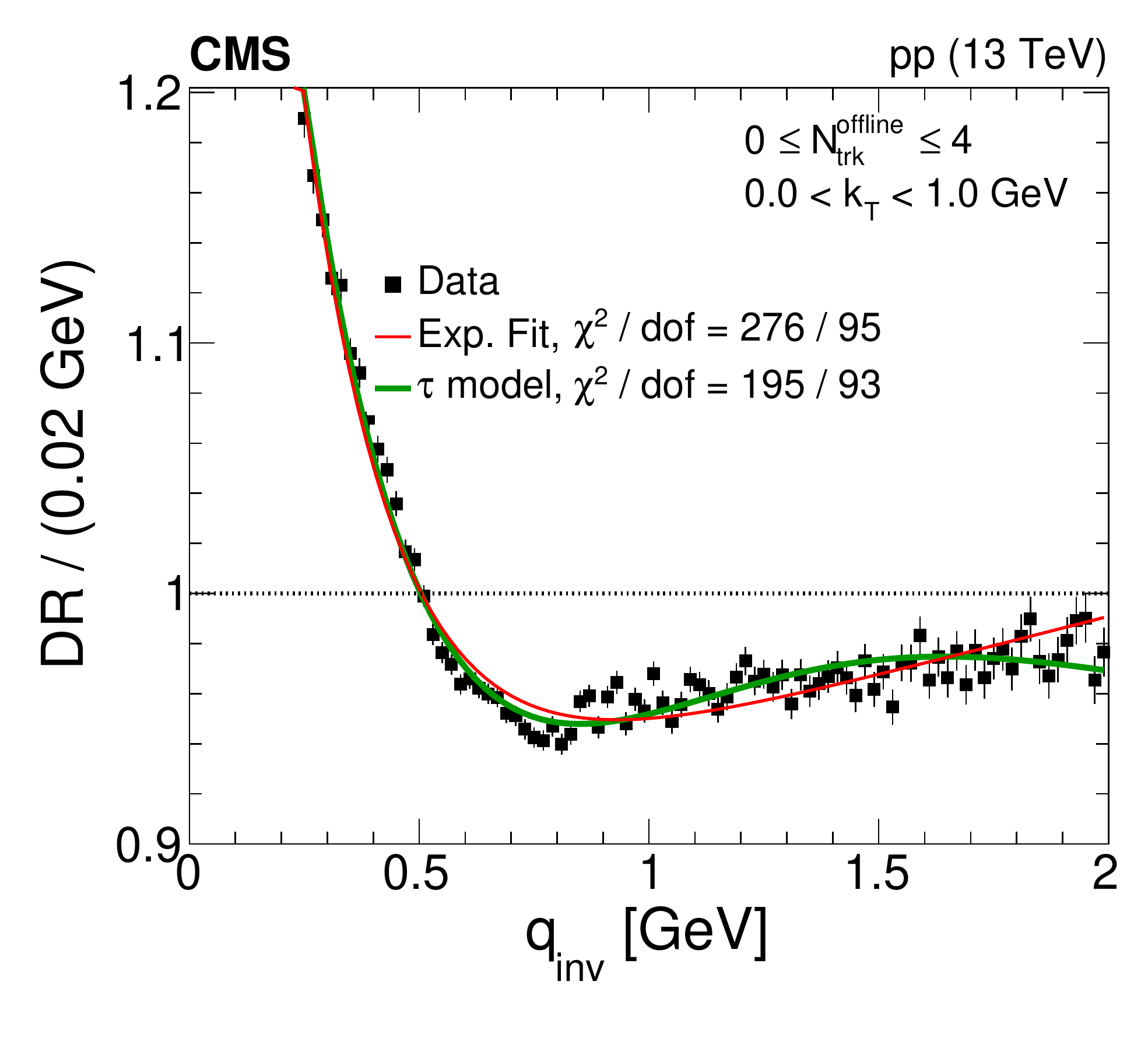}
    \includegraphics[width=0.45\textwidth]{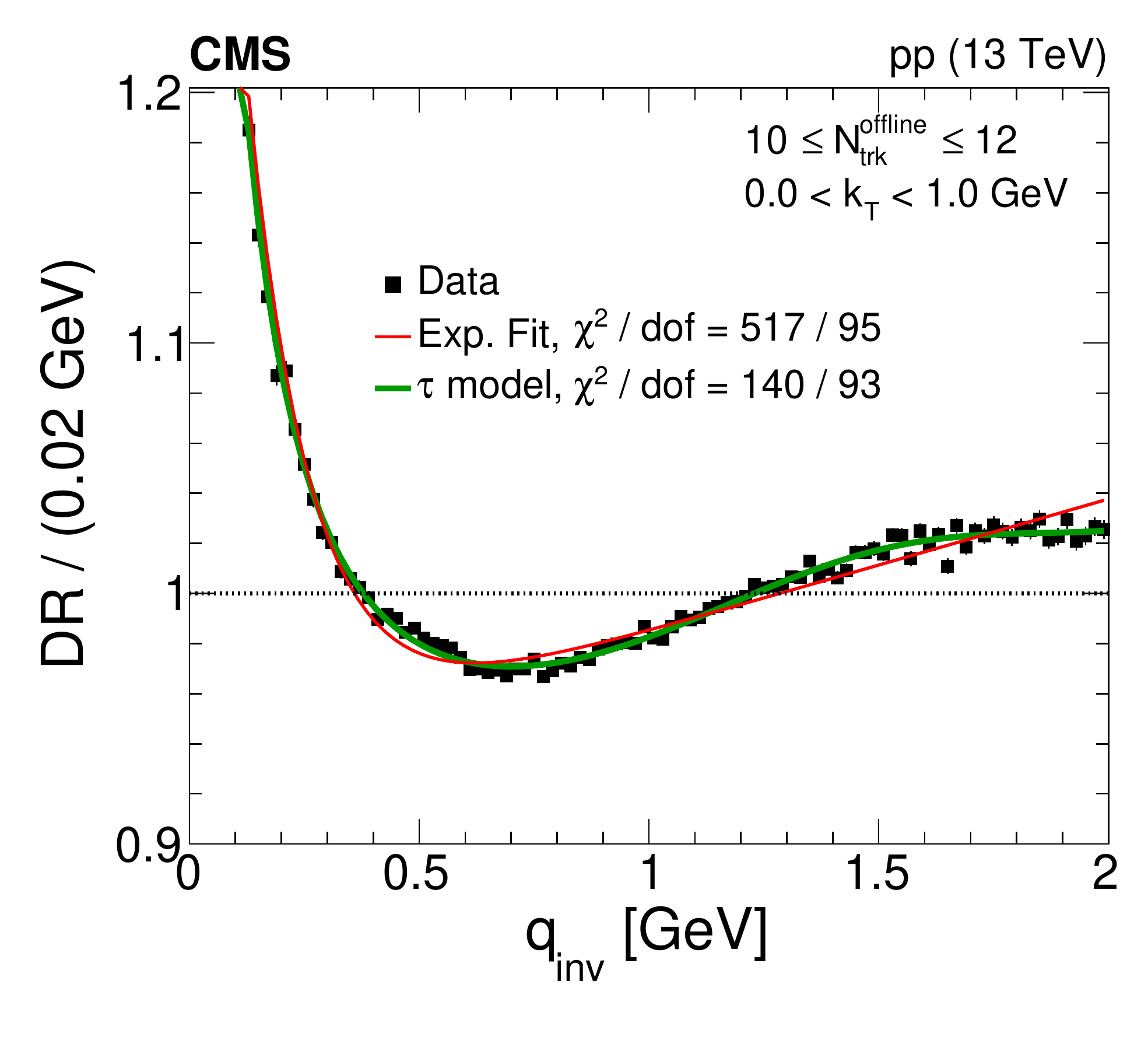} \\
    \includegraphics[width=0.45\textwidth]{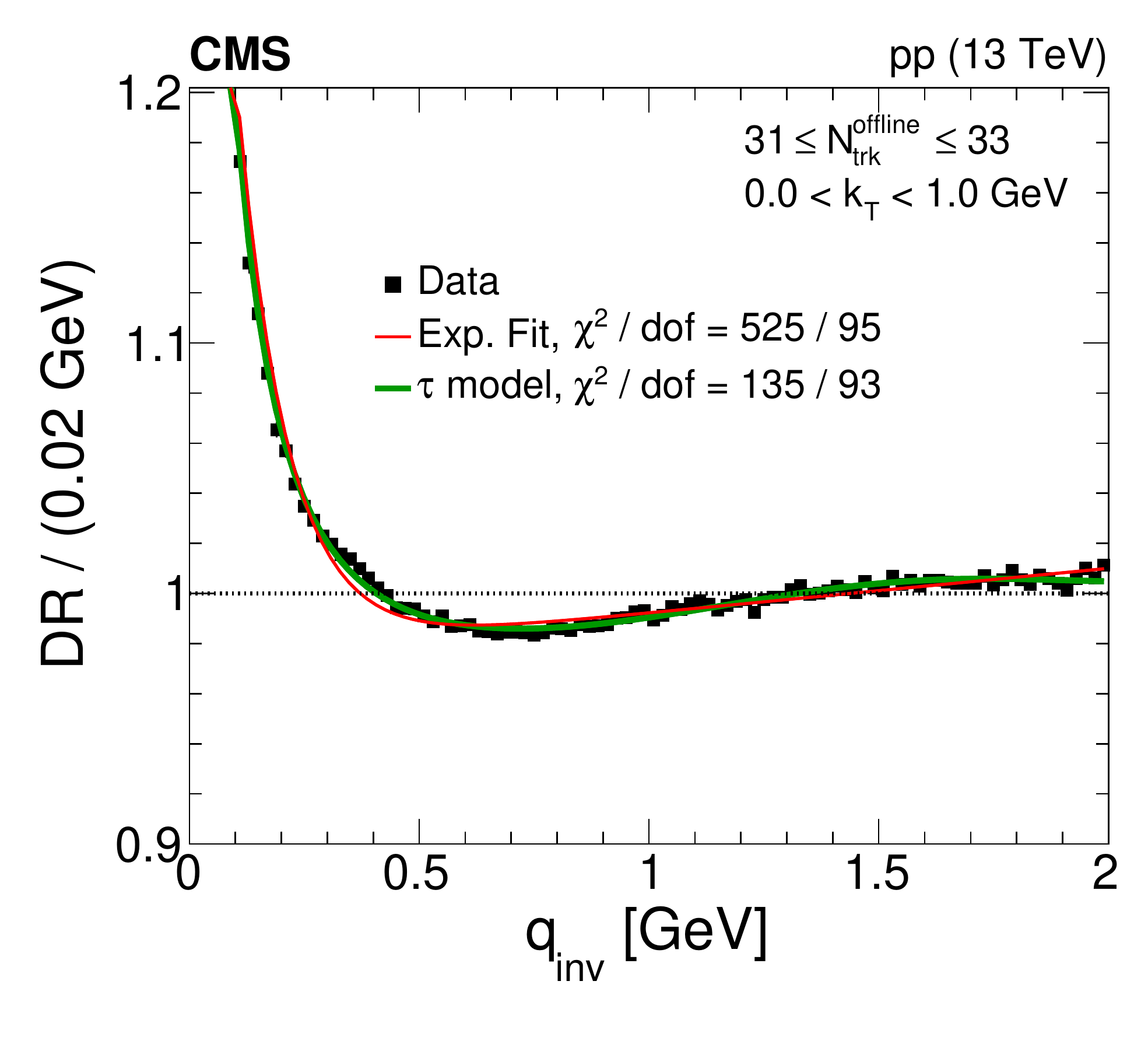}
    \includegraphics[width=0.45\textwidth]{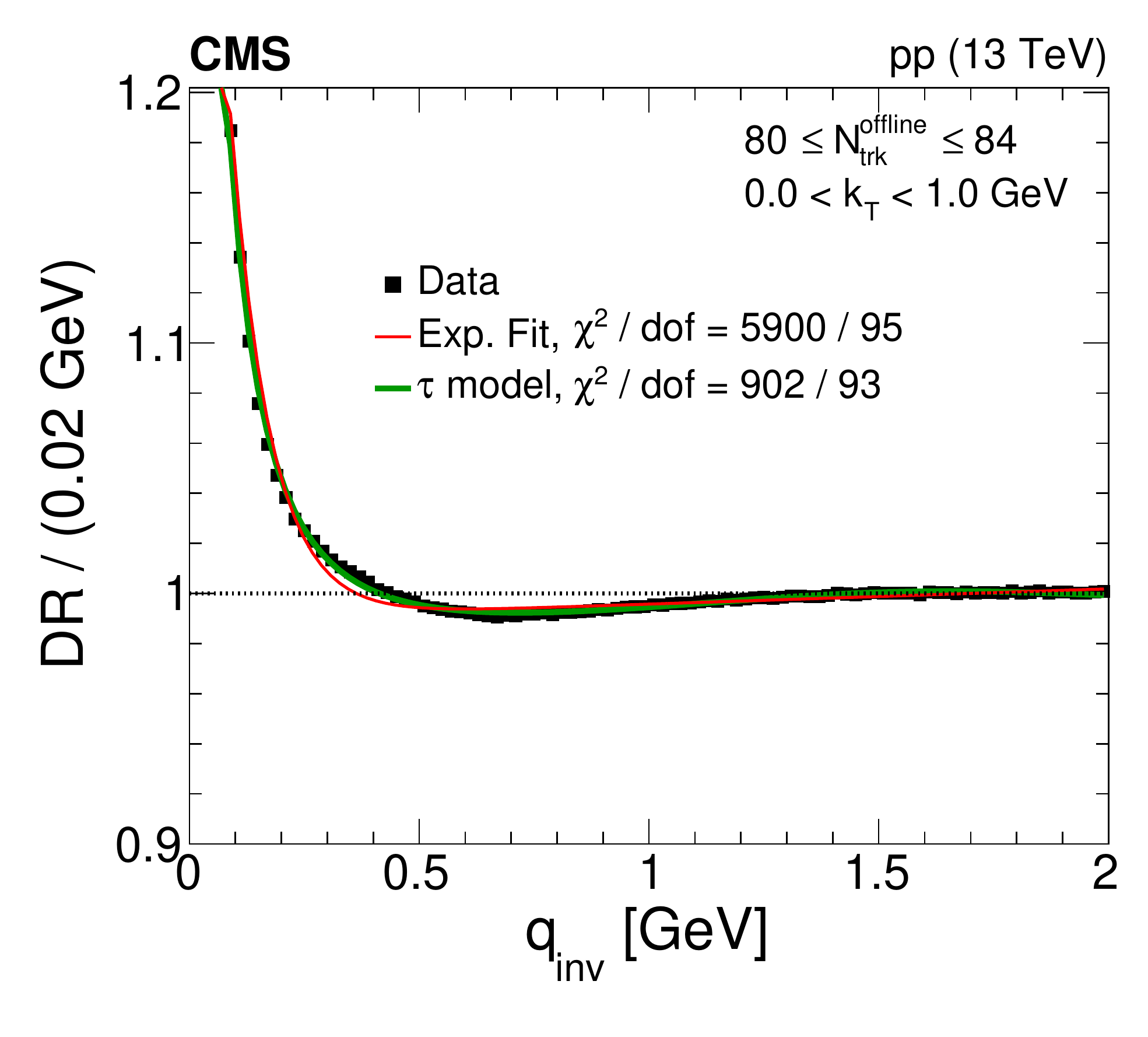} \\
    \includegraphics[width=0.45\textwidth]{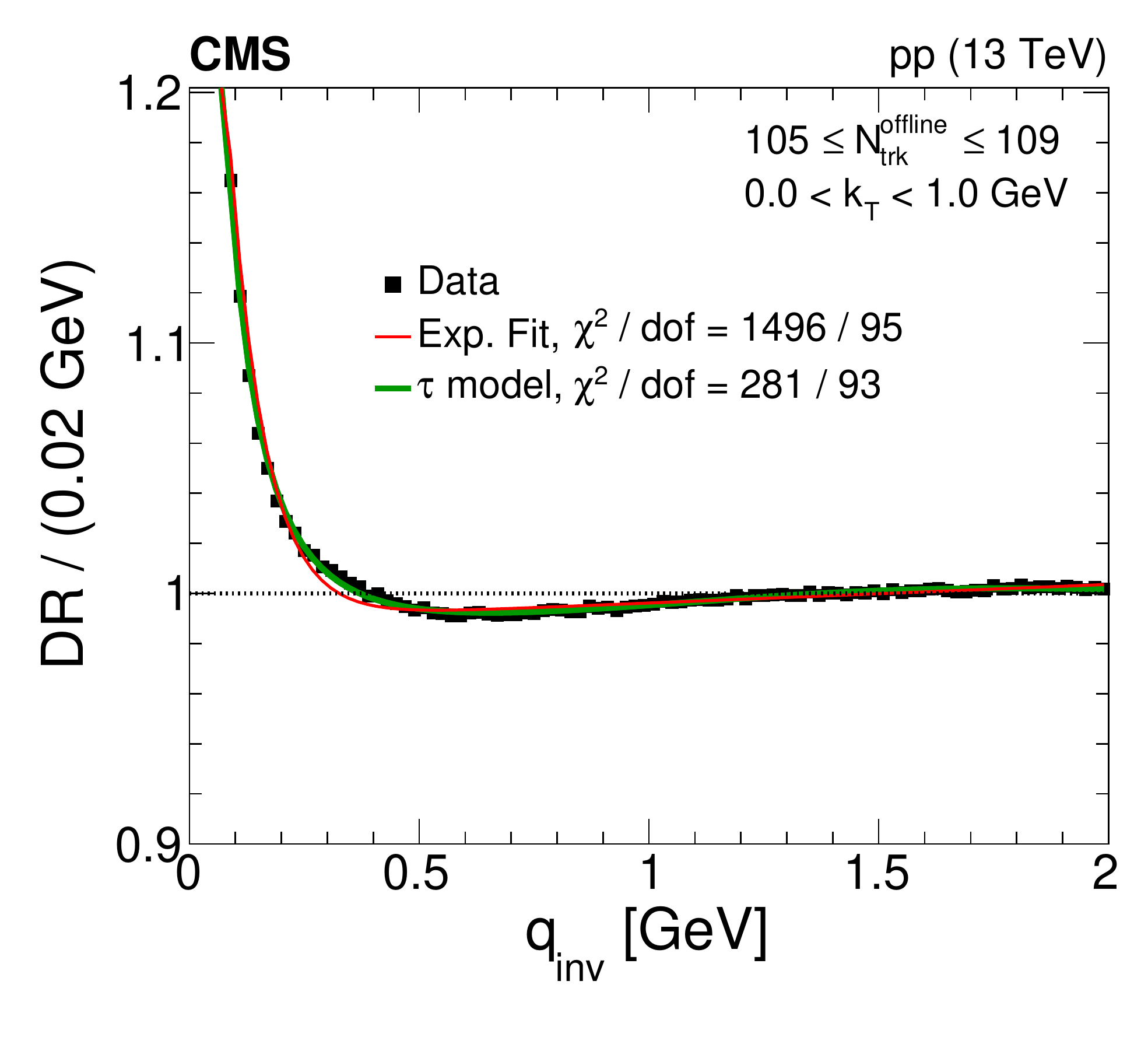}
    \includegraphics[width=0.45\textwidth]{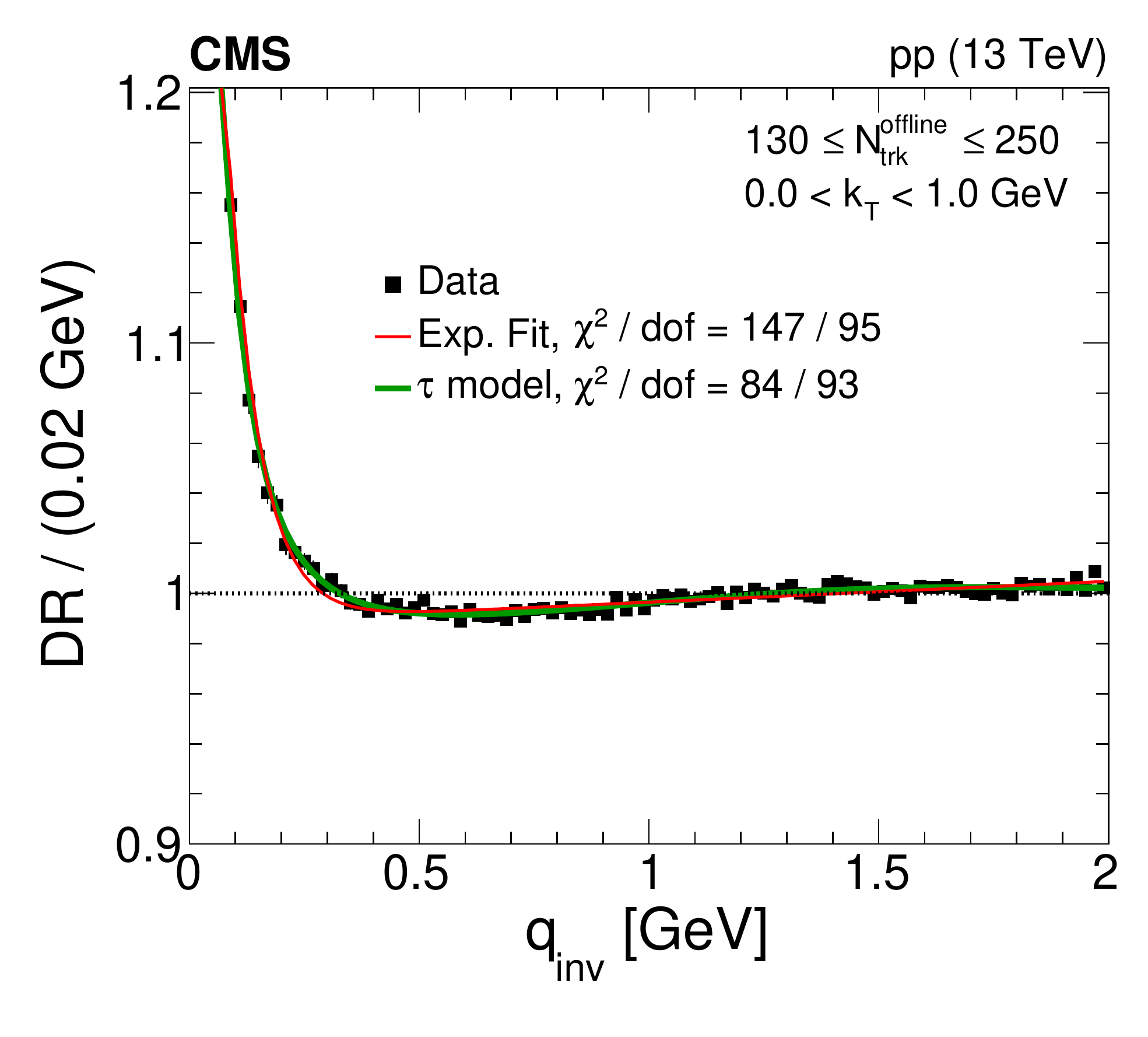}
   \caption{Correlation functions from the double ratio technique, integrated in the
  range $0< \kt < 1\GeV$, in six multiplicity bins. The results are zoomed along the vertical axis. The error bars represent statistical uncertainties and in most cases are smaller than the marker size.
   }
  \label{fig:selected-doubleratios-mb-hm}
\end{figure}

\begin{figure}[hptb]
  \centering
    \includegraphics[width=0.45\textwidth]{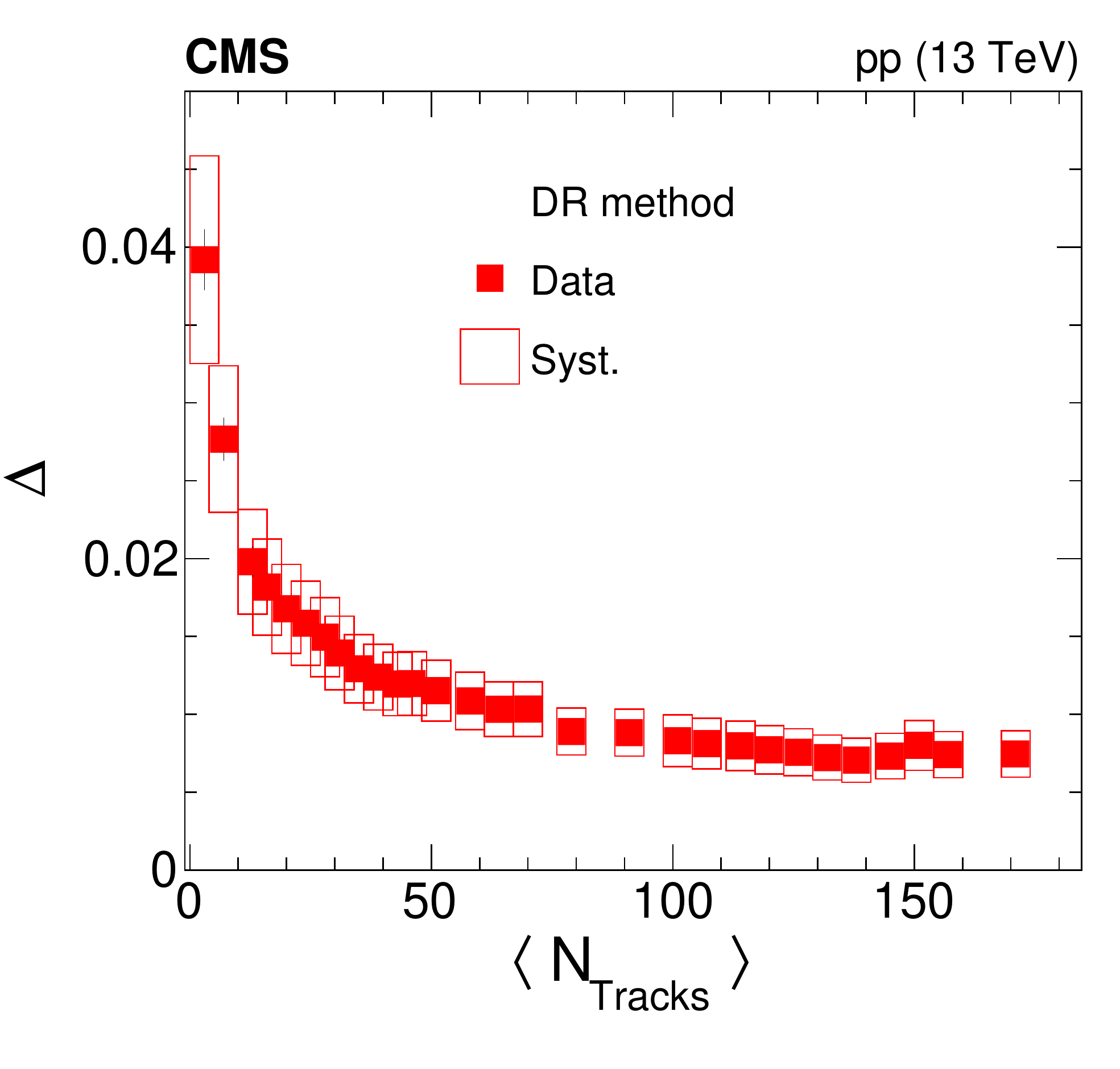}
    \includegraphics[width=0.45\textwidth]{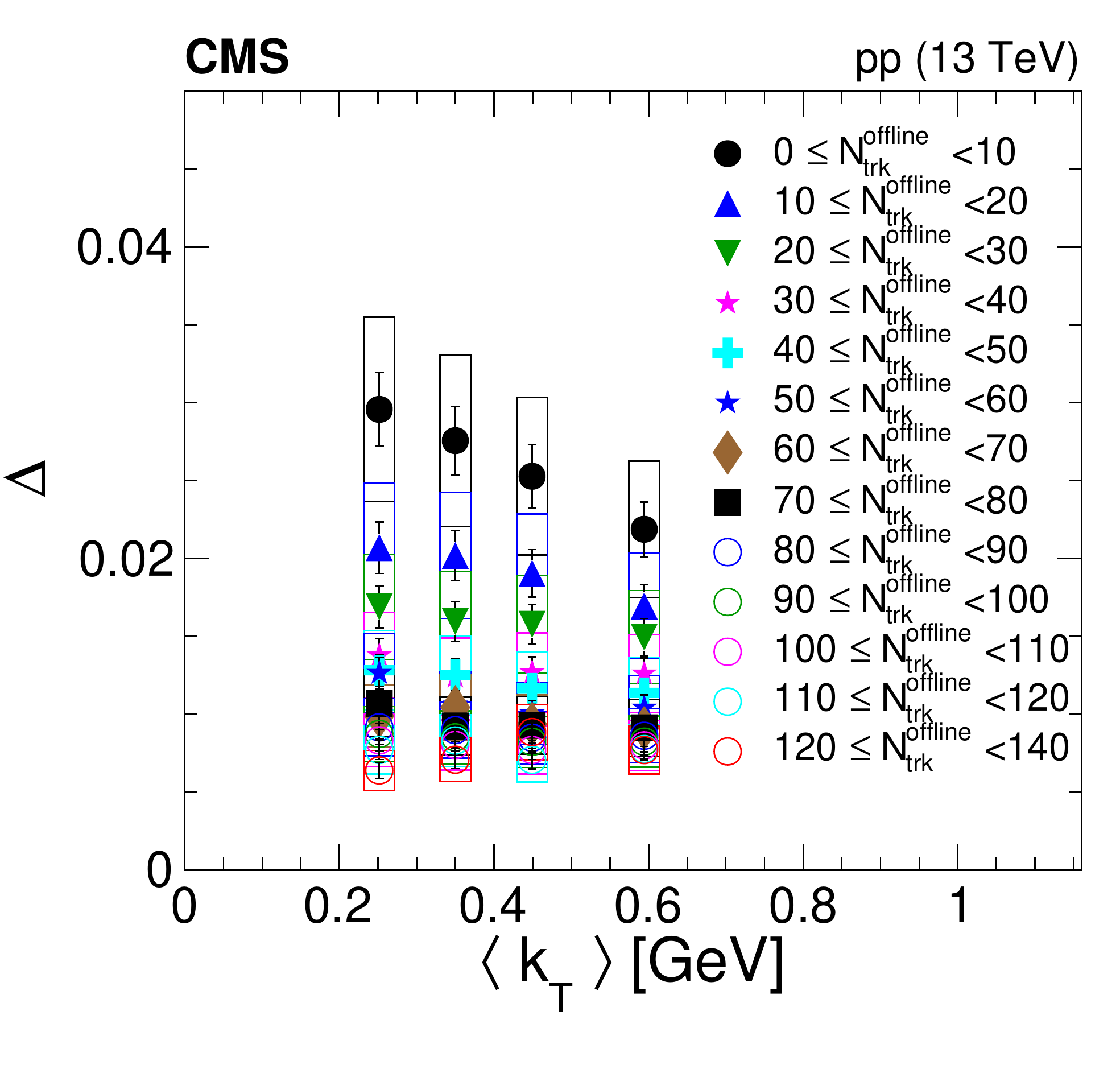}
  \caption{The depth of the anticorrelation  $\Delta$ is shown as a function of multiplicity (left) for $\kt$-integrated values. The fit parameter $\Delta$ is also shown in finer bins of $\Nt$ and $\kt$ (right). The statistical uncertainties are represented by the error bars, while the systematic ones are represented by the open boxes.
      }
  \label{fig:dr-dip-and-depth}
\end{figure}

In Fig.~\ref{fig:selected-doubleratios-mb-hm}, the DRs (zoomed along the correlation function axis) are illustrated in six ranges of multiplicity, for increasing values of $\Ntroff$, ranging from MB to HM events. An anticorrelation is also observed in this case, being more pronounced in the lower $\Ntroff$ bins. Prior to its observation in $\Pp\Pp$ collisions, this anticorrelation had been seen in $\Pep\Pem$ collisions~\cite{L3}, with features compatible with a description provided by the $\tau$-model~\cite{Htau}, in which particle production has a broad distribution in proper time and the phase space distribution of the emitted particles is dominated by strong correlations of the space-time coordinate and momentum components. Thus, this observation in MB $\Pp\Pp$ collisions suggests that such structure could be associated with small systems. More details and related discussions can be found in Ref.~\cite{fsq-14-002}.

The plots in Fig.~\ref{fig:selected-doubleratios-mb-hm} show the data points together with the exponential fit (continuous red curve) and a fit based on the $\tau$-model (continuous green curve), which better describes the shape of the dip. Nevertheless, it should be noted that the $\chi^2$ from both the $\tau$-model and the exponential fits are large for some multiplicity bins, favoring neither one of these descriptions. On the other hand, this could also reflect the fact that the uncertainties coming from the choice of MC models are not included in the fits.

The depth of the anticorrelation, $\Delta$, can be quantified \cite{cms-hbt-2nd, fsq-14-002} with respect to the baseline represented by the polynomial form $C (1 + \epsilon \; q)$, as in Eq.~(\ref{eq:1d-levy}), and the value of the curve corresponding to the $\tau$-model fit at its minimum (details can be found in Ref.~\cite{fsq-14-002}). The corresponding results are shown in Fig.~\ref{fig:dr-dip-and-depth}. The plot on the left shows the variation of $\Delta$ as a function of $\Nt$, for integrated values of $\kt$. The depth of the dip decreases with multiplicity and suggest an approach to a constant value above $\langle \Nt \rangle\sim 120 $. The behavior of the depth is shown as a function of $\kt$ in the right plot, for several $\Ntroff$ bins. In the lowest multiplicity bin, a clear decrease with $\kt$ is seen, but the slope decreases as $\Ntroff$ increases. The results for $60 < \Ntroff < 80$ show a weak $\kt$ dependence and the values of $\Delta$ are almost constant for $80 < \Ntroff < 140$.

The fact that the depth of the dip, although small, seems to tend to a constant value different from zero at the highest measured multiplicities raises the question of this effect being a consequence of the DR method or an intrinsic characteristic of the collision system that could keep memory of its initially small size, even at the highest track multiplicities produced so far in $\Pp\Pp$ collisions.

\cleardoublepage \section{The CMS Collaboration \label{app:collab}}\begin{sloppypar}\hyphenpenalty=5000\widowpenalty=500\clubpenalty=5000\vskip\cmsinstskip
\textbf{Yerevan Physics Institute, Yerevan, Armenia}\\*[0pt]
A.M.~Sirunyan, A.~Tumasyan
\vskip\cmsinstskip
\textbf{Institut f\"{u}r Hochenergiephysik, Wien, Austria}\\*[0pt]
W.~Adam, F.~Ambrogi, E.~Asilar, T.~Bergauer, J.~Brandstetter, M.~Dragicevic, J.~Er\"{o}, A.~Escalante~Del~Valle, M.~Flechl, R.~Fr\"{u}hwirth\cmsAuthorMark{1}, V.M.~Ghete, J.~Hrubec, M.~Jeitler\cmsAuthorMark{1}, N.~Krammer, I.~Kr\"{a}tschmer, D.~Liko, T.~Madlener, I.~Mikulec, N.~Rad, H.~Rohringer, J.~Schieck\cmsAuthorMark{1}, R.~Sch\"{o}fbeck, M.~Spanring, D.~Spitzbart, A.~Taurok, W.~Waltenberger, J.~Wittmann, C.-E.~Wulz\cmsAuthorMark{1}, M.~Zarucki
% ...
\vskip\cmsinstskip
% ...
1:  Also at Vienna University of Technology, Vienna, Austria\\
% ...
\end{sloppypar}
\end{document}